\documentclass[aps,twocolumn,nofootinbib,showpacs,final,balancelastpage,superscriptaddress]{revtex4}
\usepackage{graphicx}
\usepackage{tabularx}
\usepackage{amssymb}
\usepackage{amsmath}
\usepackage{amsbsy}
\usepackage{comment}
\usepackage{mathrsfs}
\usepackage{dsfont}
\usepackage{nicefrac}
\usepackage{wrapfig}
\usepackage{amsfonts}
\usepackage{url} % link package
\usepackage[nolist]{acronym}
\usepackage{nicefrac}
\usepackage{subfigure}
\usepackage{epsf,color,colordvi,pifont}
\usepackage{showkeys}
\def\deltabar{{\mathchar'26\mkern-8mu\delta}}
\def\dbar{{\mathchar'26\mkern-12mu d}}
\DeclareFontFamily{OT1}{pzc}{}
\DeclareFontShape{OT1}{pzc}{m}{it}{<-> s * [1.10] pzcmi7t}{}
\DeclareMathAlphabet{\mathpzc}{OT1}{pzc}{m}{it}
\def\dbar{{\mathchar'26\mkern-12mu d}}
\begin{document}

\title{Criticality and dominance of axion physics in highly magnetized vacuum}
\author{S. \surname{Villalba-Ch\'avez}}
\email{selym@tp1.uni-duesseldorf.de}
\affiliation{Institut f\"{u}r Theoretische Physik I, Heinrich-Heine-Universit\"{a}t D\"{u}%
sseldorf, Universit\"{a}tsstra\ss {e} 1, 40225 D\"{u}sseldorf, Germany.}
\author{A.~E.~\surname{Shabad}}
\email{shabad@lpi.ru}
\affiliation{P. N. Lebedev Physics Institute, Moscow 117924, Russia.}
\affiliation{Tomsk State University, Lenin Prospekt 36, Tomsk 634050, Russia.}
\author{C. \surname{M\"{u}ller}}
\email{c.mueller@tp1.uni-duesseldorf.de}
\affiliation{Institut f\"{u}r Theoretische Physik I, Heinrich-Heine-Universit\"{a}t D\"{u}%
sseldorf, Universit\"{a}tsstra\ss {e} 1, 40225 D\"{u}sseldorf, Germany.}
\date{\today }

\begin{abstract}
In a constant and homogeneous magnetic background, quantum vacuum fluctuations due to axion-like fields can dominate over those associated with the
electron-positron fields. Considering the framework of axion-electrodynamics, the self-energy operator for the electromagnetic field is determined 
with an accuracy to second-order in the axion-diphoton coupling. This outcome is utilized for establishing modifications to the propagation 
characteristics of photons and to the Coulomb potential of a static pointlike charge. Notably, in the magnetosphere of a  neutron star,  the effect of photon capture 
by the magnetic field, known in QED  as relating to gamma-quanta, is extended in axion electrodynamics to include X-ray photons with the result 
that a specially polarized part of the heat radiation from the surface is canalized along the magnetic field. Besides, for field strengths larger 
than the critical scale associated with this theory, the modified Coulomb potential is of Yukawa-type in the direction perpendicular to the magnetic field
at distances much smaller than the Compton wavelength of an axion,  while along the field it follows approximately the Coulomb law at any length scale. We find that at unlimitedly large magnetic fields the longstanding problem -- overcome in QED -- that 
the ground-state energy of a hydrogen atom is unbounded from below, is reinstated. However, in axion-electrodynamics this unboundedness is cut off  
because the largest magnetic field treatable within this theory is limited by the unitarity of the associated scattering matrix.
\end{abstract}

\date{\today }
\maketitle

%%%%%%%%%%%%%%%%%%%%%%%%%%%%%%%%%%%%%%%%%%%%%%%%%%%%%%%%%%%%%%%%%%%%%%%%%%%%%%%%%%%%%%%%%%%%%%%%%%%%%%%%%%%%%%%%%%%%%%%%%%%%%%%%%%%%%%
\section{Introduction}
%%%%%%%%%%%%%%%%%%%%%%%%%%%%%%%%%%%%%%%%%%%%%%%%%%%%%%%%%%%%%%%%%%%%%%%%%%%%%%%%%%%%%%%%%%%%%%%%%%%%%%%%%%%%%%%%%%%%%%%%%%%%%%%%%%%%%%%

Upon quantization, the axial chiral $\mathrm{U(1)_{A}-}$invariance is exonerated of representing a quantum symmetry in the theory of the strong interactions, 
namely Quantum Chromodynamics (QCD). The occurrence of an anomalous contribution in the QCD Lagrangian -- linked to a nontrivial topological structure of the QCD 
vacuum -- prevents the conservation of the axial current by dispossessing the theory from the Charge-Parity (CP)-symmetry. However -- with astonishing experimental 
accuracy -- QCD manifests itself as a CP-preserving framework and the anomalous ``$\theta-$term'' seems to be fully irrelevant. This famous controversy -- known as 
the strong CP-problem -- is reconciled through the Peccei-Quinn mechanism \cite{Peccei:1977hh}, where a global $\mathrm{U(1)_{PQ}}$-symmetry is promoted to compensate 
the CP-violating term once it is broken spontaneously. While the described solution is perhaps the most appealing among other possible ideas, it did not come 
free of a new challenge for the contemporary physics: the detection of the emergent Nambu-Goldstone boson, i.e. the QCD axion \cite{Wilczek:1977pj,Weinberg:1977ma}. 
Experimental endeavours toward this task are nowadays being carried out worldwide, based -- predominantly -- on the axion-diphoton coupling that results naturally in 
axion-electrodynamics (AED) \cite{Wilczek:1987pj}; a framework whose phenomenology is subject to an intense scrutiny owing to its relevance, not just in particle 
physics,  but also in research branches sharing its main features \cite{Hehl,Li,Ooguri}.

At lab scales, direct searches for the axion rely on plausible traces of axion-photon oscillations mediated by a constant magnetic field. Descriptions of the most 
popular detection methods can be found in Refs.~\cite{Jaeckel:2010ni,Ringwald:2012hr,Hewett:2012ns,Essig:2013lka}. The realization of this process could be relevant, 
indeed, in polarimetric arrangements \cite{Cameron:1993mr,BMVreport,Chen:2006cd,Mei:2010aq,DellaValle:2013xs}\footnote{The feasibility of using strong laser sources 
instead has also been investigated. For details we refer the reader to Refs.~\cite{mendonza,Gies:2008wv,Dobrich:2010hi,Villalba-Chavez:2013bda,Villalba-Chavez:2013goa,Villalba-Chavez:2016hxw}.} 
and in Light-Shining-through-a-Wall setups \cite{Chou:2007zzc,Afanasev:2008jt,Steffen:2009sc,Pugnat:2007nu,Robilliard:2007bq,Fouche:2008jk,Ehret:2010mh,Balou}. Although 
the outcomes of these ongoing experiments have not verified the existence of the QCD axion yet, valuable bounds in its parameter space have been  inferred instead, 
as well as in Axion-Like Particles (ALPs) that are predicted in string theory \cite{Witten:1984dg,Svrcek:2006yi,Lebedev:2009ag,LCicoli:2012sz} and in various Standard 
Model extensions which attempt to incorporate the dark matter of our Universe \cite{covi,Raffelt:2006rj,Duffy:2009ig,Sikivie:2009fv,Baer:2010wm}. Constraints on axions -- or, 
more general, ALPs -- are also deduced from their potential astro-cosmological consequences which are not reflected accordingly by the current observational data of our 
universe. A basic assumption underlying this line of argument is that the interplay between the axionic degrees of freedom and the well established Standard Model 
sector -- i.e., photons in first place -- is extremely feeble. As a consequence, ALPs produced copiously in the core of stars via the Primakoff effect might escape from 
there almost freely, constituting a leak of energy that accelerates the cooling of the star and, thus, shortens its lifetime. Therefore, the number of red giants in 
the helium-burning phase in globular clusters should diminish considerably. That this fact does not take place -- at least not significantly -- constraints the axion-diphoton 
coupling $g$ to lie below $g\lesssim 10^{-10}\ \mathrm{GeV^{-1}}$ for ALP masses $m$ below the $\mathrm{keV}$ scale \cite{Raffelt:1985nk,Raffelt:1999tx,Raffelt:2006cw}.

Precisely on the surface of stellar objects identified as neutron stars \cite{Manchester,Kouveliotou,Bloom} and magnetars, magnetic fields as large as 
$B\sim \mathcal{O}(10^{14}-10^{15})\ \mathrm{G}$ are predicted to exist. As these strengths are bigger than the critical scale $B_{0}=4.42\times 10^{13}\ \mathrm{G}$ of 
Quantum Electrodynamics (QED), such astrophysical scenarios are propitious for the realization of a variety of yet unobserved quantum processes, which are central in our current understanding on the 
nature and origin of the pulsar radiation. Notable among them are, the photon capture effect by the magnetic field owing to the resonant behavior \cite{shabad1972} of the vaсuum 
polarization in QED \cite{shabadnat,shabad3v,shabad2004}\footnote{See also Refs.~\cite{Herold,shabadusov1,shabadusov2}, where the photon 
gradually turns into a positronium atom via the photon-positronium polaritonic state while being captured.}, the one-photon production of electron-positron pairs and 
the photon splitting effect \cite{adler1,adler2,adler3}. Clearly, the strong-field environments provided by these compact objects can also be favorable for an ALPs 
phenomenology -- as they are in QED -- contrary to what is predicted relying on the weak coupling treatment. This occurs, because the aforementioned field strengths could 
compensate for the weakness of the coupling and significantly  stimulate  quantum vacuum fluctuations of axionlike fields. This paper is devoted to analyzing this possibility 
in the ALPs phenomenology. We study the critical regime of AED for which the polarization operator calculated within the second-order accuracy in this theory is responsible.

We shall demonstrate, for instance, that for sufficiently large magnetic fields, the axionlike fields lead to both stronger refraction and screening effects as compared 
with the predictions of QED [at least when we are kinematically far from the vacuum polarization resonance]. In particular, virtual ALPs,  might be responsible for the 
capture of $\gamma $-quanta by the magnetic field lines in the magnetosphere of neutron stars and magnetars prior than the analogous effect starts acting in QED. This might 
fall into discrepancy with the well-established mechanism of pulsar radiation. However, we find that within the allowed window for the  axion parameters, this is not the 
case  for magnetic fields larger than  $B\sim 10^{13}\ \rm G$, and for masses $m\ll 1\ \rm keV$ the axion physics does not dominate over the QED phenomenology. Furthermore, it 
is also indicated that ALPs can induce strong screening in the Coulomb potential of a static pointlike charge in the direction perpendicular to the magnetic field. Contrary to studies 
developed within QED \cite{shabad5,shabad6,Sadooghi:2007ys,Adorno2016}, our investigation reveals that along the field direction, and within the regime where the associated 
axion phenomenology could dominate over the corresponding QED effects, the screening due to ALPs is very weak, no matter how strong the field strength is (within the limits 
imposed on it by the unitarity). Hence, the modified potential we obtain follows the Coulomb law approximately. As a consequence, the plausible existence of heavy axionlike 
particles at extremely large magnetic fields might lead to a reestablishment of the formula for the unscreened ground-state energy of a nonrelativistic electron in the hydrogen 
atom \cite{elliott}. The longstanding problem linked to this formula, i.e. the fact that the energy spectrum is unbounded from below when the magnetic field grows unlimitedly, 
was solved when the screening induced by the vacuum polarization of QED was incorporated and the virtuality of ALPs was ignored \cite{shabad5,shabad6,Vysotsky:2010sz,Machet:2010yg, Popov}. 
We will see that in AED, by requiring the unitarity of the scattering matrix, a cutoff is incorporated for the largest treatable magnetic field, which may cause dramatic 
changes in the limiting ground-state energy.

The investigation carried out here relies on a quantized approach of AED developed in Sec.~\ref{polarizationtensorrenormalizedb}. It provides a nonrenormalizable framework whose quantum 
corrections are determined by using effective field theory techniques that have proved to be appropriate for extracting valuable insights from quantum gravity \cite{thooft,Stelle:1976gc,Donoghue:1994dn,Donoghue:1993eb}, 
chiral perturbation theory \cite{s.weinberg,Gasser:1983yg,Gasser:1984gg,Ecker:1995zu} and nonlinear QED \cite{Halter:1993kj,Kong:1998ic,Dicus:1997ax}, by preserving -- parallely -- the 
unitarity of the respective scattering matrices. In Sec.~\ref{polarizationtensorrenormalizedb} the main equations describing  axion-diphoton interaction in an external magnetic field are fixed, 
and the diagrams for polarization operators determining the approximation are specified. In Sec.~\ref{polarizationtensorrenormalizedbg} the approximation-independent eigenmode expansion of the  
polarization tensor is given in terms of its eigenvectors and eigenvalues, valid in a magnetic field both in QED and AED. We determine the polarization  tensor with 
an accuracy of the second order in the axion-diphoton coupling. Although the neutral nature of quantum vacuum fluctuations due to  axionlike fields contrasts with those associated 
with the charged Dirac ones, the outcomes resulting from the quantized axion-photon theory pretty much resemble those obtained in QED. This similarity is stressed in Sec.~\ref{polarizationtensorrenormalizedbg}, 
where various asymptotes of the eigenvalues of the polarization tensor are established in parallel. Also in that section the plausible regions of parameters  of undiscarded ALPs, where the 
contribution of the latter  could dominate over the QED effects, are established. 

In Sec.~\ref{polarizationtensorrenormalizedc} we define the mass shell by solving the dispersion equation for a photon of the extraordinary polarization mode interacting with an axion in the  
presence of a magnetic field to find two branches of the spectrum: the massless one, whose rest energy is zero -- responsible for the photon of that mode -- and the massive one for the axion. 
As these branches do not intersect, no axion-photon oscillation occurs unless the nonconservation of the momentum is admitted due to possible nonhomogeneity of the background. The same spectrum 
is obtained by solving the field equation in the pseudoscalar sector in Appendix \ref{Appendix}. Parallely, we find the group velocity on these branches and fix the angles between its direction 
and the wave vector, which are  nonvanishing due to the optic anisotropy. The flattening of the massless branch near small values of the transverse momentum results in vanishing of the group 
velocity component across the magnetic field. This phenomenon forms a basis for the photon capture considered in the subsequent sections~\ref{Implicationcapture1} and \ref{Implicationcapture2}. 
Throughout the investigation, we use a special normalization of the  polarization vectors that allows us to completely exclude, on the mass shell, one of the three eigenmodes, leaving two polarization 
degrees of freedom for massless excitations that exist along with one massive degree of freedom. Formulas are given for polarizations of electric and magnetic components of free massless excitations. The electric 
field in the extraordinary mode lies in the plane spanned by the external magnetic field and the wave vector, whereas the other massless mode, not affected by the axions, has its electric vector 
orthogonal to that plane. In Sec.~\ref{Implicationcapture1} we recall the effect of $\gamma $-quantum capture in QED and argue that in AED the same effect might destroy the accepted mechanism of 
electron-positron plasma production in pulsars and hence of their radio-emission, if it was not eliminated by the already existing restrictions on the axion mass and coupling constant. In 
Sec.~\ref{Implicationcapture2} we deal with the capture of soft photons that occurs in AED. We consider the refraction of radiation, when it leaves the region of strong field, and assure ourselves 
that the heat radiation from the pulsar surface belonging to the extraordinary mode is canalized parallel to the magnetic field, and it may leave it, whereas the radiation polarized in the orthogonal 
direction is not canalized and is subject to the angular distribution determined by the refraction law. We believe that the effect that the radiation of a certain polarization condenses close to 
the magnetic field direction, while the orthogonally polarized radiation is distributed over the angle maybe, if observed or eliminated by observation, might become a criterion for the  existence of 
axions.  Although the pulsar's strong magnetic field makes the axion heavy, we argue that the number of certain type of heat ALPs escaping from it could be comparable to the heat photons which leave 
the pulsar at X-ray temperatures.

Finally, in Sec.~\ref{GPa} we determine the general expression for the axion-modified Coulomb potential. Its anisotropization, short-ranging and dimensional 
reduction are revealed manifestly via numerical evaluations. We show, in particular, that for field strengths larger than the critical scale associated 
with this effective theory, the modified potential is short-ranged in the direction perpendicular to the magnetic field while it follows approximately 
the behavior of Coulomb's law along $\pmb{B}$. Sec.~\ref{GPb1} is devoted to support analytically these results. There, asymptotic expressions for the
modified Coulomb potential in the limits of strong magnetic fields are determined. In addition, the imprints that virtual ALPs leave within the ground 
state energy of a nonrelativistic electron in a hydrogen atom are evaluated. The limit of weak external field is studied in Sec.~\ref{weakfieldsection}, 
whereas in Sec.~\ref{conclus} we expose our conclusions.

%%%%%%%%%%%%%%%%%%%%%%%%%%%%%%%%%%%%%%%%%%%%%%%%%%%%%%%%%%%%%%%%%%%%%%%%%%%%%%%%%%%%%%%%%%%%%%%%%%%%%%%%%%%%
\section{Axion-electrodynamics in a constant magnetic background \label{polarizationtensorrenormalized}}
%%%%%%%%%%%%%%%%%%%%%%%%%%%%%%%%%%%%%%%%%%%%%%%%%%%%%%%%%%%%%%%%%%%%%%%%%%%%%%%%%%%%%%%%%%%%%%%%%%%%%%%%%%%%

%%%%%%%%%%%%%%%%%%%%%%%%%%%%%%%%%%%%%%%%%%%%%%%%%%%%%%%%%%%%%%%%%%%%%%%%%%%%%%%%%%%%%%%%%%%%%%%%%%%%%%%%%%%%
\subsection{Polarization-tensor approach \label{polarizationtensorrenormalizedb}}
%%%%%%%%%%%%%%%%%%%%%%%%%%%%%%%%%%%%%%%%%%%%%%%%%%%%%%%%%%%%%%%%%%%%%%%%%%%%%%%%%%%%%%%%%%%%%%%%%%%%%%%%%%%%

To leading order in the axion-diphoton coupling $\bar{g}$, the Lagrangian density describing AED combines the standard Maxwell Lagrangian with the free Lagrangian 
density of the pseudoscalar field $\bar{\phi}(x)$ and with the scalar interaction Lagrangian density involving the dual of the electromagnetic field tensor 
$\tilde{f}_{\mu \nu }=\frac{1}{2}\epsilon ^{\mu \nu \alpha \beta }f_{\alpha\beta }$ where $f_{\mu \nu }=\partial _{\mu }\bar{a}_{\nu }-\partial _{\nu } \bar{a}_{\mu }$ 
and $\epsilon ^{0123}=1$. Explicitly, 
\begin{equation}
\mathpzc{\bar{L}}=-\frac{1}{4}f^{2}+\frac{1}{2}(\partial \bar{\phi})^{2}-\frac{1}{2}\bar{m}^{2}\bar{\phi}^{2}+\frac{1}{4}\bar{g}\bar{\phi}\tilde{f}f.
\label{initialaction}
\end{equation}%
Hereafter, we use a metric with  signature $\mathrm{diag}(\mathpzc{g}_{\mu\nu })=(1,-1,-1,-1)$, and Heaviside$-$Lorentz units with the speed of light and the Planck 
constant set to unity $c=\hbar =1$. To abbreviate, the following notations have also been used: $f^{2}\equiv f_{\mu \nu }f^{\mu \nu}$, $\tilde{f}f\equiv \tilde{f}_{\mu \nu }f^{\mu \nu }$, 
$(\partial \bar{\phi})^{2}\equiv (\partial _{\mu }\bar{\phi})(\partial ^{\mu }\bar{\phi})$. As the interacting term has a mass dimension $-5$, this framework belongs 
to the class of perturbatively nonrenormalizable theories. In connection, $\mathpzc{\bar{L}}$ can be regarded as a Wilsonian effective Lagrangian parametrizing the 
leading order contribution of its ultraviolet completion, i.e. a renormalizable theory linked to physics beyond the Standard Model which might contain new heavy particles 
at the energy scale $\Lambda_{\mathrm{UV}}\sim \frac{1}{\bar{g}}$. If this hypothetical theory satisfies fundamental principles such as Lorentz and  gauge invariance, and
preserves -- in addition -- the unitarity, quantum processes due to fluctuations of modes of the fields in $\mathpzc{\bar{L}}$, taking place at energies substantially below 
$\Lambda _{\mathrm{UV}}$, should also be described in a unitary way.

\begin{figure}[tbp]
\includegraphics[width=.35\textwidth]{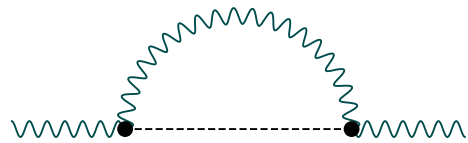}
\caption{Feynman diagram depicting the axion-modified vacuum polarization tensor. While the dashed line represents the free axion propagator $\Delta_\mathrm{F}(x,\tilde{x})$, the internal wavy 
line denotes the free photon propagator $D_{\protect\mu\protect\nu}^{(0)}(x,\tilde{x})$. The external lines stand for amputated legs associated with photons.}
\label{fig:mb000}
\end{figure}

A renormalization program for extracting finite quantities from perturbative calculations based primarily on Eq.~(\ref{initialaction}) can be carried out only by continually 
incorporating higher-dimensional operators into the Lagrangian to provide counterterms that cancel divergences. These terms are polynomial in the derivatives and preserving 
the formal invariance properties of $\mathpzc{\bar{L}}$, but on the other hand, suppressed by higher powers of $g$. Recently, this procedure has been utilized in the determination 
of the polarization tensor [see Fig.~\ref{fig:mb000}] in the one-loop approximation \cite{alinapaper}: 
\begin{equation}
\begin{split}
&\Pi _{\mathrm{vac}}^{\mu \nu}(q_{1},q_{2})=\deltabar_{q_{1},q_{2}}\left[ q_{2}^{2}\mathpzc{g}^{\mu \nu }-q_{2}^{\mu
}q_{2}^{\nu }\right] \pi (q_{2}^{2}), \\
& \pi (q^{2})=-\frac{g^{2}m^{2}}{64\pi ^{2}}\left[ 1-\frac{1}{3}\frac{q^{2}}{m^{2}}\right] \\
& \qquad \qquad \qquad -\frac{g^{2}}{16\pi ^{2}}\int_{0}^{1}ds\;\Delta(s)\ln \left(\frac{\Delta (s)}{m^{2}}\right),
\end{split}
\label{polarizationtensor}
\end{equation}
where the shorthand notation $\deltabar_{q_{1},q_{2}}\equiv (2\pi )^{4}\delta^{4}(q_{1}-q_{2})$ has been introduced. Here, $\Delta (s)=m^{2}s-q^{2}s(1-s)$ with $m$ being the
renormalized ALP mass. Likewise, $g$ must be understood as the renormalized axion-diphoton coupling. In Ref.~\cite{alinapaper}, relations between the bare and the corresponding  
renormalized parameters are given. Following our discussion below Eq.~(\ref{initialaction}), for establishing $\Pi _{\mathrm{vac}}^{\mu \nu }(q_{1},q_{2})$ the enlarged Lagrangian 
density 
\begin{equation}
\mathpzc{L}=\bar{\mathpzc{L}}+\frac{1}{2}\bar{g}^{2}\mathpzc{b}^{2}(\partial f)^{2}+\ldots
\label{intermediateaction}
\end{equation}%
is required. Here, $(\partial f)^{2}\equiv (\partial _{\mu }f^{\mu \nu })(\partial _{\lambda}f_{\ \nu }^{\lambda })$ and $\mathpzc{b}$ is  an arbitrary real parameter. It is precisely 
the last term written in Eq.~(\ref{intermediateaction}) $\sim \frac{1}{2}\bar{g}^{2}(\partial f)^{2}$ which allows us to reabsorb a divergence arising from the loop in Fig.~\ref{fig:mb000} 
that cannot be reabsorbed in the wavefunction renormalization constant $\mathpzc{Z}_{3}$ by demanding that the radiative correction should not modify the residue of the photon propagator 
at $q^{2}=0 $ \cite{Weinberg:1995mt,Schwartz}. The reason for considering Eq.~(\ref{initialaction}) as an effective Lagrangian rather than an ordinary Lagrangian density containing 
higher-order derivatives, lies in avoiding -- under quantization -- a number of undesirable properties tied to the nature of this type of theory, including the proliferation of
Pauli-Villar ghosts, the lack of a finite ground-state energy, as well as the emergence of equations of motion requiring extra boundary conditions to specify their solutions at any spacetime 
point. As shown in Ref.~\cite{alinapaper}, within the $\sim g^{2}$ approximation, no effect other than to remove the divergences that result from the loop can be assigned to the last term 
in Eq.~(\ref{intermediateaction}). This fact is connected to the equivalence theorem \cite{Arzt:1993gz,GrosseKnetter:1993td,Georgi:1991ch}.

\begin{figure}
\begin{center}
\includegraphics[width=7cm]{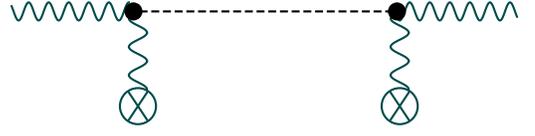}
\caption{\label{fig.001} Diagrammatic representation of the external field [vertical wavy lines] dependent contribution to the  vacuum polarization tensor  mediated by  quantum 
fluctuations of pseudoscalar fields. Here, the dashed line represents the ALP propagator $\Delta_\mathrm{F}(x,\tilde{x})$, whereas the horizontal wavy lines must be understood 
as  amputated photon legs. }
\end{center}
\end{figure}

In the following we evaluate the vacuum polarization tensor in AED under the influence of a constant external field $\mathscr{F}_{\mu \nu }=\partial _{\mu }\mathscr{A}_{\nu }(x)-\partial _{\nu }\mathscr{A}_{\mu }(x)$, 
the Lorentz invariants of which satisfy the conditions  $\mathfrak{F}=\frac{1}{4}\mathscr{F}_{\mu \nu }\mathscr{F}^{\mu \nu }>0$ and $\mathfrak{G}=\frac{1}{4}\mathscr{F}_{\mu \nu }\tilde{\mathscr{F}}^{\mu \nu }=0$
simultaneously. This kind of field configuration  is called magneticlike, because an inertial reference frame exists, where it is purely magnetic, with no admixture of electric field. Let us replace $f_{\mu \nu }$ 
in Eq.~(\ref{intermediateaction}) by $f_{\mu \nu }+\mathscr{F}_{\mu \nu }$. Because of the presence of $\mathscr{F}$, the Lagrangian in Eq.~(\ref{intermediateaction}) acquires an additional interaction 
$\mathpzc{L}_{\mathscr{F}}=\frac{1}{2}\bar{g}\phi \tilde{\mathscr{F}}f$.  This vertex is responsible for the appearance of  another Feynman diagram [see Fig.~\ref{fig.001}] which contributes to the photon radiative 
correction within the accuracy of the second order in the axion-diphoton coupling. In contrast to the loop in Fig.~\ref{fig:mb000}, the analytical expression of this additional graph is free of 
ultraviolet divergences. Hence, within the accuracy of second-order in the axion-diphoton coupling $\sim g^2$, a renormalization procedure is required for cancelling those divergences linked to the loop.  This program 
has been carried out in Ref.~\cite{alinapaper} and its final result is given in Eq.~(\ref{polarizationtensor}). All these details allow us to write the renormalized photon propagator in the following form 
\begin{eqnarray}
\begin{split}
& D_{\alpha \beta }(x,\tilde{x})=D_{\alpha \beta }^{(0)}(x,\tilde{x})+\int
d^{4}yd^{4}\tilde{y}D_{\alpha \mu }^{(0)}(x,y) \\
& \qquad \qquad \qquad \qquad \times \Pi ^{\mu \nu }(y,\tilde{y})D_{\nu
\beta }^{(0)}(\tilde{y},\tilde{x})+\mathcal{O}(g^{4}).
\end{split}
\label{perturbativeexpansionpropagator} 
\end{eqnarray}%
Here $D_{\alpha \beta }^{(0)}(x,\tilde{x})=\int \dbar^{4}p\frac{-i\mathpzc{g}_{\alpha \beta }}{p^{2}+i0}\mathrm{e}^{ip(x-%
\tilde{x})}$ denotes the free photon propagator in Feynman gauge, with the shorthand notation $\dbar^{4}p\equiv d^{4}p/(2\pi )^{4}$. The renormalized polarization tensor involved 
in Eq.~(\ref{perturbativeexpansionpropagator}) then splits into two contributions: 
\begin{equation}
\Pi ^{\mu \nu }(x,\tilde{x})=i\Pi _{\mathrm{vac}}^{\mu \nu }(x,\tilde{x}%
)+i\Pi _{\mathscr{F}}^{\mu \nu }(x,\tilde{x}).  \label{AXIONPT}
\end{equation}%
Each of these terms is in correspondence with the Feynman graphs displayed in Figs.~\ref{fig:mb000} and \ref{fig.001}, respectively. While the Fourier transform of the first contribution [$\Pi _{\mathrm{vac%
}}^{\mu \nu }(q_{1},q_{2})=\int d^{4}xd^{4}\tilde{x}e^{-iq_{1}x}\Pi _{\mathrm{vac}}^{\mu \nu }(x,\tilde{x})e^{iq_{2}\tilde{x}}$] is given in Eq.~(\ref{polarizationtensor}), the expression for the 
tree-level diagram with two external field insertions reads 
\begin{equation}
\Pi _{\mathscr{F}}^{\mu \nu }(x,\tilde{x})=-g^{2}\tilde{\mathscr{F}}^{\mu
\tau }(x)\left[ \partial _{\tau }^{x}\partial _{\sigma }^{\tilde{x}}\Delta _{%
\mathrm{F}}(x-\tilde{x})\right] \tilde{\mathscr{F}}^{\sigma \nu }(\tilde{x}).
\label{AXIONPTexternalfield}
\end{equation}%
Here $\Delta _{\mathrm{F}}(x,\tilde{x})=\int \dbar^{4}p\frac{ie^{-ip(x-\tilde{x})}}{p^{2}-m^{2}+i0}$ stands for the axion propagator with the $i0-$prescription allowing to elude the pole at $p^{2}=m^{2}$ 
on the real axis. Observe that Eq.~(\ref{AXIONPTexternalfield}) has been written in terms of the renormalized coupling $g$ and mass $m$. This is because, within a second order approximation in the coupling 
constant, no modification to the renormalized axion-diphoton coupling $g$ occurs. Actually, the renormalization constant associated with $g$ does not deviate from its classic tree-level value which is one, 
and so, no distinction between the renormalized and physical coupling is needed hereafter.

We are interested in dispersive and electrostatic effects caused primarily by the interaction that virtual ALPs  mediate  between a small-amplitude electromagnetic wave and a strong magnetic field. 
The extent to which these phenomena manifest is determined by the dependence that the photon propagator acquires on the magnetic field strength. We remark that, if the latter is sufficiently strong, the 
quantum correction in Eq.~(\ref{perturbativeexpansionpropagator}) might reach values in which the perturbative treatment is not longer justified. In such a situation, it is more suitable to deal with the 
photon propagator resulting from the sum of the infinite series containing Feynman diagrams of all orders in  perturbation theory. While there are various ways for establishing this object, we will follow a path in which   
the effective action of the theory, i.~e., the Legendre transform of the functional generating connected graphs \cite{Weinberg:1996mt}, is required. In line, we indicate that -- in contrast to Fig.~\ref{fig:mb000} -- the 
Feynman graph in Fig.~\ref{fig.001} is not a one-particle-irreducible diagram. Because of this, it would not appear within the effective action:
\begin{equation}
\Gamma \lbrack \mathpzc{a},\phi ]=\frac{1}{2}\int d^{4}x\int d^{4}\tilde{x}\;%
\pmb{\Phi}^{\mathrm{T}}(x)\pmb{\Gamma}^{(2)}(x,\tilde{x})\pmb{\Phi}(\tilde{x}%
)+\ldots
\label{effectiveaction}
\end{equation}%
where the abbreviation $+\ldots $ stands for higher-order terms in the average fields $\mathpzc{a}(x)$ and $\phi (x)$. The inverse Green's function  $\pmb{\Gamma}^{(2)}(x,\tilde{x})$ 
and the flavor field $\pmb{\Phi}(x)$ involved in this expression are given by 
\begin{equation*}
\pmb{\Gamma}^{(2)}(x,\tilde{x})=\left[ 
\begin{array}{cc}
\Gamma (x,\tilde{x}) & \Gamma _{\mu }(x,\tilde{x}) \\ 
\Gamma _{\mu }(x,\tilde{x}) & \Gamma _{\mu \nu }(x,\tilde{x})
\end{array}%
\right] ,\ \pmb{\Phi}(x)=\left[ 
\begin{array}{c}
\phi (x) \\ 
\mathpzc{a}^{\mu }(x) \\ 
\end{array}%
\right] .
\end{equation*}%
While the dia\-gonal components in $\pmb{\Gamma}^{(2)}(x,\tilde{x})$ are the respective two-point proper correlation functions associated with the axion field $\phi (x)$ and the small-amplitude 
electromagnetic waves $\mathpzc{a}^{\mu }(x)$, the off-diagonal ones allow for oscillations between each other. Within the accuracy of second order in the coupling constant [$\sim g^{2}$], these are 
\begin{equation*}
\begin{array}{c}
\Gamma (x,\tilde{x})=-\left[ \square +m^{2}\right] \delta ^{4}(x-\tilde{x}%
)+i\Sigma (x,\tilde{x}), \\ 
\\ 
\Gamma ^{\mu \nu }(x,\tilde{x})=\left[ \square \mathpzc{g}^{\mu \nu
}-\partial ^{\mu }\partial ^{\nu }\right] \delta ^{4}(x-\tilde{x})+i\Pi _{%
\mathrm{vac}}^{\mu \nu }(x,\tilde{x}), \\ 
\\ 
\Gamma ^{\mu }(x,\tilde{x})=g\mathscr{\tilde{F}}_{\ \nu }^{\mu }(x)\partial
^{\nu }\delta ^{4}(x-\tilde{x}).%
\end{array}%
\end{equation*}%
Here $\square =\frac{\partial ^{2}}{\partial t^{2}}-\nabla ^{2}$, whereas $\Sigma (x,\tilde{x})$ stands for the renormalized self-energy operator of $\phi (x)$, the expression of which can be found in 
Ref.~\cite{alinapaper}.

As outlined in Ref.~\cite{Villalba-Chavez:2016hxw}, $\Pi_{\mathscr{F}}^{\mu\nu}(x,\tilde{x})$ [see Eq.~(\ref{AXIONPT})] can be realized in the modified Maxwell's equation by integrating out the pseudoscalar  
ALP field $\phi(x)$, i.e., by setting $\phi(x)$ on the solution of the equation of motion  \cite{Villalba-Chavez:2016hxw}: 
\begin{eqnarray}
\begin{split}
&\phi(x)=\frac{1}{\left(\square+m^2\right)}\left[\frac{1}{2}g\tilde{\mathscr{F}}\mathpzc{f}+i\int d^4\tilde{x}\ \Sigma(x,\tilde{x})\phi(\tilde{x})\right],
\end{split}\label{phionshell}
\end{eqnarray}
where $\mathpzc{f}\equiv\mathpzc{f}_{\mu\nu}=\partial_\mu \mathpzc{a}_\nu-\partial_\nu \mathpzc{a}_\mu$. Inserting this expression into the effective action [see Eq.~(\ref{effectiveaction})], 
and collecting the corrections proportional to $g^2$, we end up with 
\begin{equation}  \label{bilinearEA}
\begin{split}
&\Gamma_{\mathrm{os}}=\frac{1}{2}\int d^4x d^4\tilde{x}\;\mathpzc{a}^\mu(x)\mathpzc{D}_{\mu\nu}^{-1}(x,\tilde{x})\mathpzc{a}^\nu(\tilde{x})+\mathcal{O}[\mathpzc{a}^3], \\
&\mathpzc{D}^{-1}_{\mu\nu}(x,\tilde{x})=\left[\square\mathpzc{g}_{\mu\nu}-\partial_\mu\partial_\nu\right]\delta^4(x-\tilde{x}%
)+\Pi_{\mu\nu}(x,\tilde{x})
\end{split}
\end{equation}
with $\Pi^{\mu\nu}(x,\tilde{x})$ given in Eq.~(\ref{AXIONPT}). Noteworthy, this bilinear approximation to the effective action provides the following equations of motion 
for the small-amplitude electromagnetic wave [$\partial \mathpzc{a} =0$]: 
\begin{eqnarray}
\begin{split}
&\square \mathpzc{a}^\mu(x)+\int d^4\tilde{x}\ \Pi^{\mu}_{\ \nu}(x,\tilde{x})%
\mathpzc{a}^{\nu}(\tilde{x})=0.
\end{split}
\label{DSE}
\end{eqnarray}
We emphasize that Eqs.~(\ref{bilinearEA}) and (\ref{DSE}) apply provided there is no expectation value for the axion fields permeating the universe, or that such vacuum expectation value is negligible 
in comparison with the quantum vacuum fluctuations that are induced by the ALP field.

%%%%%%%%%%%%%%%%%%%%%%%%%%%%%%%%%%%%%%%%%%%%%%%%%%%%%%%%%%%%%%%%%%%%%%%%%%%%%%%%%%%%%%%%%%%%%%%%%%%%%%%%%%%%
\subsection{Polarization tensor, asymptotes and plausible dominance over QED\label{polarizationtensorrenormalizedbg}}
%%%%%%%%%%%%%%%%%%%%%%%%%%%%%%%%%%%%%%%%%%%%%%%%%%%%%%%%%%%%%%%%%%%%%%%%%%%%%%%%%%%%%%%%%%%%%%%%%%%%%%%%%%%%

When the external field is a constant magneticlike background characterized by a four-potential $\mathscr{A}^{\mu }(x)=-\frac{1}{2}\mathscr{F}_{\ \nu}^{\mu }x^{\nu }$ and the invariants 
$\mathfrak{F}=\frac{1}{2}\pmb{B}^{2}$, $\mathfrak{G}=0$, the Fourier transform of the polarization tensor can be written in the following form 
\begin{equation}
\Pi ^{\mu \nu }(q_{1},q_{2})=\deltabar_{q_{1},q_{2}}\mathpzc{P}^{\mu \nu }(q).  \label{Fouriertransform}
\end{equation} As the theory posesses the same properties of Lorentz-, translational and gauge invariance, and spatial parity as QED, the tensorial structure involved in this expression 
can be diagonalized in the same form as the one established for QED with magnetic field in Refs.~\cite{batalin,Shabad:1975ik}  basing on these general properties: 
\begin{equation*}
\displaystyle\mathpzc{P}^{\mu \nu }(q)=-\sum_{j=1}^{3}\varkappa _{j}\frac{\flat_{j}^{\mu }\flat _{j}^{\nu }}{\flat_j^2},\quad \mathpzc{P}_{\ \nu }^{\mu }(q)\flat_{j}^{\nu }=\varkappa _{j}\flat _{j}^{\nu }.
\end{equation*}The conditions of solvability of the field equation (\ref{DSE}) are the dispersion equations 
\begin{equation}
q^2=\varkappa_i \label{dispersionequationphoton}
\end{equation} which define the mass-shells of involved  particles, photon included. The scalar eigenvalues $\varkappa _{j}$  in the approximation considered are 
\begin{equation}
\displaystyle\varkappa _{1,3}=q^{2}\pi (q^{2}),\quad \varkappa _{2}=g^{2}%
\frac{q\tilde{\mathscr{F}}^{2}q}{q^{2}-m^{2}}+\varkappa _{1,3},
\label{tensordecomposition}
\end{equation}%
where $\pi (q^{2})$ is given by  Eq.~(\ref{polarizationtensor}).  Note that the external field and hence the anisotropy is introduced through the interaction with the axion. If $g=0$, the isotropy 
is restored: $\varkappa_{1}=\varkappa_{2}=\varkappa_{3}$. In the special Lorentz frame, where the external field is purely magnetic and directed along the axis denoted as $\parallel$, the Lorentz 
scalar $q\tilde{\mathscr{F}}^{2}q/(2\mathfrak{F})$ becomes $q_{0}^{2}-q_{\parallel}^{2}$  where the photon energy is $q_{0}$ and the momentum component along the magnetic field is $\pmb{q}_{\parallel}$. 
There is also an extra Lorentz scalar $q\mathscr{F}^{2}q/(2\mathfrak{F})=q_{\perp }^{2}$, with $\pmb{q}_{\perp}$ denoting the projection of the momentum $\pmb{q}$ onto the plane perpendicular 
to $\pmb{B}$, so that $2\mathfrak{F}q^{2}=q\tilde{\mathscr{F}}^{2}q-q\mathscr{F}^{2}q$.

At this point it is worth remarking that the set of eigenvectors $\flat_{j}^{\mu }$ with $j=1,2,3$ is built from first principles, independently of any approximation used in the calculations \cite{batalin}. 
Explicitly, 
\begin{equation}
\begin{split}
& \flat _{1}^{\mu }=q^{2}\mathscr{F}_{\ \lambda }^{\mu }\mathscr{F}_{\
\nu }^{\lambda }q^{\nu }-q^{\mu }(q\mathscr{F}^{2}q), \\
& \flat _{2}^{\mu }=\tilde{\mathscr{F}}_{\ \nu }^{\mu }q^{\nu },\quad \flat _{3}^{\mu }=
\mathscr{F}_{\ \nu }^{\mu }q^{\nu }.
\end{split}
\label{decomposition}
\end{equation}%
Observe that, due to the Dirac delta involved in Eq.~(\ref{Fouriertransform}), the decomposition in Eq.~(\ref{tensordecomposition}) does not depend on which choice $q=q_{1}$ or $q=q_{2}$ is taken. The 
eigenvectors contained in this decomposition are mutually orthogonal $\flat_{j\mu }\flat_{\ell }^{\mu }=-\delta_{j\ell}\flat^2_\ell$ and transverse $q_{\mu}\flat _{j}^{\mu }=0$ with $j=1,2,3$. Besides, they satisfy 
the completeness relation $\mathpzc{g}^{\mu \nu }-q^{\mu }q^{\nu }/q^{2}=-\sum_{j=1}^{3}\flat_{j}^{\mu }\flat _{j}^{\nu }/\flat_j^2$. Formally, the diagonal expansion in Eq.~(\ref{tensordecomposition}) admits an 
additional contribution which is longitudinal by construction $\sim q^{\mu }q^{\nu }$. Owing to the gauge invariance property of the polarization tensor $\mathpzc{P}_{\ \nu }^{\mu}(q)q^{\nu }=0$, the 
eigenvalue associated with this term vanishes identically.

When AED is embedded in the  QED action, the eigenvalues of the polarization tensor [see  Eq.~(\ref{tensordecomposition})]  will contribute additively to the corresponding 
QED eigenvalues. In this subsection  we exploit this property to establish regions of ALPs parameters for which a dominance over the QED effects could take place. This seems to be particularly 
accesible because the tree correction to the vacuum polarization tensor, i.e. the first term in $\varkappa_2$ [see Eq.~(\ref{tensordecomposition})] exhibits a quadratic growth in the external  
field while the  corresponding eigenvalue in QED  grows linearly instead [$b\gg1$]:
\begin{equation}\label{qedbehavior}
\varkappa_2^{\mathrm{QED}}\approx\left\{\begin{array}{c}
                        \displaystyle \frac{\alpha b}{3\pi}(q_0^2-q_\parallel^2)\qquad\mathrm{for}\quad q_0^2-q_\parallel^2\ll m_e^2,\\ \\ \displaystyle
                          \frac{2\alpha b}{\pi}m_e^2\quad\mathrm{for}\quad m_e^2\ll q_0^2-q_\parallel^2\ll m_e^2b.
                         \end{array}\right.
\end{equation} Here  $\alpha=1/137$ denotes  the fine structure constant,   $b=B/B_0$ and  $B_0=m_e^2/e\approx4.42\times 10^{13}\ \rm G$ the critical QED scale. In this context, $m_e$ and $e$ refer
to the electron mass and  the  absolute value of its charge.\footnote{It is precisely  due to this linear dependence that an anomalous magnetic moment for mode-$2$ photons  can be defined 
\cite{VillalbaChavez:2009ia,VillalbaChavez:2012ea,PerezRojas:2008nx,Rojas:2013zga}.} It is worth mentioning that the application of these asymptotes requires  $q_\perp^2\ll m_e^2 b$.

In order to establish undiscarded ALPs parameters which might lead  to strong refraction as well as to a more pronounced screening property than the one predicted from  QED, it is required -- first -- to
derive  some asymptotic formulae of Eq.~(\ref{tensordecomposition}). Let us consider the situation in which $q^{2}\ll m^{2}$. In this case all the eigenvalues of the polarization tensor are real.\footnote{This 
statement does not hold for the QED part of polarization operator eigenvalues, because, for instance, $\varkappa_2^{\mathrm{QED}}$ has an imaginary part when $q_{0}^{2}-q_{\parallel }^{2}>4m_e^{2}$. So, 
when the condition $q_{\perp }^{2}\gg 4m_{e}^{2}-m^{2}$ is fulfilled the imaginary part is present although $q^{2}\ll m^{2}$.} Their behavior can be inferred easily by quoting the corresponding limit for 
$\pi (q^{2})$ \cite{alinapaper}: 
\begin{equation}
\begin{split}
& \left. \varkappa _{1,3}\right\vert _{q^{2}\ll m^{2}}\approx \frac{%
g^{2}q^{4}}{144\pi ^{2}}, \\
& \left. \varkappa _{2}\right\vert _{q^{2}\ll m^{2}}\approx -\mathfrak{b}%
^{2}(q_{0}^{2}-q_{\parallel }^{2})+\frac{g^{2}q^{4}}{144\pi ^{2}}.
\end{split}
\label{sdfsdfsdfasdfsdfsfs}
\end{equation}%
Here, we have introduced the parameter $\mathfrak{b}=B/B_c$ with  $B_c=m/g$ denoting a critical field scale associated with AED. When  the magnetic-field-depending contribution 
dominates the vacuum polarization, the resulting expression for $\varkappa_2\approx-\mathfrak{b}^2(q_0^2-q_\parallel^2)$ exceeds the  corresponding QED expression [first line in Eq.~(\ref{qedbehavior})] 
provided $\mathfrak{b}^2\gg \alpha b/(3\pi)$. This condition is satisfied for the QCD axion, the mass of which has been restricted by astrophysical and cosmological constraints to lie between 
$10^{-6}\ \mathrm{eV}<m<10^{-2}\ \rm eV$.\footnote{This window restricts even further if axions constitute the dominant component of dark matter, in which case $50\ \mu \mathrm{eV}<m<1.5\ \rm meV $ \cite{Borsanyi}.}   
The corresponding window for $g$ is $10^{-16}\ \mathrm{GeV}^{-1}\lesssim g \lesssim 10^{-12}\ \mathrm{GeV}^{-1}$. For the ranges  above, the critical scale of the theory would be of the order of  
$B_c\sim 10^{20}\ \rm G $. We remark that  magnetic fields of this order  of magnitude  and even larger have been speculated to exist in the cores of neutron stars \cite{Chakrabarty}. If the   
external field exceeds this critical scale by three orders of magnitude the parameter $\mathfrak{b}^2\sim 10^6$ turns out to be  of the same  order of magnitude as  $\alpha b/(3\pi)\sim 10^{6}$ 
and the effects of AED cannot be ignored any longer. Now, the best laboratory  limit for ALPs is held by the OSCAR collaboration: $g<4\times 10^{-8}\ \rm GeV^{-1}$ for $m\lesssim \rm 100\  \mu eV$ \cite{Balou}. If one takes an allowed mass of 
the order of  $m\sim 10^{-4}\ \rm eV$ and upper limit above, the critical field is  $B_c\sim 10^{14}\ \rm G$. The latter is comparable with the  magnetic fields that might 
exist in the  magnetosphere of neutron stars and magnetars. For a magnetic field exceeding this critical value by two orders of magnitude -- which could be realized at the surface of cosmological  gamma-ray 
bursters if  they are rotation-powered neutron stars \cite{usovnature,katz,ruderman} --  we obtain that $\mathfrak{b}^2\sim 10^4$ is  four orders of magnitude larger than $\alpha b/(3\pi)\sim 1$. Under such a circumstance 
the  vacuum polarization linked to virtual  ALPs  would prevail over the one associated with virtual electron-positron pairs. 

Next, let us consider $q^{\mu }$ spacelike [$q^{2}<0$] with $\vert q^{2}\vert\gg m^{2} $. Here the three eigenvalues are also real and behave asymptotically as 
\begin{equation}
\begin{split}
& \varkappa _{1,3}|_{\vert q^{2}\vert \gg m^{2}}\approx \frac{g^{2}q^{4}}{96\pi ^{2}}%
\left[ \ln \left( \frac{|q^{2}|}{m^{2}}\right) -\frac{7}{6}\right] , \\
& \varkappa _{2}|_{\vert q^{2}\vert\gg m^{2}}\approx m^{2}\mathfrak{b}^{2}\frac{%
q_{0}^{2}-q_{\parallel }^{2}}{q^{2}}+\varkappa _{1,3}|_{\vert q^{2}\vert \gg m^{2}}.
\end{split}
\label{rekappa123}
\end{equation}
Notably, if the magnetic-field-depending contribution dominates the vacuum polarization and $q_0^2-q_\parallel^2\gg q_\perp^2$,  the second eigenvalue of the polarization tensor 
$\varkappa_2\approx m^2\mathfrak{b}^2$ [compare with the corresponding QED expression given  in the  second line in Eq.~(\ref{qedbehavior})]. Hence, the vacuum polarization 
of AED would dominate over the QED contribution if $m\mathfrak{b}\gg (2\alpha b/\pi)^{\nicefrac{1}{2}} m_e$ corresponding to $g\gg m_e [2\alpha/(B B_0)]^{\nicefrac{1}{2}}$. 
For a magnetic field $B\sim 10^{22}\  \rm G$ this could occur if  $g\gg10^{-3}\ \rm GeV^{-1}$. For ALPs masses $m< 10 \  \rm MeV$, values of $g$ overpassing the previous limitation  
have been ruled out  from a variety of observations in particle physics, astro-particle physics and cosmology. A compilation of discarded areas in the  parameter space of ALPs can 
be found in Ref.~\cite{Jaeckel:2015jla,Dobrich:2015jyk}.  A small window $10^{-2}\ \mathrm{GeV}<m<1\ \rm GeV$ with $g<10^{-2} \ \rm GeV^{-1}$ remains for which $10^{19}\ \mathrm{G}<B_c<10^{21}\ \mathrm{G}$. 
Another possibility is to  consider the effects associated with a neutral pion. In such a case, the substitutions $m\to m_\pi=135 \rm \; MeV$ and $g\to\alpha/(\pi f_\pi)\approx0.025\ \rm GeV^{-1}$, 
with $f_\pi=92\rm\; MeV$ denoting the  pion decay constant  have to be made and  $B_c=2.8\times 10^{20}\ \rm G$. So, for fields exceeding this value  one can expect a dominance of 
the vacuum polarization linked to neutral pions over the one associated with QED.

Now, if $q^{\mu }$ is a timelike four-vector [$q^{2}>0$] with $\vert q^{2}\vert\gg m^{2}$, the $\Pi ^{\mu \nu }$-eigenvalues are all complex. While the asymptotes for their real parts 
coincide with the expression given in Eq.~(\ref{rekappa123}), their imaginary contributions are given by $\mathrm{Im}\left[ \varkappa _{i}\right]_{\vert q^2\vert\gg m^{2}}\approx -g^2 q^{4}/(96\pi )$ 
for $i=1,2,3$. In line with this statement, we recall that an exact expression for the imaginary parts of the eigenvalues $\varkappa_1$ and $\varkappa_3$ in Eq.~(\ref{tensordecomposition}) 
can be obtained [for details see Ref.~\cite{alinapaper}]: 
\begin{equation}
\mathrm{Im}[\varkappa _{1,3}]=-\frac{g^{2}}{96\pi q^{2}}\left[ q^{2}-m^{2}%
\right] ^{3}\Theta (q^{2}-m^{2}),  \label{imkappa13}
\end{equation}%
where $\Theta (x)$ denotes the unit-step function: $\Theta (x)=1$ at $x\geqslant 0$, $\Theta (x)=0$ at $x<0$. The formula corresponding to the second eigenvalue follows by restoring the $i0-$prescription in 
Eq.~(\ref{tensordecomposition}) [$m\rightarrow m-i0$] and applying the relation $ (x+i0)^{-1}=\mathrm{P}\frac{1}{x}-i\pi \delta (x)$, with $\mathrm{P}$ referring to the Cauchy principal value. Consequently, 
\begin{equation}
\mathrm{Im}[\varkappa_2]=- m^2\pi\mathfrak{b}^2(q_0^2-q_\parallel^2)\delta(q^2-m^2)+\mathrm{Im}[\varkappa_{1,3}].\label{imkappa2}
\end{equation}
Here the expression for $\mathrm{Im}[\varkappa _{1,3}]$ is given by Eq.~(\ref{imkappa13}). Observe that, when the condition $\vert q^{2}\vert\gg m^{2}$ is met, both Eq.~(\ref{imkappa13}) and (\ref{imkappa2}) 
reduce to the asymptotic formula given right below Eq.~(\ref{rekappa123}).

%%%%%%%%%%%%%%%%%%%%%%%%%%%%%%%%%%%%%%%%%%%%%%%%%%%%%%%%%%%%%%%%%%%%%%%%%%%%%%%%%%%%%%%%%%%%%%%%%%%%%%%%%%%%
\section{Anisotropy in refraction and capture of photons in a superstrong magnetic field }
%%%%%%%%%%%%%%%%%%%%%%%%%%%%%%%%%%%%%%%%%%%%%%%%%%%%%%%%%%%%%%%%%%%%%%%%%%%%%%%%%%%%%%%%%%%%%%%%%%%%%%%%%%%%

%%%%%%%%%%%%%%%%%%%%%%%%%%%%%%%%%%%%%%%%%%%%%%%%%%%%%%%%%%%%%%%%%%%%%%%%%%%%%%%%%%%%%%%%%%%%%%%%%%%%%%%%%%%%
\subsection{Dispersion laws, group velocity and degrees of freedom \label{polarizationtensorrenormalizedc}}
%%%%%%%%%%%%%%%%%%%%%%%%%%%%%%%%%%%%%%%%%%%%%%%%%%%%%%%%%%%%%%%%%%%%%%%%%%%%%%%%%%%%%%%%%%%%%%%%%%%%%%%%%%%%

\begin{figure}[t]
\centering
\includegraphics[width=7cm]{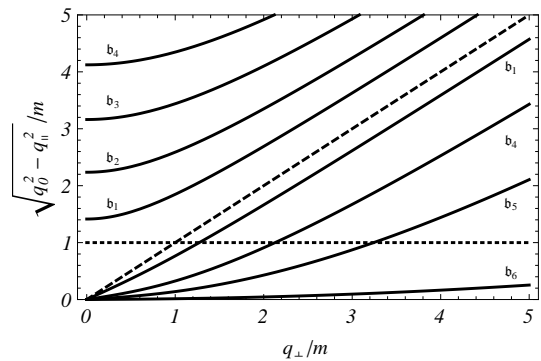}
\caption{Dipersion relations linked to the second propagation mode of the vacuum polarization tensor in quantum AED [see Eq.~(\protect\ref{dispersionrelations})]. While the dashed line 
represents the light cone law, the horizontal dotted one shows the situation for which $q_0^2-q_\parallel^2=m^2$. The four curves below the light cone line are associated with the massless branch. Conversely, 
above this one, four massive dispersion laws are depicted. These curves were determined by choosing various magnetic field parameters $\mathfrak{b}_i$ with $i=1,2,3,4,5,6 $ so that 
$\mathfrak{b}_1=1,\ \mathfrak{b}_2=2,\ \mathfrak{b}_3=3,\ \mathfrak{b}_4=4,\ \mathfrak{b}_5=10,\ \mathfrak{b}_6=100$.}
\label{fig:002}
\end{figure}

Although resulting from different nature, the singularity exhibited in the second eigenvalue at $q^{2}=m^{2}$ [see Eq.~(\ref{tensordecomposition})] somewhat resembles the known 
situation in QED, where there appears an inverse square root singularity in $\varkappa_{2}$ at the border to the  continuum of free electron and positron states produced by a
single photon: the cyclotronic resonance of the vacuum \cite{shabad1972,Shabad:1975ik}, or the pole where the electron and positron are bound to form a positronium atom \cite{Herold,shabadusov1,shabadusov2}.  
The  genuine dispersion relations associated with the mode-$2$ eigenwave neither are $q_{0}=\vert\pmb{q}\vert$ nor $q_{0}=\sqrt{\pmb{q}^{2}+m^{2}}$ but rather 
\begin{equation}
q_{0\mp }^{2}=\pmb{q}^{2}+\frac{1}{2}m_{\ast }^{2}\left[ 1\mp \sqrt{1+\frac{4%
\mathfrak{b}^{2}}{1+\mathfrak{b}^{2}}\frac{q_{\perp }^{2}}{m_{\ast }^{2}}}%
\right] +\mathcal{O}(g^{4}).
\label{dispersionrelations}
\end{equation}%
The formula above results from solving the nontrivial algebraic equation that arises when Eq.~(\ref{DSE}) is Fourier transformed, i.e. the equation $q^{2}=\varkappa _{2}$.  Observe 
that the dispersion law labeled with a negative sign describes a massless particle as $q_{0-}$ vanishes when $q_{\perp ,\parallel }\rightarrow 0$. Conversely, the 
dispersion relation $q_{0+}$ corresponds to a massive mode [$\lim_{q_{\parallel ,\perp }\rightarrow 0}q_{0+}^{2}=m_{\ast }^{2}$] whose effective mass $m_{\ast }=m\sqrt{1+\mathfrak{b}^{2}}$ 
reduces to the ALP mass $m$ only when the external field is zero. Therefore, a small-amplitude electromagnetic wave characterized by the second propagation mode is actually a state 
in which massive and massless particles coexist. Both branches of the dispersion law [see Eq.~(\ref{dispersionrelations})] are depicted in Fig.~\ref{fig:002} for various values of 
the $\mathfrak{b}$ parameter. As  $\mathfrak{b}$ grows, the massless dispersion curves tend to stick to the horizontal axis.  This behavior can be understood from Eq.~(\ref{dispersionrelations}) 
when the case $q_\perp\ll m_*\approx m\mathfrak{b}$ with $\mathfrak{b}\gg 1$ is considered:
\begin{equation}\label{infrredde}
 q_{0-}^2-\pmb{q}_\parallel^2\approx\frac{q_\perp^4}{m^2\mathfrak{b}^2}.
\end{equation}
The pattern in Fig.~\ref{fig:002} obtained by studying the resulting photon field is to be interpreted in such a way that the axion and mode-$2$ photon interact in the magnetic field 
to produce the spectrum of two disconnected branches: one massless and one massive. [The same spectrum is obtained by studying the resulting axion field in Appendix~\ref{Appendix}]. The 
massless branch may be referred to as belonging to a photon modified by its interaction with the axion, while the massive branch corresponds to the axion modified by its interaction with 
the mode-$2$ photon. No mixed state between the axion and the photon is formed and, thus, no oscillation between these particles can takes place in a strictly constant and homogeneous 
magnetic field. Here we encounter a vast difference with QED in this type of background, where the free photon dispersion curve quasi-intersects with the dispersion curve of an electron-positron 
pair -- mutually free or bound into a positronium atom -- before the interaction between the photon and the pair is taken into account. This quasi-intersection results in forming the 
photon-pair mixed state, the polariton, after this interaction comes into play \cite{shabadnat,shabad3v,Herold,shabadusov1,shabadusov2}. In such a context,  photon-positronium oscillations 
are possible \cite{Leinson1,Leinson2}.

When calculating  the components of the group velocity   $\mathpzc{v}_{\perp,\parallel}=\partial q_{0-}/\partial q_{\perp,\parallel}$ linked to  Eq.~(\ref{infrredde}),  we find 
\begin{equation}
\mathpzc{v}_{\perp}\approx \frac{2}{\mathfrak{b}^2} \frac{q_\perp^3}{m^2 q_{0-}},\qquad \mathpzc{v}_{\parallel}\approx \frac{q_\parallel}{q_{0-}}.  \label{q-par-is-zero}
\end{equation}As a consequence, the angle between the direction of  propagation of the electromagnetic energy and the external magnetic field 
\begin{equation}\label{thetabri}
\tan \theta= \frac{ \mathpzc{v}_{\perp}}{ \mathpzc{v}_{\parallel}}\approx\frac{2}{\mathfrak{b}^2}\frac{q_\perp^2}{m^2}\tan \vartheta
\end{equation} does not coincide with the one [$\tan \vartheta=q_{\perp}/ q_{\parallel}$]  between the photon  momentum $ \pmb{q}$  and  $\pmb{B}$,  [$\tan \theta<\tan \vartheta$]. 
Furthermore, we  observe that it tends to vanish when  $q_{\perp }\ll m\mathfrak{b}$ the faster, the stronger the field is, since one has  [see Eq.~(\ref{q-par-is-zero})]:
\begin{equation}
\mathpzc{v}_{\perp}\to 0\quad\mathrm{and} \quad  \mathpzc{v}_{\parallel}\to1.
\label{groupvel}
\end{equation} This implies that the group velocity becomes parallel to the magnetic field  for hard, as well as for soft photons. 

The described situation is not new. It resembles the main feature on which the capture effect of photons 
with energy much lower than the first pair creation threshold relies \cite{shabad2004}. Following our discussion  below Eq.~(\ref{sdfsdfsdfasdfsdfsfs}), we  expect that this phenomenon 
could also be realized via a certain class of  virtual ALPs rather than by virtual electron-positron pairs, provided $\mathfrak{b}\gg1$. We will see very shortly that, in addition to 
$\gamma-$quanta \cite{shabadnat,shabad3v}, and as a direct consequence of Eq.~(\ref{q-par-is-zero}), X-rays, optical light  \cite{shabad2004}, infrared radiation and electromagnetic  
microwaves could undergo the capture due to the  magnetized vacuum of AED near the surface of neutron stars.

Let us investigate the refractive properties linked to the  massive branch [lower sign in Eq.~(\ref{dispersionrelations})]. When  $\mathfrak{b}\gg1$ and  $q_\perp \ll m_*\approx m\mathfrak{b}$ 
the associated dispersion relation approaches
\begin{equation}
q_{0+}^{2}\approx 2q_{\perp}^{2}+q_{\parallel }^{2}+m_{\ast }^{2}.
\end{equation}In this case the components of the group velocity  $\mathpzc{v}_{\perp,\parallel}=\partial q_{0+}/\partial q_{\perp,\parallel}$ are
\begin{equation}
\mathpzc{v}_{\mathbf{\perp }}\approx\frac{2q_{\perp }}{q_{0+}},\quad \mathpzc{v}_{\mathbf{\parallel }}\approx\frac{q_{\parallel }}{q_{0+}}\label{speedmassive}
\end{equation}As a result, the cosine of the angle $\eta$ between the group velocity vector $\pmb{\mathpzc{v}}$ and the momentum
is\begin{equation}
\cos \eta =\frac{q_{\parallel }\mathpzc{v}_{\parallel }+q_{\perp }\mathpzc{v}_{\perp }}{\vert \pmb{q}\vert \sqrt{\mathpzc{v}_\perp^2+\mathpzc{v}_\parallel^2}}\approx\frac{q_{\parallel }^{2}+2q_{\perp }^{2}}{\vert\pmb{q}\vert\left(4q_{\perp}^{2}+q_{\parallel }^{2}\right) ^{1/2}}.
\end{equation} In terms of the angle $\vartheta$ between the photon momentum and the magnetic field [see below Eq.~(\ref{q-par-is-zero})], this expression reads
\begin{equation}
\cos \eta \approx\frac{1+2\tan ^{2}\vartheta }{\sqrt{1+4\tan ^{2}\vartheta}}\cos\vartheta.
\end{equation}Within the range $0<\vartheta <\pi/2$ the angle $\eta$ remains very small and achieves its maximum value of $\eta \approx \arccos[0.943]=0.339\ \rm rad$ at $\vartheta\approx 0.2\pi$.  
This is the deflection that the direction of propagation of the wave undergoes from its momentum vector due to the anisotropy of the medium.

Now, in QED, there are two photon degrees of freedom that propagate in the  magnetized vacuum with different dispersion laws   \cite{Shabad:1975ik}. This property of birefringence is another manifestation 
of the anisotropy induced by the external field via the polarization tensor. Let us investigate a similar issue within the framework of AED. Primarily, this theory contains three degrees of freedom: 
two (massless) photon degrees and one (massive) axion degree of freedom. The interaction cannot change the overall number of degrees of freedom. The homogeneous  free-wave equation~(\ref{DSE}) for vector-potentials 
of mode-$2$ in momentum representation is satisfied separately by each of the two terms in the expansion
\begin{equation}\label{electromagneticwave}
\mathpzc{a}_{2}^{\mu}(q)=\mathcal{A}_{2}^{(-)}\delta \left(q_{0}^2-q_{0-}^2\right) \flat_{2}^{\mu }+\mathcal{A}_{2}^{(+)}\delta \left(q_{0}^2-q_{0+}^2\right) \flat _{2}^{\mu}.
\end{equation}
Hence the two arbitrary coefficients  here $\mathcal{A}_{2}^{(\pm )}(q)$ are independent of each other, i.e. there is no way to correlate the amplitudes of excited photon and axion waves on the basis of the field
equation alone. We write a (positive frequency) solution of the free-wave equation~(\ref{DSE}) for each eigenmode as 
\begin{equation}
\mathpzc{a}_{j}^{\mu }(q)=\overline{\flat}_{j}^{\mu }\delta \left(q_{0}-q_{0}^{\left(j\right) }\right),\label{choice}
\end{equation}
where $q_{0}^{\left( 2\right) }=q_{0-}$ [see Eq.(20)]. The other two values are independent of the presence of axions and are determined by the dispersion laws of mode-$1$, which in the loop approximation of QED 
coincides with the vacuum dispersion law  $q_{0}^{\left( 1\right) }=\vert\pmb{q}\vert$, and of mode-$3$ whose dispersion law is of the general form  $q_{0}^{\left( 3\right)}=[q_{\parallel }^{2}+f(q_{\perp }^{2})]^{\frac{1}{2}}$,  
the same  as $q_{0-}$ [see Eq.~(20)].  The eigenvectors $\overline{\flat}_{j}^{\mu}$ are chosen to differ -- in the rest frame of the magnetic field -- from the vectors (\ref{decomposition}) by a factor: 
$\overline{\flat }_{j}^{\mu }=\flat _{j}^{\mu}/(q_{\perp}B^{2})$. The division by $q_{\perp}$ is crucial, as we shall see, to exclude mode-$1$ and to gather conveniently the two massless degrees of freedom 
in modes-$2$ and -$3$.

Then the electric  $\pmb{e}_{j}\sim(q_{0}\;\bar{\pmb{\flat}}_{j}-\pmb{q}\;\bar{\flat}_{j}^{0})$  and magnetic  $\pmb{h}_{j}\sim \pmb{q}\times \bar{\pmb{\flat}}_{j}$ fields of each mode behave as:
\begin{eqnarray}\label{polarization}
\begin{array}{c}\displaystyle
\pmb{e}_{1}\sim-\pmb{n}_{\perp }q_{0}q^{2},\quad \pmb{h}_1\sim-\left[\pmb{n}_{\perp}\times \pmb{q}_{\parallel }\right] q^{2},\\ \\
\displaystyle\pmb{e}_{2\perp }\sim\pmb{n}_{\perp}q_\parallel^2,\quad\pmb{e}_{2\parallel}\sim\frac{\pmb{q}_{\parallel}}{q_\perp}(q_{\parallel}^2-q_{0}^{2}),\\ \pmb{h}_2\sim q_{0}\left[\pmb{q}_{\parallel}\times \pmb{n}_{\perp}\right], \\ \\
\displaystyle\pmb{e}_{3}\sim-q_{0}\left[\pmb{n}_{\perp}\times\pmb{q}_{\parallel}\right],\\\displaystyle \pmb{h}_{3\perp}\sim-\pmb{n}_{\perp}q_{\parallel}^2,\quad \pmb{h}_{3\parallel }\sim \pmb{q}_{\parallel}q_{\perp },
\end{array}
\end{eqnarray}Here $\pmb{n}_{\perp,\parallel}=\pmb{q}_{\perp}/\vert\pmb{q}_{\perp}\vert$ is a  unit vector along the perpendicular direction of $\pmb{B}$. It may seem that $\pmb{e}_{\parallel }^{(2)}$ in 
Eq.~(\ref{polarization}) is singular in $q_{\perp }=0$. This is not the case, however, because if $q_{\perp }=0$ also $q_{\parallel }^{2}-q_{0}^{2}=0$ for a massless branch [see Eq.~(\ref{infrredde})].

Note that the set of eigenvectors in Eq.~(\ref{decomposition}) remains an orthogonal basis also when the background field is absent. Therefore, these formulae may be  used, too. In the vacuum with no background magnetic 
field there are two polarizational degrees of freedom, and these are modes $3$ and $2$. The mode-$3$ is identically transverse, $\pmb{e}_3\cdot \pmb{q}=\pmb{h}_3\cdot \pmb{q}=\pmb{e}_3\cdot \pmb{h}_3=0$, whereas 
the mode-$2$ becomes transverse under the vacuum dispersion law: $\pmb{e}_2\cdot \pmb{q}=q_{\parallel}^2\left(\pmb{q}^{2}-q_{0}^{2}\right)/q_\perp=0$ if $q_{0}^{2}=q_{\parallel }^{2}+q_{\perp }^{2}$.  In this case, 
the relations $\pmb{h}_2\cdot \pmb{q}=\pmb{e}_2\cdot \pmb{h}_2=0$ are fulfilled identically. As for mode-$1$, it is seen from Eq.~(\ref{polarization}) that it nullifies in the vacuum as long as the dispersion
law is $q^{2}=0$.

In the magnetic field, mode-$2$ is the so-called extraordinary wave, whose longitudinal electric component is generally nonzero, $\pmb{e}_2\cdot\pmb{q}\neq 0$, since now the dispersion law implies $q^{2}\neq 0$. 
The exclusion is provided in the background magnetic field by the momentum parallel to the latter: $\pmb{q}_{\perp }=\pmb{0}$. In this case the longitudinal component of mode-$2$,  
$\pmb{e}_2\cdot \pmb{q}=q_{\parallel}^2(q_{\parallel}^{2}-q_{0}^{2})/q_\perp \rightarrow 0$  due to the dispersion law $q_{0}^{2}-q_{\parallel }^{2}=f(q_{\perp }^{2})$ [see Eq.~(\ref{dispersionlawsgeneral})], since 
for the massless branch the function $f(q_{\perp }^{2})$ turns to zero as fast as $q_{\perp}^{2}$ in QED or faster than $q_{\perp}^{2}$ in the axion theory according to Eq.~(\ref{groupvel}). The identically transverse mode-$3$  
continues to represent the second degree of freedom in the magnetic field, too. As for mode-$1$, in the background magnetic field, at least as long as the one-loop approximation of QED is used, its dispersion law  
remains $q^{2}=0$ \cite{Shabad:1975ik}.  Therefore the electric and magnetic fields in mode-$1$ vanish according to Eq.~(\ref{polarization}), leaving again only two polarizational degrees of freedom on the 
mass-shell $q^2=\varkappa_j$.\footnote{The vanishing of mode-$1$ may more directly be  seen already from expression for $\flat_{1}^{\mu}$ in Eq.~(\ref{decomposition}). The four-gradient term $\sim q^{\mu}$ in it gives
no contribution to the field strength, whereas the $q^{2}$-term nullifies thanks to the dispersion law $q^{2}=0$.} Therefore, with the choice of the amplitude of the free waves fixed as in Eq.~(\ref{choice}), mode-$1$ 
remains unphysical for every direction of propagation. For this reason we shall not take it hitherto into account.

%%%%%%%%%%%%%%%%%%%%%%%%%%%%%%%%%%%%%%%%%%%%%%%%%%%%%%%%%%%%%%%%%%%%%%%%%%%%%%%%%%%%%%%%%%%%%%%%%%%%%%%%%%%%%%%%%%%%%%%%%%%%%%%%%%%%%%%%%%%%%%%%%
\subsection{Implications of a plausible capture of $\pmb \gamma-$quanta mediated by virtual axions \label{Implicationcapture1}}
%%%%%%%%%%%%%%%%%%%%%%%%%%%%%%%%%%%%%%%%%%%%%%%%%%%%%%%%%%%%%%%%%%%%%%%%%%%%%%%%%%%%%%%%%%%%%%%%%%%%%%%%%%%%%%%%%%%%%%%%%%%%%%%%%%%%%%%%%%%%%%%%%

Let us now consider the photon capture effect in different terms. In understanding that the group velocity specifies the direction of propagation of the wave envelope, the  suppression of 
its perpendicular component implies the effect of capture of the  mode-$2$ photon by the strong magnetic field, as described previously in QED. We will see, however, that this phenomenon 
has quite different features in the present context. Observe that  Eqs.~(\ref{q-par-is-zero}) and (\ref{speedmassive}) apply for $q_{\perp}\ll m_*$  with $\mathfrak{b}\gg1$. Beyond this limit, i.e. for $q_\perp\gg m_*$, 
the components of the group velocity approach
\begin{equation}
\mathpzc{v}_\perp\approx\frac{q_\perp}{q_{0\mp}},\quad \mathpzc{v}_\parallel=\frac{q_\parallel}{q_{0\mp}},
\end{equation}and the angle $\theta=\arctan(\mathpzc{v}_\perp/\mathpzc{v}_\parallel)$ between the direction of propagation of the electromagnetic energy and $\pmb{B}$ does not deviate substantially 
from the one formed by the momentum of the small-amplitude wave and the external field [$\theta\approx \vartheta=\arctan(q_\perp/q_\parallel)$].

Suppose that we have a magnetic field with its lines of force curved. When propagating in such a field a photon changes its momentum components $q_{\perp}$ and $q_{\parallel }$, while 
its energy $q_{0}$ is kept constant, because the external field does not depend on time. Thereby the photon evolves along its dispersion curve. Let primarily the so-called curvature 
$\gamma $-quantum with $q_{0}>2m_{e}$ be injected in the direction parallel to the magnetic field, as it is the case within the widely accepted scenario of the events that take place in 
the pulsar magnetosphere. The curvature gamma-quantum is emitted by an electron propagating along a curved magnetic line of force. As the photon propagates inside the dipole-shaped magnetic field it gains 
a crucial pitch angle with the direction of this field, kinematically sufficient for the photon to create an electron-positron pair. This takes place when $q_{0}^{2}-q_{\parallel}^{2}=4m_{e}^2$. 
A ``classical'' on-shell photon is characterized by the  vacuum dispersion law  $q_{0}^2-q_{\parallel}^2=q_{\perp}^2$ [dashed line in Fig.~\ref{fig:002}] unless the magnetic field exceeds 
too much the critical scale  $B_0=0.1\; m_{e}^2/e\approx4.42\times 10^{12}\ \rm G$, in which case the resonant phenomena \cite{shabadnat} must be  taken into account.\footnote{When taking into account the resonant 
phenomena owing to the singularity of the polarization tensor in momentum space in the point, where the photon creates a free or bound pair (the positronium atom), it was shown that the 
capture effect is present in QED. Its role in QED was reduced to a certain increase of the interval where the photon freely propagates before it turns into the pair. We shall not take this 
correction into account in the consideration that follows.} Therefore, before reaching the necessary pitch angle -- which happens for mode-$2$ photons at $q_{\perp}=2m_{e}$ -- the photon propagates 
straightforwardly in the coordinate space and then it produces the pair.  This is the mechanism of the formation of the electron-positron plasma by mode-$2$ photons responsible for the 
further directed electromagnetic emission of the pulsar. 

Now, in accordance with Eq.~(\ref{groupvel})  the admission of the axion-photon  interaction makes the curvature photon of mode-$2$  be captured by the magnetic line of force practically immediately 
after the photon is  emitted,  provided $q_{\perp}\ll m_*$. This condition might  lead to the capture of massless mode-$2$ photons  of the order of the characteristic energy 
scale associated with the electron mass,  i.e. for $m_{*}\gg m_{e}$ with $\mathfrak{b}\gg1$. However, if this was the case,  a single sufficiently hard -- belonging to the $\gamma $-range -- photon in 
the magnetic field  has no chance to produce a pair for long, the established mechanism of the pulsar radiation being destroyed.\footnote{As for  mode-$3$ photons not affected by the 
presence of axions these could contribute to the plasma creation only if their energy is larger than $m_{e}\left[1+\left(1+b\right)^{1/2}\right] >2m_{e}$. For sufficiently large magnetic field they 
cannot save the necessary plasma creation mechanism resting on the less energetic mode-$2$ photons, once the latter ones are excluded from the game by the presence of axions, as described above.} Indeed, 
following our preceding  discussion,  in the magnetic field the  production of an electron-positron pair by a $\gamma$-photon is possible if $q_{0-}^{2}-q_{\parallel }^{2}\geqslant 4m_{e}^{2}$. 
Substituting $q_{0-}^{2}-q_{\parallel }^{2}=4m_{e}^{2}$ into Eq.~(\ref{dispersionrelations}) we obtain the equation [$\mathfrak{b}\gg1$] 
\begin{equation}
4m_{e}^{2}=\left( q_{\perp }^{2}\right) _{\text{thr}}+\frac{1}{2}m_{\ast}^{2}\left[ 1-\sqrt{1+4\frac{\left( q_{\perp }^{2}\right) _{\text{thr}}}{m_{\ast }^{2}}}\right]
\end{equation}for the threshold value of $q_{\perp }^{2}$ that borders this process. Its positive solution is 
\begin{equation*}
\left( q_{\perp }^{2}\right) _{\text{thr}}=4m_{e}^{2}+m_{\ast }^{2}+\frac{1}{%
2}m_{\ast }\sqrt{m_{\ast }^{2}+4m_{e}^{2}.}
\end{equation*}Hence, the accepted mechanism of plasma generation in neutron stars is destroyed if it may take place at the values $\left( q_{\perp }^{2}\right) _{\text{thr}}$ essentially exceeding the 
threshold value of $\left( q_{\perp }^{2}\right) _{\text{thr}}=4m_{e}^{2}$ characteristic of QED in a magnetic field. This occurs if the two conditions 
\begin{equation}\label{constraint123}
m_{\ast }^{2}>\frac{4}{3} m_{e}^{2}\quad \mathrm{and}\quad \mathfrak{b\gg }1
\end{equation} are met. In such a case the photon is captured by the magnetic line of force, and it cannot create a pair before it propagates far enough along the line to leave the region of the strong dipole 
field of the pulsar where the one-photon pair creation is no longer possible. In other words, if in this highly magnetized object  AED dominates over QED,  mechanisms other than the one recognized so far 
would be  required for explaining the plasma generation in neutron star surfaces. Let us examine the axion parameters for which  this phenomenon could take place. Bearing in mind that in the surface of  neutron 
stars the  magnetic field strength is of  the order of $B\sim 10^{13}\  \rm G$ ,  we find from Eq.~(\ref{constraint123})  that the  established  mechanism of radiation is detroyed provided the axion-diphoton 
coupling constant lies  above the upper limit $g>10^3\ \mathrm{GeV}^{-1}$ for masses  $m\ll 1\ \rm keV$. It is worth remarking that this combination of  axion parameters is  ruled out by  various 
experimental observations. In fact, for ALPs masses  $m\ll 1\ \rm keV$, the OSCAR collaboration has constrained  $g$ to lie below $g<4\times 10^{-8}\ \rm GeV^{-1}$ \cite{Balou}. Thus, we verify that for fields as large as 
$B\sim 10^{13}\ \rm G$  and for  ALPs masses $m\ll 1\ \rm keV$, the current bounds on $g$ forbid a dominance of AED over QED phenomenology. 

We are to comment also on the possibility of pair creation by the "photon" belonging to the massive branch. Kinematically this process is allowed if  $q_{0}^{2}-q_{\parallel}^{2}\geqslant 4m_e^2$.  
However, the curvature $\gamma $-quantum, i.e. the one emitted in the pulsar magnetosphere by an electron moving along a curved magnetic line of force, is the standard zero-mass photon. The option of creating 
excitations belonging to the masive branch is not seen. Thereby, the plasma creation by the massive-branch ``photon'' is left beyond discussion.

%%%%%%%%%%%%%%%%%%%%%%%%%%%%%%%%%%%%%%%%%%%%%%%%%%%%%%%%%%%%%%%%%%%%%%%%%%%%%%%%%%%%%%%%%%%%%%%%%%%%%%%%%%%%%%%%%%%%%%%%%%%%%%%%%
\subsection{Capture of lower-energy photons and polarization of thermal radiation of pulsars \label{Implicationcapture2}}
%%%%%%%%%%%%%%%%%%%%%%%%%%%%%%%%%%%%%%%%%%%%%%%%%%%%%%%%%%%%%%%%%%%%%%%%%%%%%%%%%%%%%%%%%%%%%%%%%%%%%%%%%%%%%%%%%%%%%%%%%%%%%%%%%

The direct thermal radiation emitted by the hot surface of neutron stars is directly observable in the X-ray range [see, e.g., Ref.~\cite{Potekhin}]. Let us inspect the consequence of the capture 
of their mode-$2$ part by the strong magnetic field due to the presence of axions as pointed out below Eq.~(\ref{thetabri}). To this end we first consider the refraction law of electromagnetic waves when they leave the region 
of strong magnetic field and enter the space free of it. This consideration will be independent of what mechanisms are responsible for formation of the vacuum polarization and dispersion laws, in 
other words of whether there are ALPs or we just face the usual QED with magnetic field.

Let an electromagnetic wave be created at the surface of a neutron star inside a strong magnetic field, which gradually disappears at the distance $\ell$ far from the surface. We accept that 
the wave length $\lambda$ is much less than $\ell$, and hence adiabatic approximation is valid, wherein the dielectric properties of the medium may be considered as depending on the coordinate. 
Conventionally, the magnetic field changes only along its own direction, but is constant in the direction transverse to it. Then the photon energy (frequency) $q_{0}$ and the photon momentum component 
$q_{\perp}$ across the magnetic field do not change, when the wave leaves the region occupied by the magnetic field, due to the corresponding  space-time translational invariance. On the contrary, 
the momentum component along the magnetic field $q_{\parallel}$ does change, because the magnetic field varies in this direction. We shall supply this component with the subscript ``mag'', when it 
relates to the region occupied by the magnetic field, and with ``vac'' ouside of it. There are two different dispersion laws:
\begin{eqnarray}\label{dispersionlawsgeneral}
q_{0}^{2}=q_{\parallel\mathrm{vac}} ^{2}+q_{\perp }^{2}, \quad q_{0}^{2}=q_{\parallel\mathrm{mag}}^{2}+f(q_{\perp}^{2}).  
\end{eqnarray}The last relation is a consequence of relativistic covariance, the function $f(q_{\perp }^{2})$ is model-dependent and it  is different for each propagating mode.  According to 
Eq. (\ref{infrredde})  one has
\begin{equation}
f(q_{\perp }^{2})=\frac{q_{\perp }^{4}}{m^2\mathfrak{b^2}}
\label{f(kperp)}
\end{equation}provided the condition $q_{\perp }\ll m\mathfrak{b}$ is fulfilled.\footnote{Note that in mode-3 the function $f(q_\perp^2)$ is proportional to $q^2_\perp$ at small $q_\perp$. Therefore, 
this region of momentum is an arena for utmost manifestation of birefringence.} Let $\vartheta_{\rm vac}$ be the refraction angle, i.e. the angle between the outgoing momentum $\pmb{q}_{\rm vac}$ and the magnetic field 
\begin{equation}\label{analysis1}
\sin \vartheta_{\rm vac}=\frac{q_{\perp }}{\vert\pmb{q}_{\rm vac}\vert}=\frac{q_{\perp }}{\text{ }q_{0}},
\end{equation}
and $\vartheta_{\rm mag}$ the incident angle, i.e. the angle between the ingoing momentum $\pmb{q}_{\rm mag}$ and the magnetic field. With the use of the dispersion laws, Eq.~(\ref{dispersionlawsgeneral}), 
it is expressed in terms of the refraction angle as 
\begin{equation}
\begin{split}\label{analysis2}
&\sin \vartheta_{\rm mag}=\frac{q_{\perp }}{\vert\pmb{q}_{\rm mag}\vert}=\sqrt{\frac{q_{\perp }^{2}}{q_{\perp }^{2}+q_{0}^{2}-f(q_{\perp }^{2})}}\\
&\qquad\quad=\frac{1}{\sqrt{1+\frac{1}{\sin^{2}\vartheta_{\rm vac}}-\frac{1}{q_{0}^{2}}f(q_{0}^{2}\sin^{2}\vartheta_{\rm vac})}}.
\end{split}
\end{equation}This can  also be  written as 
\begin{equation}
\begin{split}
&\tan ^{2}\vartheta_{\rm mag}=\frac{\sin ^{2}\vartheta_{\rm vac}}{1-\frac{1}{q_{0}^{2}}f(q_{0}^{2}\sin^{2}\vartheta_{\rm vac})}\\
&\qquad\qquad\qquad \qquad\approx \frac{\sin ^{2}\vartheta_{\rm vac}}{1-\frac{q_{0}^{2}}{m_*^{2}}\sin^{4}\vartheta_{\mathrm{vac}}},
\end{split}\label{refraction3}
\end{equation}provided that $q_{0}\sin \vartheta_{\rm vac}\ll m_*$.  The critical angle $\vartheta _{\rm cr}$  for total internal reflection is achieved when $\vartheta_{\rm vac}=\pi/2$. It depends on the energy: 
\begin{equation}
\vartheta _{\mathrm{cr}}=\arctan \left(1-2\frac{q_{0}^{2}}{m_*^{2}}\right)^{-\frac{1}{2}}\gtrsim\frac{\pi }{4} 
\end{equation}Only approximately one half of the isotropic long-wave radiation has a chance to leave the magnetic field. 

The capture effect further modifies the situation. We have found [see Eq.~(\ref{q-par-is-zero})] 
that on the massless photon branch for $q_{\perp }^{2}$ as small as $q_{\perp }^{2}=q_{0}^{2}\sin ^{2}\vartheta_{\rm vac}\ll m_*^{2}=m^{2}\mathfrak{b}^{2}$ all photons are canalized towards the direction 
of the magnetic field. The fulfillment of this strong inequality is guaranteed for all incident angles provided that the photon energy $q_{0}$ is much less than $m_*$. That is $q_{0}\ll gB$. On the other 
hand, the dispersion curve passes through the point $\vartheta_{\rm vac}=\vartheta_{\rm mag}=0$, which means that the wave parallel to the magnetic field is not deflected, when it goes out of it. Therefore 
the whole mode-$2$ part of the radiation would gather in a beam going out of the magnetic field parallel to it. 

As the mode-$2$, in the  axion-dominating regime, is canalized parallelly to the background magnetic field,  most of its energy is concentrated at $\pmb{q}_{\perp}=\pmb{0}$.  As argued above, according to 
Eq.~(\ref{polarization}) it is a transverse wave with its electric field lying in the plane spanned by the vectors $\pmb{B}$ and $\pmb{q}$. In accord with the analysis above  [see  Eq.~(\ref{analysis1}) 
or (\ref{refraction3})], it leaves  the background field with the refraction angle close to the incident angle, when both are small.  On the other hand, the mode-$3$ component of the heat radiation, 
whose electric field is normal to the plane formed by the background magnetic field and the direction of propagation, is not canalized. (Recall, that it is not influenced by the presence of axions and 
remains as calculated via QED [see Ref.~\cite{shabad2004}]).  Its original angular distribution is only modulated by the refraction law  Eq.~(\ref{refraction3}), and it is not concentrated along any 
direction. Therefore, if the axion exists, the heat photons with  energies $q_{0}\ll g B$ polarized in the plane to which the vectors  $\pmb{B}$ and $\pmb{q}$ belong, peak along the direction of the 
magnetic field against the smooth background of photons polarized orthogonal to this plane.

This means that if the observation direction deviates from being parallel to the magnetic field we would face the smooth background of photons polarized orthogonal to this plane and the smooth background 
of axions -- that are not subject to the capture. The registered radiation would be scant of photons polarized in the said plane.

As long as we deal with the thermal mechanism of radiation we can judge upon how many accompanying axions may be present together with photons. The distribution of heat particles is characterized by the 
Boltzmann exponent $\sim e^{-\frac{q_{0+}}{k_B T}}$, where $q_{0+}$ is given in Eq.~(\ref{dispersionrelations}) and $k_B=8.6\times 10^5\;\rm eV/K$ denotes the Boltzmann constant. For the massive 
branch (\ref{dispersionrelations}),  $f(q_{\perp }^{2})$ is a growing function with  $f(0)=m_*^{2}$. So, we have 
\begin{equation}
\text{max }e^{-\frac{q_{0}}{k_{B}T}}=e^{-\frac{m_{\ast }}{k_{B}T}}
\end{equation}Consider now a light ALP with $m\sim(10^{-3}- 10^2)\;\rm \mu eV$ and  $g\sim 4\times10^{-8}\;\rm GeV^{-1}$ [see  discussion below Eq.~(\ref{sdfsdfsdfasdfsdfsfs})], for which the critical field in AED, 
$B_c\sim 10^{9}-10^{14}\;\rm G$,  lies  within the characteristic strengths  of pulsars. If the critical field is overpassed by two orders of magnitude then the effective axion mass $m_*\approx m\mathfrak{b}\sim 
(10^{-1}-10^{-4})\;\mu\mathrm{eV}$ becomes heavy due to the external magnetic field dressing [$m_*\gg m$]. However, for X-ray temperatures $T=10^{6}\;\rm K$  corresponding to  $k_{B}T=86\;\rm eV$ the exponential  in 
the right hand side  of the expression above is close to unity. This implies that the density of heat-created ALPs inside the magnetic field might be comparable to the density of photons, and a large amount of axions 
might escape from the pulsar as well. This fact constitutes a loss of energy which should accelerate the cooling of the pulsar and shorten its lifetime. In similarity to the case dealing with red giants, this property  could be 
exploited to establish new constraints on the axion parameter space. However, this is beyond the scope of the present investigation.

The peaking (or not peaking) of photons in the parallel direction is the best way to make sure that the axions exist (or do not). Besides, there exists already significant progress in high-purity polarimetric techniques for X-ray probes \cite{Marx2011,Marx2013} which are expected to be exploited in the envisaged experiment at HIBEF collaboration 
\cite{HIBEF,Schlenvoigt}. These techniques are planned to be used in polarimeters designed to investigate the soft X-ray emission in the strong magnetic fields of neutron stars and white dwarfs 
\cite{Soffitta:2013hla,Bernard:2013jea,Iwakiri:2016ioo,Caiazzo:2018evl}.\footnote{In the optical regime, polarization measurements  are within reach. They currently  provide evidences for 
vacuum birefringence in radiation coming from  isolated neutron stars \cite{Mignani:2016fwz}.} Eventually, these  measurements could allow us to probe the dominance and criticality of AED over QED, and provide new 
constraints on the ALP parameter space \cite{Wang:2009sg}. Indeed, if the proposed effect could be extinguished from the future observations one will be able to ban the values of the axion coupling $g>q_{0}/B$, where 
the pulsar value is taken for the magnetic field $B$ and the x-ray value for the photon energy $q_{0}$.

%%%%%%%%%%%%%%%%%%%%%%%%%%%%%%%%%%%%%%%%%%%%%%%%%%%%%%%%%%%%%%%%%%%%%%%%%%%%%%%%%%%%%%%%%%%%%%%%%%%%%%%%%%%%%%%%%%%
\section{The modified Coulomb potential in axion-electrodynamics\label{GP}}
%%%%%%%%%%%%%%%%%%%%%%%%%%%%%%%%%%%%%%%%%%%%%%%%%%%%%%%%%%%%%%%%%%%%%%%%%%%%%%%%%%%%%%%%%%%%%%%%%%%%%%%%%%%%%%%%%%%

%%%%%%%%%%%%%%%%%%%%%%%%%%%%%%%%%%%%%%%%%%%%%%%%%%%%%%%%%%%%%%%%%%%%%%%%%%%%%%%%%%%%%%%%%%%%%%%%%%%%%%%%%%
\subsection{General formulae\label{GPa}}
%%%%%%%%%%%%%%%%%%%%%%%%%%%%%%%%%%%%%%%%%%%%%%%%%%%%%%%%%%%%%%%%%%%%%%%%%%%%%%%%%%%%%%%%%%%%%%%%%%%%%%%%%

Aside from influencing the propagation of photons, the polarization tensor [see Eq.~(\ref{tensordecomposition})] also induces  modifications of Coulomb's law. This distortion is determined here through the temporal 
component of the electromagnetic four-potential 
\begin{equation}
\mathpzc{a}_\mu(x)=-i\int \mathpzc{D}_{\mu\nu}(x,\tilde{x})\mathpzc{j}^\nu(%
\tilde{x})d^4\tilde{x},  \label{fourpotential}
\end{equation}
where $\mathpzc{j}^\nu(\tilde{x})=\mathpzc{q}\delta^\nu_{\ 0}\delta^3(\tilde{x})$ is the four-current density of a point-like static charge $\mathpzc{q}$ placed at the origin $\tilde{\pmb{x}}=\pmb{0}$ of our reference 
frame. In this expression $\mathpzc{D}_{\mu\nu}(x,\tilde{x})$ denotes the effective photon propagator, i.e. the result of summing up the infinite series that determines the renormalized photon propagator [see 
Eq.~(\ref{perturbativeexpansionpropagator})]  when diagrams of all orders are included \cite{Gabrielli:2006im}. Consequently, the problem of determining how the quantum vacuum fluctuation of a pseudoscalar axion field 
$\phi(x)$ modifies the Coulomb potential $\mathpzc{a}_C(\pmb{x})=\mathpzc{q}/(4\pi \vert\pmb{x}\vert)$ reduces, to a large extent, to finding the explicit expression for $\mathpzc{D}_{\mu\nu}(x,\tilde{x})$. Here this 
is accomplished by using the identity $\int d^4\tilde{x}\mathpzc{D}_{\mu\lambda}^{-1}(x,\tilde{x})\mathpzc{D}^{\lambda\nu}(\tilde{x},x^\prime)=i\delta^\nu_{\ \mu}\delta^4(x-x^\prime)$, where 
$\mathpzc{D}_{\mu\lambda}^{-1}(x,\tilde{x})$ is given in Eq.~(\ref{bilinearEA}) which corresponds to summing up the geometric series of one-photon-reducible diagrams. However, the close analogy between our problem 
and the one previously analyzed in QED allows us to apply the model- and approximation-independent expression for $\mathpzc{D}^{\mu\nu}(x,\tilde{x})$ found in Ref.~\cite{shabad5} directly. This general formula links 
the previous object with the corresponding eigenvalues and eigenvectors of the polarization  tensor in QED. It only remains to carry out an appropriate replacement of these eigenvalues by those resulting from the photon-ALP 
oscillations. Hence, up to an inessential longitudinal contribution, the
photon propagator reads 
\begin{equation}
\mathpzc{D}^{\mu\nu}(x,\tilde{x})=\sum_{j=1}^3\int \frac{\dbar^4 q\;i}{q^2-\varkappa_j}\frac{\flat_j^\mu\flat_j^\nu}{\flat_j^2} e^{iq (x-%
\tilde{x})}.  \label{greencircular}
\end{equation}%
We insert this formula into Eq.~(\ref{fourpotential}) bearing in mind the details explained below. As a consequence, the axion-modified Coulomb
potential is given by 
\begin{equation}
\begin{split}
\mathpzc{a}_0(\pmb{x})=\mathpzc{q}\int  \dbar^3q\frac{(\pmb{q}^2+m^2) e^{-i\pmb{q}\cdot\pmb{x}}}{\pmb{q}^2\left(\pmb{q}%
^2+m^2\right)[1-\pi(\pmb{q}^2)]+g^2B^2q_\parallel^2}.
\end{split}
\label{initialpotential}
\end{equation}When establishing this expression we used the fact that $\flat_{1}^0=\flat_{3}^0=0$, i.e. that the  first and third eigenmodes do not contribute to the electrostatic interaction. 
This feature could be anticipated because, in the stationary limit $q_0=0$, ``virtual'' photons associated with  $\flat_{1,3}^\mu$ carry magnetic fields only, while off-shell photons of mode-$2$ are purely electric.

Observe that, in absence of an external field [$B=0$], the form factor $\pi(\pmb{q}^2)$ dominates the vacuum polarization. Since so far no deviations from the standard QED predictions 
have been observed, we can assume that $\pi(\pmb{q}^2)$ is very small [$\pi(\pmb{q}^2)\ll1$]. There are no reasons to expect a change in this condition when a strong magnetic field is
present. Thus, after  a linear expansion in $\pi(\pmb{q}^2)$, Eq.~(\ref{initialpotential}) can be approached conveniently by 
\begin{equation}
\mathpzc{a}_0(\pmb{x})\approx\mathpzc{q}\int{\dbar}^3q\frac{[1+\pi(\pmb{q}^2)]\left(1+\frac{\pmb{q}^2}{m^2}\right)}{\pmb{q}^2\left(1+\frac{\pmb{q}^2}{m^2}\right)+\mathfrak{b}^2q_\parallel^2}
e^{-i\pmb{q}\cdot\pmb{x}}.  \label{initialpotential}
\end{equation}%
In the limit of $B\to0$, corresponding to $\mathfrak{b}\to0$, this formula reduces to the expression for the axion-Coulomb potential found in Ref.~\cite{alinapaper}: 
\begin{equation}
\begin{split}
&\left.\mathpzc{a}_0(\pmb{x})\right\vert_{\mathfrak{b}\to0}=\frac{\mathpzc{q}}{4\pi\vert\pmb{x}\vert}\left[1+\frac{g^2m^2}{48\pi^2}\int_{1}^\infty \frac{du}{u^5}\right.\\ &\qquad\qquad\qquad\qquad\qquad\left.\times\left[u^2-1\right]^3e^{-m\vert\pmb{x}\vert u}\right] \\
&\qquad\ =\left\{
\begin{array}{cc}
\displaystyle \frac{\mathpzc{q}}{4\pi\vert\pmb{x}\vert}\left[1+\frac{g^2m^2}{%
\pi^2}\frac{e^{-m\vert\pmb{x}\vert}}{(m\vert\pmb{x}\vert)^4}\right] & \vert%
\pmb{x}\vert\gg\lambda, \\ 
&  \\ 
\displaystyle\frac{\mathpzc{q}}{4\pi\vert\pmb{x}\vert}\left[1+\frac{g^2}{%
48\pi^2}\frac{1}{\vert\pmb{x}\vert^2}\right] & \vert\pmb{x}\vert\ll\lambda%
\end{array}
\right..
\end{split}
\label{generalpotenatiallwithoutb}
\end{equation}%
This expression has been quoted for establishing a comparison with the situation where the magnetic field is weak [$\mathfrak{b}\ll1$], in which case the correction to 
Eq.~(\ref{generalpotenatiallwithoutb}) results from an expansion in $\mathfrak{b}$ of the portion in Eq.~(\ref{initialpotential}) that does not contain $\pi(\pmb{q}^2)$. 
When the field $B$ is strong enough [$\mathfrak{b}\gg1$], the weakness of the axion-diphoton coupling is compensated, and the quantum effects could enhance notably. 
Also in this situation, the term containing $\pi(\pmb{q}^2)$ can be ignored as it turns out to be a small correction of the order $\sim g^2$. Consequently, we will deal 
in the following with 
\begin{equation}
\begin{split}
\mathpzc{a}_0(\pmb{x})&\approx\mathpzc{q} \int_0^\infty\dbar q_\perp q_\perp J_0(q_\perp x_\perp) \\
&\qquad\qquad\times\int_{-\infty}^{\infty}\dbar q_\parallel \frac{\left(1+\frac{\pmb{q}^2}{m^2}\right) e^{-iq_\parallel
x_\parallel}}{\pmb{q}^2\left(1+\frac{\pmb{q}^2}{m^2}\right)+\mathfrak{b}%
^2q_\parallel^2},
\end{split}\label{preliminarypotential}
\end{equation}%
where the original integration variables have been changed to cylindrical ones, and the integral representation of the Bessel function of order zero 
$J_0(z)=\int_0^{2\pi}\dbar\varphi e^{-iz \cos(\tilde{\varphi}-\varphi)}$ has been used \cite{NIST}. Here $\varphi$ ($\tilde{\varphi}$) is the polar angle associated with 
$\pmb{q}_\perp$ ($\pmb{x}_\perp$) and the square of the momentum $\pmb{q}$ must be understood as $\pmb{q}^2=q_\perp^2+q_\parallel^2$.

\begin{figure*}[t]
\centering
\includegraphics[width=7cm]{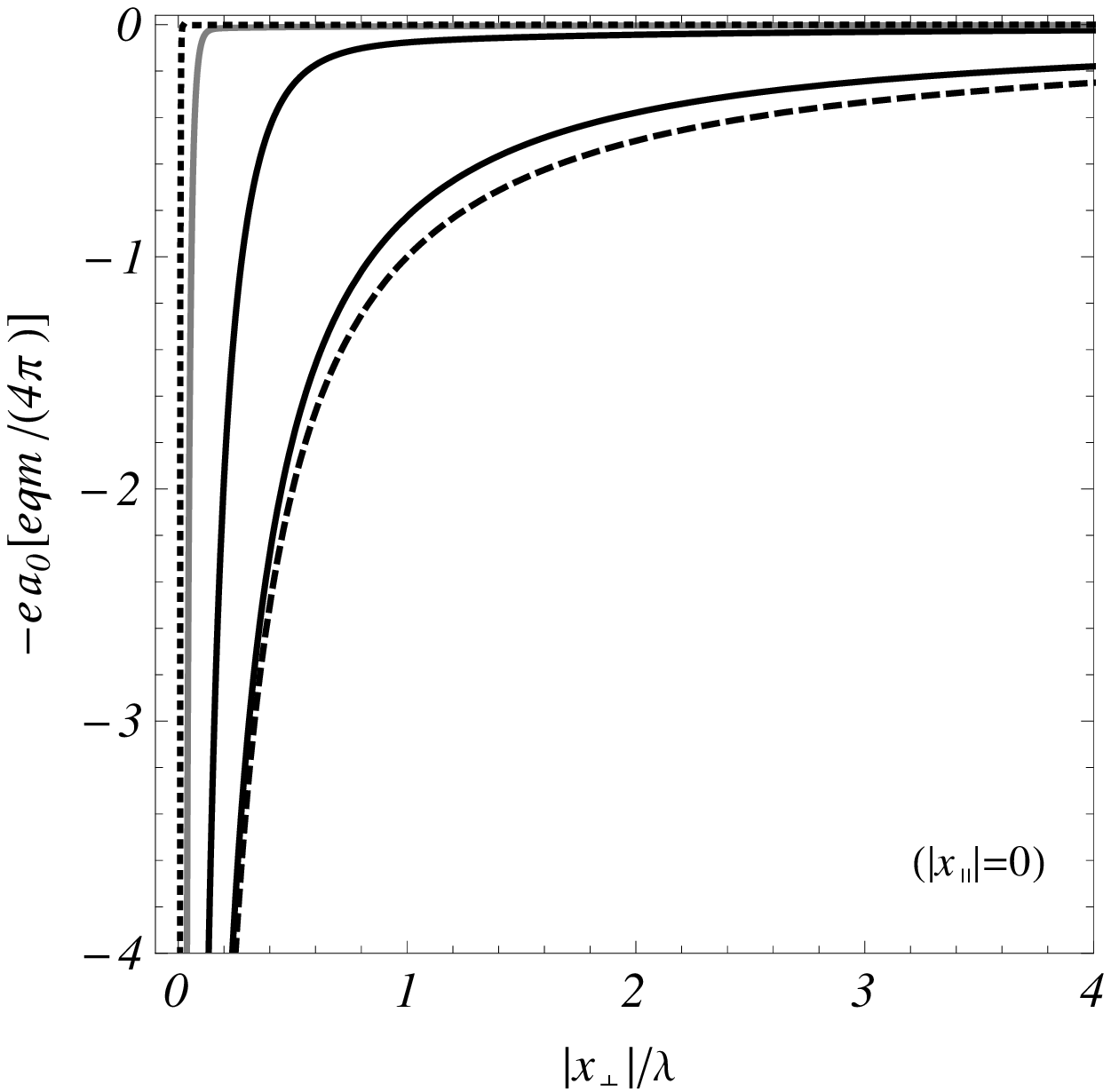} %
\includegraphics[width=7cm]{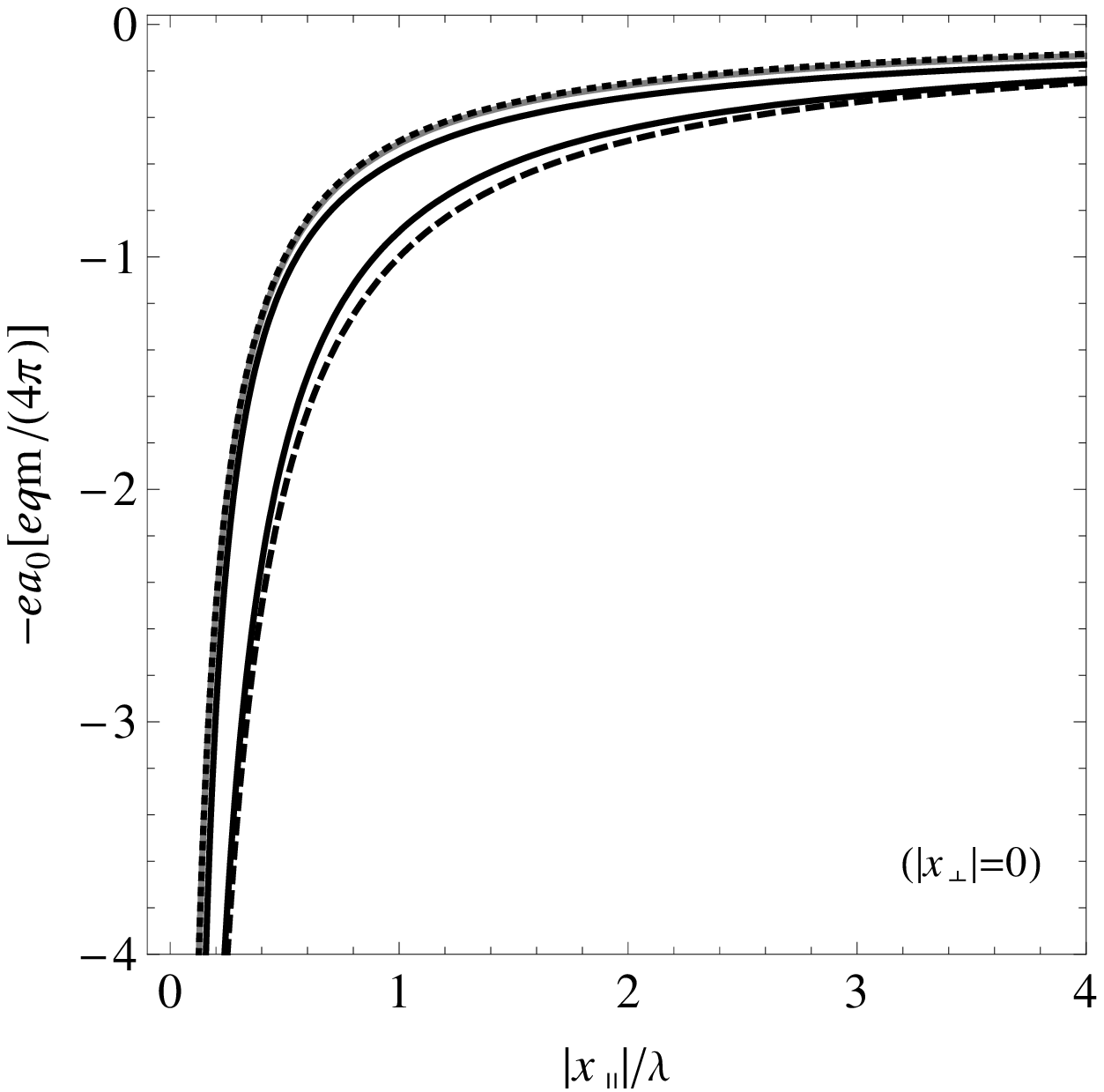}
\caption{Electrostatic energy of an electron $-e\mathpzc{a}_0(\pmb{x})$ due to the anisotropic Coulomb potential $\mathpzc{a}_0(\pmb{x})$ [see Eq.~(\protect\ref{generalpotential})] 
of a pointlike static charge induced by quantum vacuum fluctuations due to an axionlike field over a constant magneticlike background. Here the energy is shown in units of the electron
energy at the Compton wavelength of an ALP $e\mathpzc{q}m/(4\protect\pi)$. The left [right] panel depicts the behavior along the axis $\pmb{x}_\perp$ [$\pmb{x}_\parallel$] passing 
through the charge $\mathpzc{q}$ perpendicular [parallel] to the magnetic field, $\vert\pmb{x}_\parallel\vert=0$ [$\vert \pmb{x}_\perp\vert=0$]. In both panels, the dashed curve is 
the Coulomb potential, whereas solid lines in gray and black were determined from the following values (left to right) $\mathfrak{b}=100$, $\mathfrak{b}=10$, and $\mathfrak{b}=1$. 
The dotted curve in each panel must be understood as the limiting case when $\mathfrak{b}\to\mathfrak{b}_{\mathrm{max}}$.}
\label{fig:mb002}
\end{figure*}

The integration over $q_\parallel$ can be performed by using the residues theorem. For this, we point out that the integrand in Eq.~(\ref{preliminarypotential}) has four poles on the 
imaginary axis: 
\begin{equation}
q_\parallel=\left\{%
\begin{split}
&\pm i\sqrt{q_\perp^2+\frac{1}{2}m_*^2\left[1+\sqrt{1+\frac{4\mathfrak{b^2}}{%
1+\mathfrak{b}^2}\frac{q_\perp^2}{m_*^2}}\right]} \\
\\
&\pm i\sqrt{q_\perp^2+\frac{1}{2}m_*^2\left[1-\sqrt{1+\frac{4\mathfrak{b^2}}{%
1+\mathfrak{b}^2}\frac{q_\perp^2}{m_*^2}}\right]} \\
\end{split}%
\right.
\end{equation}%
with the positive roots taken for $x_\parallel<0$ and the negative ones for $x_\parallel>0$. As before, $m_*=m\sqrt{1+\mathfrak{b}^2}$ with $\mathfrak{b}=gB/m$ is the ALP 
mass dressed by the external field [see below Eq.~(\ref{dispersionrelations})]. With these details in mind, and after developing the change of variable $u=2\mathfrak{b}q_\perp/(m_*\sqrt{1+\mathfrak{b}^2})$, 
it turns out that 
\begin{equation}
\begin{split}
&\mathpzc{a}_0(\pmb{x})=\frac{1}{2}\mathpzc{a}_{\lambda_*}\sqrt{1+\frac{1}{%
\mathfrak{b}^2}}\int_0^\infty\frac{du}{2u}\frac{J_0(\tilde{x}_\perp u)}{%
\sqrt{1+u^2}} \\
&\quad\times\left\{\left[\sqrt{1+u^2}-1\right]\Lambda_+(u)e^{- \tilde{x}%
_\parallel \Lambda_+(u)}\right. \\
&\quad+\left.\left[\sqrt{1+u^2}+1\right]\Lambda_-(u)e^{-\tilde{x}_\parallel
\Lambda_-(u)}\right\}.
\end{split}
\label{generalpotential}
\end{equation}%
In this formula, $\mathpzc{a}_{\lambda_*}=\mathpzc{q}/(4\pi\lambda_*)$ is the Coulomb potential evaluated at the effective wavelength $\lambda_*=m_*^{-1}$ of an ALP. Depending on whether the magnetic 
field strength is smaller or larger than the critical scale of the theory $B_c=m/g$, this quantity will reduce either to the Compton wavelength of the axion $\lambda=m^{-1}$ or to a sort of ``Larmour'' 
length [$\lambda_*\xrightarrow{\mathfrak{b}\gg1} (gB)^{-1}$], respectively. As the framework under consideration allows for exploring distances larger than the natural length scale of the theory 
$g\sim\Lambda_{\mathrm{UV}}^{-1}$ [see discussion below Eq.~(\ref{initialaction})], the effective wavelengths of interest are those for which the condition $\lambda_*\gg g$ is fullfilled. Observe that,
in the strong field limit [$\mathfrak{b}\gg1$], the latter translates into an upper bound for the largest magnetic field that can be considered in the theory $B_{\mathrm{max}}=g^{-2}\sim \Lambda_{\mathrm{UV}}^2$, 
corresponding to a parameter $\mathfrak{b}_{\mathrm{max}}=(mg)^{-1}$ with $1\gg mg$.

Eq.~(\ref{generalpotential}) has been written in terms of the dimensionless coordinates $\tilde{x}_{\perp ,\parallel }=\frac{\vert\pmb{x}_{\perp ,\parallel}\vert}{2\lambda _{\ast }}\sqrt{1+\frac{1}{^{2}}}$ and functions which grow
monotonically with the increasing of $u$: 
\begin{equation}
\Lambda _{\pm }(u)=\sqrt{u^{2}+\frac{2^{2}}{1+\mathfrak{b}^{2}}\left[ 1\pm 
\sqrt{1+u^{2}}\right] }.
\label{complementgeneral}
\end{equation}%
For large values of their argument [$u\gg 1$], they behave as 
\begin{equation}
\left. \Lambda _{\pm }(u)\right\vert _{u\gg 1}\approx u\pm \frac{\mathfrak{b}%
^{2}}{1+\mathfrak{b}^{2}},  \label{complemnetlarge}
\end{equation}%
whereas for small values of $u\ll 1$ they approach 
\begin{equation}
\begin{split}
& \left. \Lambda _{+}(u)\right\vert _{u\ll 1}\approx \sqrt{u^{2}\left[ \frac{%
1+2\mathfrak{b}^{2}}{1+\mathfrak{b}^{2}}\right] +\frac{4\mathfrak{b}^{2}}{1+%
\mathfrak{b}^{2}}}, \\
& \left. \Lambda _{-}(u)\right\vert _{u\ll 1}\approx \frac{u}{\sqrt{1+%
\mathfrak{b}^{2}}}.
\end{split}
\label{complementsmall}
\end{equation}

The formula found above [Eq.~(\ref{generalpotential}) with (\ref{complementgeneral}) included] is our starting point for further considerations. Its consequences on the electrostatic energy of an electron 
$-e\mathpzc{a}_0(\pmb{x})$ is depicted in Fig.~\ref{fig:mb002} for various values of the parameter $\mathfrak{b}$. The curves in the left [right] panel have been determined by setting $\vert\pmb{x}_\parallel\vert=0$ 
[$\vert\pmb{x}_\perp\vert=0$]. For a common value of $\mathfrak{b}\geqslant1$, the corresponding curves in the left and right panels display different shapes and each panel reveals a different trend of the 
curves with the growing of $\mathfrak{b}$. These features constitute clear manifestations of the anisotropy induced by the external field via the vacuum polarization. In the left panel [$\vert\pmb{x}_\parallel\vert=0$], 
the curves tend to stick closer to each other, falling down rather sharply as the magnetic field grows gradually and the distance from the particle location is much smaller than the Compton wavelength of an 
ALP [$\vert \pmb{x}_\perp\vert\ll\lambda$]. Conversely, when $\vert\pmb{x}_\perp\vert=0$, the growing of the external field makes the curves slightly deviate from the standard Coulomb potential, and for 
$\mathfrak{b}\to\mathfrak{b}_{\mathrm{max}}$ the curves converge rather fast to a certain asymptotic function [dotted curve].

%%%%%%%%%%%%%%%%%%%%%%%%%%%%%%%%%%%%%%%%%%%%%%%%%%%%%%%%%%%%%%%%%%%%%%%%%%%%%%%%%%%%%%%%%%%%%%%%%%%%%%%%%%%%%%%%%%%%%%%%%%%%%%%%%
\subsection{Leading behavior for small and large distances relative to $\protect \lambda_*(1+1/\mathfrak{b}^2)^{-\nicefrac{1}{2}}$ \label{GPb1}}
%%%%%%%%%%%%%%%%%%%%%%%%%%%%%%%%%%%%%%%%%%%%%%%%%%%%%%%%%%%%%%%%%%%%%%%%%%%%%%%%%%%%%%%%%%%%%%%%%%%%%%%%%%%%%%%%%%%%%%%%%%%%%%%%%

Let us investigate the behavior of $\mathpzc{a}_0(\pmb{x})$ [see Eq.~(\ref{generalpotential})] on the axis $\pmb{x}_\parallel=\pmb{0}$. In such a situation, 
the integrand that remains combines $J_0(\tilde{x}_\perp u)$ with a function that tends asymptotically to $1$ with the growing of $u$, whatever be the value 
of $\mathfrak{b}$. Hence, the whole integrand reaches the maximum value as the argument of the Bessel function $J_0(\tilde{x}_\perp u)$ satifies the condition 
$\tilde{x}_\perp u\ll1$. Following this analysis we infer that, when $\vert\pmb{x}_\perp\vert$ is small on the scale of $\lambda_*/(1+1/\mathfrak{b}^2)^{\nicefrac{1}{2}}$, 
the main contribution to the integral in Eq.~(\ref{generalpotential}) results from the region in which $1\ll u\ll \tilde{x}_\perp^{-1}$, provided $\mathfrak{b}\gg \vert%
\pmb{x}_\perp\vert/\lambda$. As a consequence, one can exploit Eq.~(\ref{complemnetlarge}) and integrate over $u$ by using formula (6.616.1) of Ref.~\cite{Gradshteyn}.\footnote{\label{footonotesgjf} The integral tabulated in Ref.~\cite{Gradshteyn} reads 
\begin{equation*}
\int_0^\infty dx\; e^{-\alpha x } J_0(\beta \sqrt{x^2+2\gamma x})=e^{\alpha\gamma }\frac{e^{-\gamma\sqrt{\alpha^2+\beta^2}}}{\sqrt{\alpha^2+\beta^2}}.
\end{equation*}
Changing the variable to $t=(x^2+2\gamma x)^{\nicefrac{1}{2}}$, it becomes 
\begin{equation*}
\int_0^\infty \frac{dt\; t}{\sqrt{t^2+\gamma^2}}J_0(\beta t)e^{-\alpha \sqrt{t^2+\gamma^2}}=\frac{e^{-\gamma\sqrt{\alpha^2+\beta^2}}}{\sqrt{\alpha^2+\beta^2}}
\end{equation*}%
which is the desired representation  for our calculations.} With these
details in mind, the modified potential approaches 
\begin{equation}
\mathpzc{a}_0(\pmb{x}_\perp,0)\approx \frac{\mathpzc{q}}{4\pi\vert\pmb{x}_\perp\vert}e^{-\frac{1}{2}m_*\sqrt{1+\frac{1}{\mathfrak{b}^2}}\vert\pmb{x}_\perp\vert}.  \label{xperp}
\end{equation}%
This formula looks like a pure Yukawa potential in which the scale $\sim\lambda_*(1+1/\mathfrak{b}^2)^{-\nicefrac{1}{2}}$ characterizes the short-range behavior. This kind of 
dependence is somewhat expected, since the second propagation mode coexists with a massive branch determined by the ALP sector.

For magnetic fields $B\gg B_c$ [$\mathfrak{b}\gg1$], $\lambda_*\approx (m\mathfrak{b})^{-1}=1/(gB)$ is independent of the ALP mass and $\sqrt{1+1/\mathfrak{b}^2}\sim1$. The 
resulting expression can be read off directly from Eq.~(\ref{xperp}), and tends to be tiny as $\mathfrak{b}$ grows progressively for any nonzero distance $\vert\pmb{x}_\perp\vert\ll \lambda_*$. 
The sharp trend exhibited by the curves in the left panel of Fig.~\ref{fig:mb002}, nearby the particle location [$\vert\pmb{x}_\perp\vert=0$], can be understood qualitatively 
by considering the generalized limit of Eq.~(\ref{xperp}) when $\mathfrak{b}\to\infty$: 
\begin{equation}  \label{perlargeborigin}
\begin{split}
&\lim_{\lambda_*\to0}\frac{1}{ \vert \pmb{x}_\perp\vert}e^{-\frac{1}{2\lambda_*} \vert \pmb{x}_\perp\vert }=-2\delta(x_\perp)\mathrm{Ei}(-1/2).
\end{split}%
\end{equation}%
Here, $\mathrm{Ei}(z)=-\int_{-z}^\infty dt\; t^{-1}e^{-t}$ is the exponential integral function \cite{NIST}. With this result at hand, Eq.~(\ref{xperp}) becomes a magnetic field 
independent delta potential 
\begin{equation}
\mathpzc{a}_0(x_\perp,0)\vert_{\mathfrak{b}\to\infty}\approx \frac{\mathpzc{q}}{2\pi}0.56\;\delta(x_\perp).  \label{deltapotential}
\end{equation}%
Although this result indicates that the curves tend to collapse toward the vertical axis with the growing of $\mathfrak{b}$ [see Fig.~\ref{fig:mb002}], it cannot be considered as 
trustworthy because this parameter is actually limited from above by $\mathfrak{b}_{\mathrm{max}}$ and $x_\perp$ must be larger than $g$ [see discussion below Eq.~(\ref{generalpotential})]. 
It is worth pointing out that, in the limit of $\mathfrak{b}\to\mathfrak{b}_{\mathrm{max}}$ and $\vert\pmb{x}_\perp\vert\to g$ the axion-Coulomb potential is regular and reads 
$\lim_{\mathfrak{b}\to\mathfrak{b}_{\mathrm{max}}}\mathpzc{a}_0(g,0)=\mathpzc{q}\exp[-1/2]/(4\pi g)$. As pointed out  in Sec.~\ref{polarizationtensorrenormalizedc}, at this length scale, a dominance 
of AED over QED might take place only if heavy axions with $g\sim 10^{-2}\ \mathrm{GeV^{-1}}$ are considered, but not for light ALPs. Then, if the axion is thought of as a neutral pion [see 
discussion below Eq.~(\ref{rekappa123})], and the external field is $B=10^2 B_c$, the electrostatic 
energy of an electron in a vicinity of the point charge's location $\vert \pmb{x}_\perp\vert\sim g$ will be $-e\mathpzc{a}_0(g,0)=-18.9 e\mathpzc{q}m_\pi$.

When $\pmb{x}_\perp=\pmb{0}$, the Bessel function $J_0(\tilde{x}_\perp u)=1$, and the exponentials which remain in Eq.~(\ref{generalpotential}) decrease very fast for large 
values of $\tilde{x}_\parallel u$. Hence, for distances $\vert\pmb{x}_\parallel\vert\ll\lambda_*(1+1/\mathfrak{b}^2)^{-\nicefrac{1}{2}}$, the main contribution to the integral 
in Eq.~(\ref{generalpotential}) results from the region in which $1\ll u\ll \tilde{x}_\parallel^{-1}$. Once again, we exploit Eq.~(\ref{complemnetlarge}) and keep the factor 
$u/\sqrt{1+u^2}$ as it stands. These steps reduce the integral to one in which (6.616.1) of Ref.~\cite{Gradshteyn} can be used. Consequently, 
\begin{equation}
\begin{split}
&\mathpzc{a}_0(0,\pmb{x}_\parallel)\approx\frac{\mathpzc{q}}{4\pi\vert\pmb{x}%
_\parallel\vert}\cosh\left[\frac{1}{2}\frac{m_*\vert\pmb{x}_\parallel\vert}{%
\sqrt{1+\frac{1}{\mathfrak{b}^2}}}\right]e^{-\frac{1}{2}m_*\sqrt{1+\frac{1}{%
\mathfrak{b}^2}}\vert\pmb{x}_\parallel\vert}.
\end{split}
\label{strongfieldsamllxparallel12}
\end{equation}%
The outcome above applies whenever $\mathfrak{b}\gg \vert\pmb{x}%
_\parallel\vert/\lambda$. In the limit of $\mathfrak{b}\gg1$, this
expression approaches 
\begin{equation}
\begin{split}
&\mathpzc{a}_0(0,\pmb{x}_\parallel)\approx\frac{\mathpzc{q}}{8\pi\vert\pmb{x}%
_\parallel\vert}\left[1+e^{-m_*\vert\pmb{x}_\parallel\vert}\right].
\end{split}
\label{strongfieldsamllxparallel1}
\end{equation}%
We note that, as the exponent involved in this formula is much smaller than unity [$m_*\vert\pmb{x}_\parallel\vert\ll1$], the screening induced by the vacuum polarization does 
not deviate Eq.~(\ref{strongfieldsamllxparallel1}) from the Coulomb potential $\mathpzc{a}_0(0,\pmb{x}_\parallel)\approx\mathpzc{q}/(4\pi\vert\pmb{x}_\parallel\vert)$ substantially. 
This fact agrees with the behavior shown in the right panel of Fig.~\ref{fig:mb002}, which exhibits that -- for distances smaller than the ALP wavelength -- the behavior of the curves 
is  very similar to the one associated with  Coulomb's law. The limit of $\vert\pmb{x}_\parallel\vert\to g$ is readable from Eq.~(\ref{strongfieldsamllxparallel1}) and gives 
$\mathpzc{a}_0(0,g)\approx \mathpzc{q}/(4\pi g)$. 

Although Eq.~(\ref{strongfieldsamllxparallel1}) has been derived under the restriction $x_\parallel/\lambda\ll\mathfrak{b}^{-1}$ with $\mathfrak{b}\gg1$, it can be used to explore  
regions of $\vert\pmb{x}_\parallel \vert$ for which the condition $\vert\pmb{x}_\parallel\vert\sim \lambda$ holds. Indeed, at $\vert \pmb{x}_\parallel\vert=\lambda$ and $\mathfrak{b}=10^2$, 
this expression differs from the exact formula [see Eq.~(\ref{generalpotential})]  by $2.6\%$ only. The accuracy increases even further as $\mathfrak{b}$ grows, reaching values down to $0.4\%$ 
if $\mathfrak{b}\geqslant10^3$.  Noteworthy, when both  $\vert\pmb{x}_\parallel\vert\sim \lambda$ and $\mathfrak{b}\gg1$ are met, the second term in Eq.~(\ref{strongfieldsamllxparallel1}) 
drops and 
\begin{equation}
\mathpzc{a}_0(0,\pmb{x}_\parallel)\approx\frac{1}{2}\frac{\mathpzc{q}}{4\pi\vert\pmb{x}_\parallel\vert}.
\end{equation}Observe that this  formula differs  from the standard Coulomb law by a factor $1/2$. Thus, the behavior  of the axion-modified Coulomb potential in the longitudinal direction  
saturates in the sense that -- unlike the direction  perpendicular to $\pmb{B}$ [see Eq.~(\ref{xperp})] -- it reaches a universal shape, independent of the external field strength. This fact 
helps to understand why the curves in the right panel of Fig.~\ref{fig:mb002} converge to one whose behavior differs from the case in which $\pmb{x}_\parallel=0$ [left panel] as $\mathfrak{b}\gg1$. 

The procedure for obtaining Eq.~(\ref{strongfieldsamllxparallel12}) can be extended easily to the case in which 
$\vert \pmb{x}_\perp\vert\ll\lambda_*/(1+1/\mathfrak{b}^2)^{\nicefrac{1}{2}}$. This gives rise to the following short-range anisotropic potential [$\mathfrak{b}\gg1$] 
\begin{equation}
\begin{split}
&\mathpzc{a}_0(\pmb{x}_\perp,\pmb{x}_\parallel)\approx\frac{\mathpzc{q}}{4\pi\vert\pmb{x}\vert}\cosh\left[\frac{1}{2}m_*\vert\pmb{x}_\parallel\vert%
\right]e^{-\frac{1}{2}m_*\vert\pmb{x}\vert},
\end{split}
\label{shortrangestrongfield}
\end{equation}%
where $\vert\pmb{x}\vert=(\vert\pmb{x}_\perp\vert^2+\vert\pmb{x}_\parallel^2)^{\nicefrac{1}{2}}$ and $m_*\approx m\mathfrak{b}$. In the limit of $\pmb{x}_\parallel\to0$, 
Eq.~(\ref{shortrangestrongfield}) reproduces the result given in Eq.~(\ref{xperp}) with the particularization $\mathfrak{b}\gg1$. Conversely, if $\pmb{x}_\perp\to0$, the 
former reduces to Eq.~(\ref{strongfieldsamllxparallel1}) with the condition $\mathfrak{b}\gg1$ taken into account.

Let us now consider the case  $\vert\pmb{x}_\parallel\vert\gg\lambda_*(1+1/\mathfrak{b}^2)^{-\nicefrac{1}{2}}$ while $\vert \pmb{x}_\perp\vert$ is arbitrary. Since 
the functions $\Lambda_\pm(u)$ increase monotonically with the growing of $u$, the main countribution to the integral in Eq.~(\ref{generalpotential}) results from the region 
in which $u\ll1$. It is then justified to Taylor expand all instances in the integrand depending only on this variable [see Eq.~(\ref{complementsmall})]. After applying once 
again the integral (6.616.1) of Ref.~\cite{Gradshteyn} the modified Coulomb potential reads 
\begin{equation}
\begin{split}
&\mathpzc{a}_0(\pmb{x})\approx\frac{\mathpzc{q}}{4\pi\sqrt{\vert\pmb{x}_\perp\vert^2(1+\mathfrak{b}^2)+\vert\pmb{x}_\parallel\vert^2}}.
\end{split}
\label{strongfieldlargexparallel2}
\end{equation}%
This formula is an anisotropic Coulomb's law which tends to vanish as the external field grows unless $\vert\pmb{x}_\perp\vert=0$. It resembles closely the longe-range behavior 
found within a pure QED context, when the external magnetic field exceeds the critical scale associated with this framework $B_0=m_e^2/e\approx4.42\times 10^{13}\ \rm G$ 
\cite{shabad5,shabad6,Adorno2016}. However, in QED, $\mathfrak{b}^2$ is replaced by the factor $\alpha b/(3\pi)$ [see first line in Eq.~(\ref{qedbehavior})]. As it has been 
pointed out below Eq.~(\ref{sdfsdfsdfasdfsdfsfs}), we expect that  for a certain class of yet undiscarded light axions, the $\mathfrak{b}^2$ term prevails over the QED counterpart, 
making the ALPs physics dominant. Observe that, at $\vert\pmb{x}_\perp\vert=0$, Eq.~(\ref{strongfieldlargexparallel2}) reduces to a pure Coulomb form 
$\mathpzc{a}_0(0,\pmb{x}_\parallel)\approx \mathpzc{q}/(4\pi\vert\pmb{x}_\parallel\vert)$. 

Now, if $\vert\pmb{x}_\parallel\vert=0$, and $\tilde{x}_\perp\gg1$, the integral in Eq.~(\ref{generalpotential}) is dominated by the region where  $u\ll \tilde{x}_\perp^{-1}\ll1$. 
In such a case, the part of the integrand which does not include the Bessel function 
can be Taylor expanded in $u$. As a consequence, the long-range behavior of the axion-Coulomb potential, perpendicular to the direction of the field, reads [$\mathfrak{b}\gg1$] 
\begin{equation}  \label{strongfieldlargexperp}
\mathpzc{a}_0(\pmb{x}_\perp,0)\approx\frac{\mathpzc{q}}{4\pi\vert\pmb{x}%
_\perp\vert \mathfrak{b}}.
\end{equation}

That the axion-Coulomb potential along the magnetic field [$\mathfrak{b}\gg1$] resembles the $1/\vert \pmb{x}_\parallel\vert$ law approximately leads to an important implication: 
the ground-state energy of a nonrelativistic electron in the field of the nucleus of a hydrogen atom [$\mathpzc{q}=e$] can be approached by the same expression obtained by Elliott and 
Loudon \cite{elliott}: 
\begin{equation}
\begin{split}
&\varepsilon_0 =-2\alpha^2 m_e \ln^2\left(\frac{\sqrt{b}}{\alpha}\right).
\label{LE}
\end{split}%
\end{equation}
This formula is unbounded from below  as the magnetic field grows unlimitedly [$b \to\infty$].  When AED is not considered, this problem finds a natural solution by incorporating the 
screening induced by the polarization tensor of QED \cite{shabad5,shabad6}.  At this point, it should be understood that the reappearance of Eq.~(\ref{LE}) in a scenario dominated by 
quantum vacuum fluctuations due to  axionlike fields is unrealistic because, in AED, the limit of $\mathfrak{b}\to\infty$ cannot be taken formally as there is a natural cutoff custodying 
the unitarity of the scattering matrix $b<\mathfrak{b}_{\rm max}$.

%%%%%%%%%%%%%%%%%%%%%%%%%%%%%%%%%%%%%%%%%%%%%%%%%%%%%%%%%%%%%%%%%%%%%%%%%%%%%%%%%%%%%%%%%%%%%%%%%%%%%%%%%%
\subsection{The weak field limit $\mathfrak{b}\ll1$\label{weakfieldsection}}
%%%%%%%%%%%%%%%%%%%%%%%%%%%%%%%%%%%%%%%%%%%%%%%%%%%%%%%%%%%%%%%%%%%%%%%%%%%%%%%%%%%%%%%%%%%%%%%%%%%%%%%%%

For $\mathfrak{b}\ll1$, the background field constitutes a small perturbation and the first Born approximation applies. In this case, $\lambda_*$ reduces to the Compton wavelength of 
an ALP [$\lambda_*\to\lambda=m^{-1}$]. Prior to carrying out an expansion in $\mathfrak{b}$, it is rather convenient to develop in Eq.~(\ref{generalpotential}) the change of variable 
$s=u/(2\mathfrak{b})$, in which case the modified Coulomb's
potential reads 
\begin{equation}
\begin{split}
&\mathpzc{a}_0(\pmb{x})=\mathpzc{a}_{\lambda_*}\int_0^\infty\frac{ds}{2s}%
\frac{J_0(m_*\vert\pmb{x}_\perp\vert\sqrt{1+\mathfrak{b}^2} s)}{\sqrt{1+4%
\mathfrak{b}^2s^2}} \\
&\quad\times\left\{\left[\sqrt{1+4\mathfrak{b}^2s^2}-1\right]\hat{\Lambda}%
_+(s)e^{- m_*\vert\pmb{x}_\parallel\vert\hat{\Lambda}_+(s)}\right. \\
&\quad+\left.\left[\sqrt{1+4\mathfrak{b}^2s^2}+1\right]\hat{\Lambda}%
_-(s)e^{-m_*\vert\pmb{x}_\parallel\vert \hat{\Lambda}_-(s)}\right\}, \\
&\hat{\Lambda}_{\pm}(s)=\sqrt{s^2(1+\mathfrak{b}^2)+\frac{1}{2}\left[1\pm%
\sqrt{1+4\mathfrak{b}^2s^2}\right]}.
\end{split}
\label{generalpotentialalternative}
\end{equation}
After expanding the expression above up to $\mathfrak{b}^2$, and incorporating the pure vacuum contribution [see Eq.~(\ref{generalpotenatiallwithoutb}) and the subsequent discussion] we end up with 
\begin{equation}
\mathpzc{a}_0(\pmb{x})\approx \frac{\mathpzc{q}}{4\pi\vert\pmb{x}\vert} \left[1+\frac{g^2m^2}{48\pi^2}\tau(m\vert\pmb{x}\vert)-\frac{1}{2}\mathfrak{b}^2\mathpzc{G}(\pmb{x})\right],  \label{potentialsmallbfinal}
\end{equation}where the expression for $\tau(m\vert \pmb{x}\vert)$ can be read off from  Eq.~(\ref{generalpotenatiallwithoutb}):
\begin{equation}
\tau(m\vert\pmb{x}\vert)=\int_{1}^\infty \frac{du}{u^5}\left[u^2-1\right]^3e^{-m\vert\pmb{x}\vert u}.
\end{equation}In Eq.~(\ref{potentialsmallbfinal}), the function which determines the next-to-leading order 
correction reads 
\begin{equation}
\begin{split}
&\mathpzc{G}(\pmb{x})=\mathpzc{G}_1(m\vert\pmb{x}\vert)+m^2\vert \pmb{x}%
_\parallel\vert^2\mathpzc{G}_2(m\vert\pmb{x}\vert), \\
&\mathpzc{G}_1(m\vert\pmb{x}\vert)=1-\frac{2}{m^2\vert\pmb{x}\vert^2}\left[%
1-(1+m\vert\pmb{x}\vert)e^{-m\vert\pmb{x}\vert}\right], \\
&\mathpzc{G}_2(m\vert\pmb{x}\vert)=\frac{1}{m^4 \vert\pmb{x}\vert^4}\left\{6%
\left[1-(1+m\vert\pmb{x}\vert)e^{-m \vert\pmb{x}\vert}\right]\right. \\
&\qquad\qquad\qquad\qquad\qquad-\left.m^2 \vert\pmb{x}\vert^2\left[1+2
e^{-m\vert\pmb{x}\vert}\right]\right\}.
\end{split}
\label{complementsmalls}
\end{equation}%
It is worth mentioning that, when deriving Eq.~(\ref{potentialsmallbfinal}), the Eq.~(6.616.1) in Ref.~\cite{Gradshteyn} -- as well as its first and second order 
derivatives with respect to $\alpha$ -- have been used [see also footnote~\ref{footonotesgjf}].

We remark that, for distances $\vert\pmb{x}\vert$ larger than $\lambda$, the modi\-fied Coulomb potential [see Eq.~(\ref{potentialsmallbfinal})] acquires the form 
\begin{equation}
\begin{split}
&\mathpzc{a}_0(\pmb{x})\approx\frac{\mathpzc{q}}{4\pi\vert\pmb{x}\vert}\left[1+\frac{g^2m^2}{\pi^2}\frac{e^{-m\vert\pmb{x}\vert}}{(m\vert\pmb{x}\vert)^4}\right. \\
&\qquad\qquad\qquad\qquad\qquad\qquad-\left.\frac{1}{2}\mathfrak{b}^2\left(1-\frac{\vert\pmb{x}_\parallel\vert^2}{\vert\pmb{x}\vert^2}\right)\right].
\end{split}%
\end{equation}%
This expression is also obtained from Eq.~(\ref{strongfieldlargexparallel2}) when it is Taylor expanded in the small parameter $\mathfrak{b}\ll1$. Now, for distances 
$\vert\pmb{x}\vert\ll \lambda$, but larger than the natural scale of the theory $\vert\pmb{x}\vert\gg g$, the modified Coulomb potential approaches to 
\begin{equation}
\begin{split}
&\mathpzc{a}_0(\pmb{x})\approx \frac{\mathpzc{q}}{4\pi\vert\pmb{x}\vert}%
\left[1+\frac{g^2}{48\pi^2}\frac{1}{\vert\pmb{x}\vert^2}-\frac{1}{3}%
\mathfrak{b}^2m \vert\pmb{x}\vert\right. \\
&\qquad\qquad\qquad\qquad\times\left.\left(1-\frac{3}{8} m\vert\pmb{x}\vert -%
\frac{3m\vert\pmb{x}_\parallel\vert^2}{8\vert\pmb{x}\vert}\right)\right].
\end{split}
\label{correctedpotentialsmalldistances}
\end{equation}
The contribution depending on $\mathfrak{b}^2$ coincides -- up to a constant term $\frac{1}{2}\mathpzc{q}\mathfrak{b}^2m$ -- with the one derived in Ref.~\cite{Jentschura:2014yla} 
from the amplitude associated with electron-positron scattering. Observe that, in the limit of $\vert \pmb{x} \vert \to g$, the correction to the Coulomb potential is dominated 
by the factor $\mathpzc{q}/(192\pi^3 g)$, which is independent of the axion mass. The question whether virtul ALPs or neutral mesons can explain the proton radius anomaly in 
muonic hydrogen is revisited in Ref.~\cite{alinapaper}.

%%%%%%%%%%%%%%%%%%%%%%%%%%%%%%%%%%%%%%%%%%%%%%%%%%%%%%%%%%%%%%%%%%%%%%%%%%%%%%%%%%%%%%%%%%%%%%%%%%%%%%
\section{Conclusion\label{conclus}}
%%%%%%%%%%%%%%%%%%%%%%%%%%%%%%%%%%%%%%%%%%%%%%%%%%%%%%%%%%%%%%%%%%%%%%%%%%%%%%%%%%%%%%%%%%%%%%%%%%%%%%

We have considered the interaction between axionlike particles  and  photons inside the background of a superstrong constant magnetic field beyond the classical field equations 
by including quantum corrections up to the second order in the axion-diphoton coupling. Only the extraordinary photon mode -- mode-2 -- whose electric field lies in the plane spanned 
by the external magnetic field and the photon momentum  interacts with the axion, while the ordinary mode -- mode-3 -- remains the same as it was in strong-magnetic-field QED.  
We have obtained that the axion contribution to the polarization operator grows quadratically with the magnetic field, while the QED part of the polarization operator of the 
extraordinary mode is known \cite{skobelev,shabad-lebedev,melrose,heyl} to grow only linearly (while the ordinary-mode part grows much more slowly). In this axion-dominating regime
we established the common axion-extraordinary-photon dispersion law determining their mass shell. It consists of two nonintersecting branches, with no mixed state forming -- in 
contrast to QED,  where the photon mixes with the electron-positron state to form a polariton. One branch is massless in the sense that the frequency as a function of spatial 
momentum turns to zero when all components of the latter are zero. This implies that the mass defined as the rest energy is zero. This branch is associated with the photon 
modified by its interaction with the axion. The other branch is massive and it is attributed to an axionlike particle, modified by its interaction with the photon. The modified 
mass of the axion grows with the magnetic field, and when the latter is of the order of the  critical QED value $B_0=m_{e}/e=4.42\times 10^{13}\ \rm G$ or larger -- which is characteristic 
of some pulsars or magnetars -- a certain class of light ALPs [$10^{-3}\;\mu\mathrm{eV}<m<10^{2}\;\mu\rm eV$] that allows for the axion-dominating regime might escape from the star,  
this way contributing to the pulsar cooling.

The massless branch exhibits the important feature of flattening in the vicinity of zero momentum, leading to strong birefringence, because the dispersion curve of the mode-3 photon,
which remains the same as in QED, does not show such feature. The flattening results in bending the direction of the wave envelope propagation (the group velocity) towards the external 
field and in the capture of mode-2 photons by curved lines of force of the magnetic field of pulsars. Contrary to the analogous effect known in QED,  in the present scenario not only 
$\gamma$-quanta of curvature radiation  undergo the capture , but also heat photons in the X-ray range. The capture of gamma-quanta is more efficient in the axion 
electrodynamics and its potential consequence might be the destruction of the well established mechanism of formation of pulsar radiation. This does not happen, however, since the parameters 
of the axion necessary for such destruction are eliminated by already existing estimates. On the contrary, when applied to the softer thermal radiation of the pulsar surface the
capture causes that the mode-$2$ X-ray quanta all gather together to peak near the direction of the magnetic field, while beyond this direction all photons are polarized 
orthogonal to the plane spanned by the magnetic field and the observation direction. Their angular distribution is determined by the refraction at the border of the strong 
magnetic field area, studied in the paper, and by the mechanisms of formation of the soft radiation (see their review in Ref.~\cite{Potekhin}).

Despite the charge-neutrality of quantum vacuum fluctuations of axionlike fields, we have seen that the presence of a supercritical magnetic field could induce a strong 
screening of static electric charges. At distances from the source much smaller than the Compton wavelength of the axion, and for magnetic field strengths larger than the 
critical scale $m/g$ associated with the axion electrodynamics, the Coulomb potential of a point-like charge is replaced by a Yukawa-type potential in the direction 
perpendicular to the magnetic field, while along the field the Coulomb law is preserved in a modified form at any length scale. We find that at unlimitedly large magnetic 
fields the longstanding problem -- overcome in QED -- that the ground-state energy of a hydrogen atom is unbounded from below, is reinstated in axion electrodynamics. However, 
in the latter theory this unboundedness is cut off because the largest magnetic field treatable within this theory is limited by the unitarity of the associated scattering 
matrix.

In the weak-field regime, the anisotropization of the axion-Coulomb potential is also manifested. However, in contrast to the supercritical magnetic field regime, the one-loop 
contribution turns out to be the leading order correction to the potential. We therefore infer that it will distort the motion of electrons in atoms pretty much in the same 
way as the vacuum polarization  does in QED. Hence, the lines of atomic emission spectra of red giants and pulsars such as white dwarfs should be shifted by this sort of 
axion-Lamb shift. Plausible distortions of this nature should be revisited as they could provide valuable information about the existence of ALPs. An investigation on this 
subject is in preparation.

\begin{acknowledgments}
S.~Villalba-Ch\'avez and  C.~M\"{u}ller thank  A.~Golub for useful discussions. Both authors gratefully acknowledge the funding by the German Research Foundation (DFG) under Grant No. MU 3149/2-1. 
A. Shabad is thankful to the  University of D\"{u}sseldorf for the kind hospitality extended to him during his research visit in  $2016$. He acknowledges the support of the Russian Foundation for Basic Research under the 
project No.~$18-02-00149\rm a$ and by the TSU Competitiviness Improvement Program through a grant from  ``The Tomsk State University D.~I.~Mendeleev Foundation Program''.
\end{acknowledgments}

\appendix
%%%%%%%%%%%%%%%%%%%%%%%%%%%%%%%%%%%%%%%%%%%%%%%%%%%%%%%%%%%%%%%%%%%%%%%%%%%%%%%%%%%%%%%%%%%%%%%%%%%%%%%%%%%%%%%%%%%%%%%%%%%%%%%%%%%%%%%%%%%%%%%%%
\section{Dispersion equation from the axion prospect\label{Appendix}}
%%%%%%%%%%%%%%%%%%%%%%%%%%%%%%%%%%%%%%%%%%%%%%%%%%%%%%%%%%%%%%%%%%%%%%%%%%%%%%%%%%%%%%%%%%%%%%%%%%%%%%%%%%%%%%%%%%%%%%%%%%%%%%%%%%%%%%%%%%%%%%%%%

Let us consider the system of linearized equations describing the dynamics of our axion-photon system. According to Eq.~(\ref{effectiveaction}) it reads
\begin{eqnarray}
\int d^4\tilde{x}\;\pmb{\Gamma}^{(2)}(x,\tilde{x})\pmb{\Phi}(\tilde{x})=0.
\end{eqnarray}Taking into account the quantities defined between Eq.~(\ref{effectiveaction}) and (\ref{phionshell}) we obtain [$\partial\mathpzc{a}=0$]
\begin{eqnarray}
\begin{split}
&\square \mathpzc{a}^\mu(x)-g\tilde{\mathscr{F}}^{\mu}_{\ \nu}(x)\partial^\nu\phi(x)\\&\qquad\qquad\qquad+i\int d^4\tilde{x}\;\Pi^{\mu\nu}_{\mathrm{vac}}(x,\tilde{x})\mathpzc{a}_\nu(\tilde{x})=0,\\  
&\left(\square+m^2\right)\phi(x)+\frac{1}{2}g\tilde{\mathscr{F}}^{\mu\nu}(x)\mathpzc{f}_{\mu\nu}(x)\\&\qquad\qquad\qquad\quad\ -i\int d^4\tilde{x}\;\Sigma(x,\tilde{x})\phi(\tilde{x})=0.
\end{split}
\label{DSEkk}
\end{eqnarray}When ignoring the contribution determined by  $\Pi^{\mu\nu}_{\mathrm{vac}}(x,\tilde{x})$,  the solution of the resulting  equation for $\mathpzc{a}^\mu(x)$ is 
$\mathpzc{a}^\mu(x)=g\square^{-1}\tilde{\mathscr{F}}^{\mu}_{\ \nu}(x)\partial^{\nu}\phi(x)$. Its insertion into the equation of motion for the axion field leads to write
\begin{equation}
\begin{split}
&\left[\square+m^2-g^2\tilde{\mathscr{F}}^{\alpha\beta}\partial_{\alpha}\frac{1}{\square}\tilde{\mathscr{F}}^{\nu}_{\ \beta}\partial_{\nu}\right]\phi(x)\\
&\qquad\qquad\qquad-i\int d^4\tilde{x}\;\Sigma(x,\tilde{x})\phi(\tilde{x})=0.
\end{split}\label{PHOTONEQMOTION}
\end{equation}We Fourier transform this formula. If the magnetic field is strong enough as to disregard the effect coming from the axion self-energy operator,  one ends up with
\begin{equation}
\begin{split}
&\left[-q^2+m^2+g^2\frac{q\tilde{\mathscr{F}}^2q}{q^2}\right]\phi(q)=0.
\end{split}\label{axionNEQMOTION}
\end{equation}This dispersion equation coincides with that of  mode-$2$ photons [see Eq.~(\ref{dispersionequationphoton})  and (\ref{tensordecomposition})], 
provided the loop contribution $\sim q^2\pi(q^2)$ is ignored. Hence, in strong magnetic fields, the axionlike  spectrum is the same as that found for  the second propagation mode when the former  is integrated out.


\begin{thebibliography}{28}

\expandafter\ifx\csname
natexlab\endcsname\relax\def\natexlab#1{#1}\fi
\expandafter\ifx\csname bibnamefont\endcsname\relax
  \def\bibnamefont#1{#1}\fi
\expandafter\ifx\csname bibfnamefont\endcsname\relax
  \def\bibfnamefont#1{#1}\fi
\expandafter\ifx\csname citenamefont\endcsname\relax
  \def\citenamefont#1{#1}\fi
\expandafter\ifx\csname url\endcsname\relax
  \def\url#1{\texttt{#1}}\fi
\expandafter\ifx\csname urlprefix\endcsname\relax\def\urlprefix{URL
}\fi \providecommand{\bibinfo}[2]{#2}
\providecommand{\eprint}[2][]{\url{#2}}

\bibitem{Peccei:1977hh}
\bibinfo{author}{\bibfnamefont{R.~D.}~\bibnamefont{Peccei}} \bibnamefont{and}
\bibinfo{author}{\bibfnamefont{H.~R.}~\bibnamefont{Quinn}}, 
\emph{\bibinfo{Title}{CP Conservation in the Presence of Pseudoparticles}},
\bibinfo{journal}{Phys.  Rev.  Lett.} \pmb{\bibinfo{volume}{38}},
\bibinfo{pages}{1440} (\bibinfo{year}{1977}).

\bibitem{Wilczek:1977pj}
\bibinfo{author}{\bibfnamefont{F.}~\bibnamefont{Wilczek}}, 
\emph{\bibinfo{Title}{Problem of Strong p and t Invariance in the Presence of Instantons}},
\bibinfo{journal}{Phys.  Rev.  Lett.} \pmb{\bibinfo{volume}{40}},
\bibinfo{pages}{279} (\bibinfo{year}{1978}).

\bibitem{Weinberg:1977ma}
\bibinfo{author}{\bibfnamefont{S.}~\bibnamefont{Weinberg}}, 
\emph{\bibinfo{Title}{A New Light Boson?}}
\bibinfo{journal}{Phys.  Rev.  Lett.} \pmb{\bibinfo{volume}{40}},
\bibinfo{pages}{223} (\bibinfo{year}{1978}).

\bibitem{Wilczek:1987pj}
\bibinfo{author}{\bibfnamefont{F.}~\bibnamefont{Wilczek}}, 
\emph{\bibinfo{Title}{Two applications of axion electrodynamics}},
\bibinfo{journal}{Phys.  Rev.  Lett.} \pmb{\bibinfo{volume}{58}},
\bibinfo{pages}{1799} (\bibinfo{year}{1987}).

\bibitem{Hehl}
\bibinfo{author}{\bibfnamefont{F.~W.}~\bibnamefont{Hehl}},
\bibinfo{author}{\bibfnamefont{Y.~N.}~\bibfnamefont{Obukhov}}, 
\bibinfo{author}{\bibfnamefont{J.~P.}~\bibfnamefont{Rivera}},
\bibnamefont{and}
\bibinfo{author}{\bibfnamefont{H.}~\bibnamefont{Schmid}},
\emph{\bibinfo{Title}{Relativistic nature of a magnetoelectric modulus of $\rm Cr_2O_3$ crystals: A four-dimensional pseudoscalar and its measurement}},
\bibinfo{journal}{Phys. Rev. A} \pmb{\bibinfo{volume}{77}},
\bibinfo{pages}{022106} (\bibinfo{year}{2008}).

\bibitem{Li}
\bibinfo{author}{\bibfnamefont{R.}~\bibnamefont{Li}},
\bibinfo{author}{\bibfnamefont{J.}~\bibfnamefont{Wang}}, 
\bibinfo{author}{\bibfnamefont{X.~L.}~\bibfnamefont{Qi}},
\bibnamefont{and}
\bibinfo{author}{\bibfnamefont{S.~C.}~\bibnamefont{Zhang}},
\emph{\bibinfo{Title}{Dynamical axion field in topological magnetic insulators}},
\bibinfo{journal}{Nature Phys.} \pmb{\bibinfo{volume}{6}},
\bibinfo{pages}{284} (\bibinfo{year}{2010}).

\bibitem{Ooguri}
\bibinfo{author}{\bibfnamefont{H.}~\bibnamefont{Ooguri}},
\bibnamefont{and}
\bibinfo{author}{\bibfnamefont{M.}~\bibnamefont{Oshikawa}},
\emph{\bibinfo{Title}{Instability in magnetic materials with a dynamical axion field}},
\bibinfo{journal}{Phys. Rev. Lett.} \pmb{\bibinfo{volume}{108}},
\bibinfo{pages}{161803} (\bibinfo{year}{2012}).

%\bibitem{Dine:1981rt}
%\bibinfo{author}{\bibfnamefont{M.}~\bibnamefont{Dine}}, 
%\bibinfo{author}{\bibfnamefont{W.}~\bibnamefont{Fischler}}  
%\bibnamefont{and}
%\bibinfo{author}{\bibfnamefont{M.}~\bibnamefont{Srednicki}}, 
%\emph{\bibinfo{Title}{A simple solution to the strong CP problem with a harmless axion}},
%\bibinfo{journal}{Phys. Lett. B} \pmb{\bibinfo{volume}{104}},
%\bibinfo{pages}{199} (\bibinfo{year}{1981}).

\bibitem{Jaeckel:2010ni}
\bibinfo{author}{\bibfnamefont{J.}~\bibnamefont{Jaeckel}}  \bibnamefont{and}
\bibinfo{author}{\bibfnamefont{A.}~\bibnamefont{Ringwald}},  
\emph{\bibinfo{Title}{The Low-Energy Frontier of Particle Physics}}, 
\bibinfo{journal}{ Ann.\ Rev.\ Nucl.\ Part.\ Sci.}  \textbf{\bibinfo{volume}{60}},   
\bibinfo{pages}{405} (\bibinfo{year}{2010}).
 
\bibitem{Ringwald:2012hr}
\bibinfo{author}{\bibfnamefont{A.}~\bibnamefont{Ringwald}}, 
\emph{\bibinfo{Title}{Exploring the Role of Axions and Other WISPs in the Dark Universe}},
\bibinfo{journal}{Phys.\ Dark Univ.} \textbf{\bibinfo{volume}{1}},
\bibinfo{pages}{116} (\bibinfo{year}{2012}). 
 
\bibitem{Hewett:2012ns}
\bibinfo{author}{\bibfnamefont{J.~L.}~\bibnamefont{Hewett}} {\it et al.}, 
\emph{Fundamental Physics at the Intensity Frontier},
The Proceedings of the 2011 workshop on Fundamental Physics at the Intensity Frontier;
  arXiv:1205.2671 [hep-ex].
  
\bibitem{Essig:2013lka}
\bibinfo{author}{\bibfnamefont{R.}~\bibnamefont{Essig}}  {\it et al.},
\emph{Working Group Report: New Light Weakly Coupled Particles},  arXiv:1311.0029 [hep-ph].

\bibitem{Cameron:1993mr}
\bibinfo{author}{\bibfnamefont{R.}~\bibnamefont{Cameron}  {\it et al.} [BFRT Collaboration]},
\emph{\bibinfo{Title}{Search for nearly massless, weakly coupled particles by optical techniques}},
\bibinfo{journal}{Phys.  Rev.  D } \pmb{\bibinfo{volume}{47}},
\bibinfo{pages}{3707 } (\bibinfo{year}{1993}).

\bibitem{BMVreport}
\bibinfo{author}{\bibfnamefont{A.}~\bibfnamefont{Cadène}  {\it et al.} },
\emph{\bibinfo{Title}{Vacuum magnetic linear birefringence using pulsed fields: status of the BMV experiment}},
\bibinfo{journal}{Eur. Phys. J.  D} \textbf{\bibinfo{volume}{68}},
\bibinfo{pages}{16} (\bibinfo{year}{2014}).
  
\bibitem{Chen:2006cd}
\bibinfo{author}{\bibfnamefont{S.}~\bibfnamefont{J.}~\bibnamefont{Chen} {\it et al.} [Q$\&$A Collaboration]},
\emph{\bibinfo{Title}{Q $\&$ A experiment to search for vacuum dichroism, pseudoscalar-photon interaction and millicharged fermions}},
\bibinfo{journal}{Mod.\ Phys.\ Lett.\ A} \pmb{\bibinfo{volume}{22}},
\bibinfo{pages}{2815} (\bibinfo{year}{2007}).

\bibitem{Mei:2010aq}
\bibinfo{author}{\bibfnamefont{H.}~\bibfnamefont{H.}~\bibnamefont{Mei} {\it et al.} [Q$\&$A Collaboration]},
\emph{\bibinfo{Title}{Axion Search with Q \& A Experiment}},
\bibinfo{journal}{Mod.\ Phys.\ Lett.\ A} \pmb{\bibinfo{volume}{25}},
\bibinfo{pages}{983} (\bibinfo{year}{2010}).

\bibitem{DellaValle:2013xs}
\bibinfo{author}{\bibfnamefont{F.~Della}~\bibnamefont{Valle}  {\it et al.}  [PVLAS Collaboration]},
\emph{\bibinfo{Title}{First results  from  the new PVLAS apparatus: a new limit on vacuum magnetic  birefringence}}, 
\bibinfo{journal}{Phys. Rev. D} \pmb{\bibinfo{volume}{90}}, 
\bibinfo{pages}{092003} (\bibinfo{year}{2014}).

\bibitem{mendonza}
\bibinfo{author}{\bibfnamefont{J.}~\bibfnamefont{T.}~\bibnamefont{Mendon\c{c}a}},
\emph{\bibinfo{Title}{Axion excitation by intense laser fields}},
\bibinfo{journal}{Eurphys. Lett.} \textbf{\bibinfo{volume}{79}},
\bibinfo{pages}{21001} (\bibinfo{year}{2007}).

\bibitem{Gies:2008wv}
\bibinfo{author}{\bibfnamefont{H.}~\bibnamefont{Gies}},
\emph{\bibinfo{Title}{Strong laser fields as a probe for fundamental physics}},
\bibinfo{journal}{Eur. Phys. J.  D} \pmb{\bibinfo{volume}{55}},
\bibinfo{pages}{311} (\bibinfo{year}{2009}).

\bibitem{Dobrich:2010hi}
\bibinfo{author}{\bibfnamefont{B.}~\bibnamefont{D\"obrich}} \bibnamefont{and}
\bibinfo{author}{\bibfnamefont{H.}~\bibnamefont{Gies}}, 
\emph{\bibinfo{Title}{Axion-like-particle search with high-intensity lasers}},
 \bibinfo{journal}{JHEP} \pmb{\bibinfo{volume}{1010}},
\bibinfo{pages}{022} (\bibinfo{year}{2010}).

\bibitem{Villalba-Chavez:2013bda}
\bibinfo{author}{\bibfnamefont{S.}~\bibnamefont{Villalba-Ch\'avez}} \bibnamefont{and}
\bibinfo{author}{\bibfnamefont{A.}~\bibnamefont{Di~Piazza}},
\emph{\bibinfo{Title}{Axion-induced birefringence effects in laser driven nonlinear vacuum interaction}},
\bibinfo{journal}{JHEP} \pmb{\bibinfo{volume}{1311}},
\bibinfo{pages}{136} (\bibinfo{year}{2013}).

\bibitem{Villalba-Chavez:2013goa}
\bibinfo{author}{\bibfnamefont{S.}~\bibnamefont{Villalba-Ch\'avez}},
\emph{\bibinfo{title}{Laser-driven search of axion-like particles including vacuum polarization effects}},   
\bibinfo{journal}{Nucl. Phys. B} \textbf{\bibinfo{volume}{881}}, 
\bibinfo{pages}{1} (\bibinfo{year}{2014}).

\bibitem{Villalba-Chavez:2016hxw}
\bibinfo{author}{\bibfnamefont{S.}~\bibnamefont{Villalba-Ch\'avez}}, \bibinfo{author}{\bibfnamefont{T.} \bibnamefont{Podszus}},   \bibnamefont{and}
\bibinfo{author}{\bibfnamefont{C.} \bibnamefont{M\"{u}ller}},
\emph{\bibinfo{title}{Polarization-operator approach to optical signatures of axion-like particles in strong laser pulses}},   
\bibinfo{journal}{Phys. Lett. B} \pmb{\bibinfo{volume}{769}},
\bibinfo{pages}{233} (\bibinfo{year}{2017}).



\bibitem{Chou:2007zzc}
\bibinfo{author}{\bibfnamefont{A.}~\bibfnamefont{S.}~\bibfnamefont{Chou}  {\it et al.}  [GammeV (T-969) Collaboration]},
\emph{\bibinfo{Title}{Search for axion-like particles using a variable baseline photon regeneration technique}},
\bibinfo{journal}{Phys. Rev. Lett.} \textbf{\bibinfo{volume}{100}},
\bibinfo{pages}{080402} (\bibinfo{year}{2008}).

\bibitem{Steffen:2009sc}
\bibinfo{author}{\bibfnamefont{J.}~\bibnamefont{H.}~\bibnamefont{Steffen}}
\bibnamefont{and}
\bibinfo{author}{\bibfnamefont{A.}~\bibnamefont{Upadhye}},
\emph{\bibinfo{Title}{The GammeV suite of experimental searches for axion-like particles}},
\bibinfo{journal}{Mod. Phys. Lett. A} \textbf{\bibinfo{volume}{24}},
\bibinfo{pages}{2053} (\bibinfo{year}{2009}).

\bibitem{Afanasev:2008jt}
\bibinfo{author}{\bibfnamefont{A.}~\bibfnamefont{Afanasev}  {\it et al.}},
\emph{\bibinfo{Title}{New Experimental limit on Optical Photon Coupling to Neutral, Scalar Boson}},
\bibinfo{journal}{Phys. Rev.  Lett.} \textbf{\bibinfo{volume}{101}},
\bibinfo{pages}{120401} (\bibinfo{year}{2008}).

\bibitem{Pugnat:2007nu}
\bibinfo{author}{\bibfnamefont{P.}~\bibfnamefont{Pugnat}  [OSQAR Collaboration]},
\emph{\bibinfo{Title}{First results from the OSQAR photon regeneration experiment: No light shining through a wall}},
\bibinfo{journal}{Phys. Rev. D} \textbf{\bibinfo{volume}{78}},
\bibinfo{pages}{092003} (\bibinfo{year}{2008}).

\bibitem{Robilliard:2007bq}
\bibinfo{author}{\bibfnamefont{C.}~\bibnamefont{Robilliard} {\it et al.}},
\emph{\bibinfo{Title}{No light shining through a wall}},
\bibinfo{journal}{Phys.  Rev.  Lett.} \textbf{\bibinfo{volume}{99}},
\bibinfo{pages}{190403} (\bibinfo{year}{2007}).

\bibitem{Fouche:2008jk}
\bibinfo{author}{\bibfnamefont{M.}~\bibfnamefont{~Fouche}  {\it et al.}},
\emph{\bibinfo{Title}{Search for photon oscillations into massive particles}},
\bibinfo{journal}{Phys.  Rev.  D.} \textbf{\bibinfo{volume}{78}},
\bibinfo{pages}{032013} (\bibinfo{year}{2008}).

\bibitem{Ehret:2010mh}
\bibinfo{author}{\bibfnamefont{K.}~\bibfnamefont{Ehret}  {\it et al.}  [ALPS collaboration]},
\emph{\bibinfo{Title}{New ALPS Results on Hidden-Sector Lightweights}},
\bibinfo{journal}{Phys. Lett.   B } \pmb{\bibinfo{volume}{689}},
\bibinfo{pages}{149} (\bibinfo{year}{2010}).

\bibitem{Balou}
\bibinfo{author}{\bibfnamefont{R.}~\bibfnamefont{~Balou}  {\it et al.} [OSCAR collaboration]},
\emph{\bibinfo{Title}{New exclusion limits on scalar and pseudoscalar axionlike particles from light shining through a wall}},
\bibinfo{journal}{Phys.  Rev.  D.} \textbf{\bibinfo{volume}{92}},
\bibinfo{pages}{092002} (\bibinfo{year}{2015}).

\bibitem{Witten:1984dg}
\bibinfo{author}{\bibfnamefont{E.}~\bibnamefont{Witten}},
\emph{\bibinfo{Title}{Some Properties of O(32) Superstrings}},
\bibinfo{journal}{ Phys.\ Lett.\ B } \pmb{\bibinfo{volume}{149}},
\bibinfo{pages}{351} (\bibinfo{year}{1984}).

\bibitem{Svrcek:2006yi}
\bibinfo{author}{\bibfnamefont{P.}~\bibnamefont{Svrcek}} \bibnamefont{and}
\bibinfo{author}{\bibfnamefont{E.}~\bibnamefont{Witten}}, 
\emph{\bibinfo{Title}{Axions In String Theory}},
\bibinfo{journal}{JHEP} \pmb{\bibinfo{volume}{06}},
\bibinfo{pages}{051} (\bibinfo{year}{2006}).

\bibitem{Lebedev:2009ag}
\bibinfo{author}{\bibfnamefont{O.} \bibnamefont{Lebedev}}  \bibnamefont{and}
\bibinfo{author}{\bibfnamefont{S.~Ramos}~\bibnamefont{Sanchez}}, 
\emph{\bibinfo{Title}{The NMSSM and String Theory}},
\bibinfo{journal}{ Phys.\ Lett.\ B} \pmb{\bibinfo{volume}{684}},
\bibinfo{pages}{48} (\bibinfo{year}{2010}).

\bibitem{LCicoli:2012sz}
\bibinfo{author}{\bibfnamefont{M.}~\bibnamefont{Cicoli}}, \bibinfo{author}{\bibfnamefont{M.}~\bibnamefont{Goodsell}}   \bibnamefont{and}
\bibinfo{author}{\bibfnamefont{A.}~\bibnamefont{Ringwald}}, 
\emph{\bibinfo{Title}{The type IIB string axiverse and its low-energy phenomenology}},
\bibinfo{journal}{JHEP} \pmb{\bibinfo{volume}{1210}},
\bibinfo{pages}{146} (\bibinfo{year}{2012}).

\bibitem{covi}
\bibinfo{author}{\bibfnamefont{L.}~\bibnamefont{Covi}},
\bibinfo{author}{\bibfnamefont{J.~E.}~\bibnamefont{Kim}} \bibnamefont{and}
\bibinfo{author}{\bibfnamefont{L.}~\bibnamefont{Roszkowski}}, 
\emph{\bibinfo{Title}{Axinos as Cold Dark Matter}},
\bibinfo{journal}{Phys. Rev. Lett.} \pmb{\bibinfo{volume}{82}},
\bibinfo{pages}{4180} (\bibinfo{year}{1999})

\bibitem{Raffelt:2006rj}
\bibinfo{author}{\bibfnamefont{G.~G.}~\bibnamefont{Raffelt}},
\emph{\bibinfo{Title}{Axions: Motivation, limits and searches}},
\bibinfo{journal}{J. Phys. A} \pmb{\bibinfo{volume}{40}},
\bibinfo{pages}{6607} (\bibinfo{year}{2007}).

\bibitem{Duffy:2009ig}
\bibinfo{author}{\bibfnamefont{L.~D.}~\bibnamefont{Duffy}} \bibnamefont{and}
\bibinfo{author}{\bibfnamefont{K.~van}~\bibnamefont{ Bibber}},
\emph{\bibinfo{Title}{Axions as Dark Matter Particles}},
\bibinfo{journal}{New J. Phys.} \pmb{\bibinfo{volume}{11}},
\bibinfo{pages}{105008} (\bibinfo{year}{2009}).

\bibitem{Sikivie:2009fv}
\bibinfo{author}{\bibfnamefont{P.}~\bibnamefont{Sikivie}},
\emph{\bibinfo{Title}{Dark matter axions}},
\bibinfo{journal}{Int. J. Mod. Phys.  A} \pmb{\bibinfo{volume}{25}},
\bibinfo{pages}{554} (\bibinfo{year}{2010}).

\bibitem{Baer:2010wm}
\bibinfo{author}{\bibfnamefont{H.}~\bibnamefont{Baer}},
\bibinfo{author}{\bibfnamefont{A.}~\bibfnamefont{D.}~\bibnamefont{Box}}
\bibnamefont{and}
\bibinfo{author}{\bibfnamefont{H.}~\bibnamefont{Summy}},
\emph{\bibinfo{Title}{Neutralino versus axion/axino cold dark matter in the 19 parameter SUGRA model}},
\bibinfo{journal}{JHEP} \pmb{\bibinfo{volume}{1010}},
\bibinfo{pages}{023} (\bibinfo{year}{2010}).

\bibitem{Raffelt:1985nk}
\bibinfo{author}{\bibfnamefont{G.~G.}~\bibnamefont{Raffelt}},
\emph{\bibinfo{title}{Astrophysical Axion Bounds Diminished By Screening Effects}},
\bibinfo{journal}{Phys. Rev. D} \textbf{\bibinfo{volume}{33}},
\bibinfo{pages}{897} (\bibinfo{year}{1986}).

\bibitem{Raffelt:1999tx}
\bibinfo{author}{\bibfnamefont{G.~G.}~\bibnamefont{Raffelt}},
\emph{\bibinfo{title}{Particle physics from stars}},
\bibinfo{journal}{Ann. Rev. Nucl. Part. Sci.} \textbf{\bibinfo{volume}{49}},
\bibinfo{pages}{163} (\bibinfo{year}{1999}).

\bibitem{Raffelt:2006cw}
\bibinfo{author}{\bibfnamefont{G.~G.}~\bibnamefont{Raffelt}},
\emph{\bibinfo{title}{Astrophysical axion bounds}},
\bibinfo{journal}{Lect. Notes Phys.} \textbf{\bibinfo{volume}{741}},
\bibinfo{pages}{51} (\bibinfo{year}{2008}).

\bibitem{Manchester} 
\bibinfo{author}{\bibfnamefont{R.}~\bibfnamefont{M.}~\bibnamefont{Manchester}},
\bibinfo{author}{\bibfnamefont{G.~B.}~\bibnamefont{Hobbs}}, \bibinfo{author}{\bibfnamefont{A.}~\bibnamefont{Teoh}}
and  \bibinfo{author}{\bibfnamefont{M.}~\bibnamefont{Hobbs}},
\emph{\bibinfo{title}{The Australia Telescope National Facility Pulsar Catalogue}},
\bibinfo{journal}{Astron. J.} \textbf{\bibinfo{volume}{129}}, 
\bibinfo{pages}{1993} (\bibinfo{year}{2005}). 

\bibitem{Kouveliotou}
\bibinfo{author}{\bibfnamefont{C.}~\bibnamefont{Kouveliotou} \emph{et al}},
\emph{\bibinfo{tilte}{An X-ray pulsar with a superstrong magnetic field in the soft gamma-ray}},
\bibinfo{journal}{Nature} \textbf{\bibinfo{volume}{393}},
\bibinfo{pages}{235} (\bibinfo{year}{1998}). 

\bibitem{Bloom}
\bibinfo{author}{\bibfnamefont{J.}~\bibfnamefont{S.}~\bibnamefont{Bloom}}, 
\bibinfo{author}{\bibfnamefont{S.~R.}~\bibnamefont{Kulkarni}}, 
\bibinfo{author}{\bibfnamefont{F.~A.}~\bibnamefont{Harrison}}, 
\bibinfo{author}{\bibfnamefont{T.}~\bibnamefont{Prince}},
\bibinfo{author}{\bibfnamefont{E.~S.}~\bibnamefont{Phinney}} 
and \bibinfo{author}{\bibfnamefont{D.~A.}~\bibnamefont{Frail}},
\emph{\bibinfo{title}{Expected characteristics of the subclass of Supernova Gamma-ray Bursts (S-GRBs)}},
\bibinfo{journal}{Astrophys. J.} \textbf{\bibinfo{volume}{506}}, 
\bibinfo{pages}{L105} (\bibinfo{year}{1998}).

\bibitem{shabad1972} 
\bibinfo{author}{\bibfnamefont{A.~E.~}\bibnamefont{Shabad}}, 
\emph{\bibinfo{title}{Cyclotronic Resonance in the Vacuum Polarization}},
\bibinfo{journal}{Lett. Nuovo  Cimento}, \pmb{\bibinfo{volume}{3}}, %
\bibinfo{pages}{457}, (\bibinfo{year}{1972}).

\bibitem{shabadnat} 
\bibinfo{author}{\bibfnamefont{A.~E.~}\bibnamefont{Shabad}} and
\bibinfo{author}{\bibfnamefont{V.~V.~}\bibnamefont{Usov}}.
\emph{\bibinfo{title}{Gamma-quanta capture by magnetic field and pair creation suppression in pulsars}},
\bibinfo{journal}{Nature (London)}, \pmb{\bibinfo{volume}{295}}, %
\bibinfo{pages}{215}, (\bibinfo{year}{1982}).

\bibitem{shabad3v}
\bibinfo{author}{\bibfnamefont{A.~E.}~\bibnamefont{Shabad}} \bibnamefont{and}
\bibinfo{author}{\bibfnamefont{V.~V.} \bibnamefont{Usov}},
\emph{\bibinfo{title}{Propagation of $\gamma$-radiation in strong magnetic fields of pulsars}},
\bibinfo{journal}{Astrophys. Space Sci.},
\textbf{\bibinfo{volume}{102}},
\bibinfo{pages}{327}, (\bibinfo{year}{1984}).

\bibitem{shabad2004}
\bibinfo{author}{\bibfnamefont{A.~E.}~\bibnamefont{Shabad}},
\emph{\bibinfo{title}{Photon Propagation in a Supercritical Magnetic Field}},
\bibinfo{journal}{Sov. Phys. JETP},
\textbf{\bibinfo{volume}{98}},
\bibinfo{pages}{186}, (\bibinfo{year}{2004}).

\bibitem{Herold}
\bibinfo{author}{\bibfnamefont{H.}~\bibnamefont{Herold}}, 
\bibinfo{author}{\bibfnamefont{H.}~\bibnamefont{Ruder}} and \bibinfo{author}{\bibfnamefont{G.}~\bibnamefont{Wunner}}.
\emph{\bibinfo{title}{Can $\gamma$ quanta Really Be Captured by Pulsar Magnetic Fields?}},
\bibinfo{journal}{Phys. Rev. Lett.}, \pmb{\bibinfo{volume}{54}}, %
\bibinfo{pages}{1452}, (\bibinfo{year}{1985}).

\bibitem{shabadusov1}
\bibinfo{author}{\bibfnamefont{A.~E.}~\bibnamefont{Shabad}} \bibnamefont{and}
\bibinfo{author}{\bibfnamefont{V.~V.} \bibnamefont{Usov}},
\emph{\bibinfo{title}{Gamma-quanta conversion into positronium atoms in a strong magnetic field}},
\bibinfo{journal}{Astrophys. Space Sci.},
\textbf{\bibinfo{volume}{117}},
\bibinfo{pages}{309}, (\bibinfo{year}{1985}).

\bibitem{shabadusov2}
\bibinfo{author}{\bibfnamefont{A.~E.}~\bibnamefont{Shabad}} \bibnamefont{and}
\bibinfo{author}{\bibfnamefont{V.~V.} \bibnamefont{Usov}},
\emph{\bibinfo{title}{Photon dispersion in a strong magnetic field with positronium formation: Theory}},
\bibinfo{journal}{Astrophys. Space Sci.},
\textbf{\bibinfo{volume}{128}},
\bibinfo{pages}{377}, (\bibinfo{year}{1986}).

\bibitem{adler1} 
\bibinfo{author}{\bibfnamefont{S.}~\bibnamefont{L.}~\bibnamefont{Adler}},
\bibinfo{author}{\bibfnamefont{J.~N.}~\bibnamefont{Bahcall}},
\bibinfo{author}{\bibfnamefont{C.~G.}~\bibnamefont{Callan}} and
\bibinfo{author}{\bibfnamefont{M.~N.~}\bibnamefont{Rosenbluth}},
\emph{\bibinfo{title}{Photon Splitting in a Strong Magnetic Field}},
\bibinfo{journal}{Phys. Rev. Lett.} \pmb{\bibinfo{volume}{25}}, 
\bibinfo{pages}{1061} (\bibinfo{year}{1970}). 

\bibitem{adler2} 
\bibinfo{author}{\bibfnamefont{S.}~\bibnamefont{L.}~\bibnamefont{Adler}},
\emph{\bibinfo{title}{Photon splitting and photon dispersion in a strong magnetic field}},
\bibinfo{journal}{ Ann. Phys.} \pmb{\bibinfo{volume}{67}}, \bibinfo{pages}{599} (\bibinfo{year}{1971}). 

\bibitem{adler3} 
\bibinfo{author}{\bibfnamefont{S.}~\bibnamefont{L.}~\bibnamefont{Adler}} and 
\bibinfo{author}{\bibfnamefont{C.}~\bibnamefont{Schubert}},
\emph{\bibinfo{title}{Photon splitting in a strong magnetic field: Recalculation and comparison with previous calculations}},
\bibinfo{journal}{Phys. Rev. Lett.} \pmb{\bibinfo{volume}{77}}, 
\bibinfo{pages}{1695} (\bibinfo{year}{1996}). 

\bibitem{shabad5}
\bibinfo{author}{\bibfnamefont{A.~E.} \bibnamefont{Shabad}} \bibnamefont{and}
\bibinfo{author}{\bibfnamefont{V.~V.} \bibnamefont{Usov}},
\emph{\bibinfo{title}{Modified Coulomb Law in a Strongly Magnetized Vacuum}},
\bibinfo{journal}{Phys, Rev. Lett.} \textbf{\bibinfo{volume}{98}},
\bibinfo{pages}{180403} (\bibinfo{year}{2007}).

\bibitem{shabad6}
\bibinfo{author}{\bibfnamefont{A.~E.} \bibnamefont{Shabad}} \bibnamefont{and}
\bibinfo{author}{\bibfnamefont{V.~V.} \bibnamefont{Usov}},
\emph{\bibinfo{title}{Electric field of a point-like charge in a strong magnetic field and ground state of a hydrogen-like atom}},
\bibinfo{journal}{Phys, Rev. D} \textbf{\bibinfo{volume}{77}},
\bibinfo{pages}{025001} (\bibinfo{year}{2008}).

\bibitem{Adorno2016}
\bibinfo{author}{\bibfnamefont{T.~C.} \bibnamefont{Adorno}}, \bibinfo{author}{\bibfnamefont{D.~M.} \bibnamefont{Gitman}}  \bibnamefont{and}
\bibinfo{author}{\bibfnamefont{A.~E.} \bibnamefont{Shabad}},
\emph{\bibinfo{title}{Coulomb field in a constant electromagnetic background}},
\bibinfo{journal}{Phys, Rev. D} \textbf{\bibinfo{volume}{93}},
\bibinfo{pages}{125031} (\bibinfo{year}{2016}).

\bibitem{Sadooghi:2007ys}
\bibinfo{author}{\bibfnamefont{N.} \bibnamefont{Sadooghi}} \bibnamefont{and}
\bibinfo{author}{\bibfnamefont{A.}~\bibnamefont{Sodeiri} \bibnamefont{Jalili}},
\emph{\bibinfo{title}{New look at the modified Coulomb potential in a strong magnetic field}},
\bibinfo{journal}{Phys, Rev. D.} \textbf{\bibinfo{volume}{76}},
\bibinfo{pages}{065013} (\bibinfo{year}{2007}).

\bibitem{elliott} 
\bibinfo{author}{\bibfnamefont{R.~J.}~\bibnamefont{Elliott}} \bibnamefont{and}
\bibinfo{author}{\bibfnamefont{R.}~\bibnamefont{Loudon}}, 
\emph{\bibinfo{title}{Theory of the absorption edge in semiconductors in a high magnetic field}},
\bibinfo{journal}{J. Phys. Chem. Solids}  \pmb{\bibinfo{volume}{15}},
\bibinfo{pages}{196} (\bibinfo{year}{1960}).

\bibitem{Vysotsky:2010sz}
\bibinfo{author}{\bibfnamefont{M.~I.}~\bibnamefont{Vysotsky}},
\emph{\bibinfo{title}{Atomic levels in superstrong magnetic fields and D=2 QED of massive electrons: Screening}},
\bibinfo{journal}{JETP Lett.} \textbf{\bibinfo{volume}{92}},
\bibinfo{pages}{15} (\bibinfo{year}{2010}).

\bibitem{Machet:2010yg}
\bibinfo{author}{\bibfnamefont{B.~}\bibnamefont{Machet}} \bibnamefont{and}
\bibinfo{author}{\bibfnamefont{M.~I.}~\bibnamefont{Vysotsky}},
\emph{\bibinfo{title}{Modification of Coulomb law and energy levels of the hydrogen atom in a superstrong magnetic field}},
\bibinfo{journal}{Phys, Rev. D.} \textbf{\bibinfo{volume}{83}},
\bibinfo{pages}{025022} (\bibinfo{year}{2011}).

\bibitem{Popov}
\bibinfo{author}{\bibfnamefont{V.~S.}~\bibnamefont{Popov}} \bibnamefont{and}
\bibinfo{author}{\bibfnamefont{B.~M.}~\bibnamefont{Karnakov}},
\emph{\bibinfo{title}{On the spectrum of the hydrogen atom in an ultrastrong magnetic field}},
\bibinfo{journal}{JETP} \textbf{\bibinfo{volume}{114}},
\bibinfo{pages}{1} (\bibinfo{year}{2012}); \bibinfo{journal}{Zh. Eksp. Teor. Fiz.} \textbf{\bibinfo{volume}{141}},
\bibinfo{pages}{5} (\bibinfo{year}{2012}).

\bibitem{thooft} 
\bibinfo{author}{\bibnamefont{G.}~\bibfnamefont{'tHooft}} \bibnamefont{and}
\bibinfo{author}{\bibfnamefont{M.}~\bibnamefont{Veltman}}, 
\emph{\bibinfo{title}{One-loop divergencies  in theory of gravitation}},
\bibinfo{journal}{Ann.\ Inst.\ H.\ Poincare}  \pmb{\bibinfo{volume}{A20}},
\bibinfo{pages}{69} (\bibinfo{year}{1974}).

\bibitem{Stelle:1976gc} 
\bibinfo{author}{\bibnamefont{K.~S.}~\bibfnamefont{Stelle}},
\emph{\bibinfo{title}{Renormalization of Higher Derivative Quantum Gravity}},
\bibinfo{journal}{Phys.\ Rev.\ D }  \pmb{\bibinfo{volume}{16}},
\bibinfo{pages}{953} (\bibinfo{year}{1977}).

\bibitem{Donoghue:1994dn} 
\bibinfo{author}{\bibnamefont{J.~F.}~\bibfnamefont{Donoghue}},
\emph{\bibinfo{title}{General relativity as an effective field theory: The leading quantum corrections}},
\bibinfo{journal}{Phys.\ Rev.\ D }  \pmb{\bibinfo{volume}{50}},
\bibinfo{pages}{3874} (\bibinfo{year}{1994}).

\bibitem{Donoghue:1993eb} 
\bibinfo{author}{\bibnamefont{J.~F.}~\bibfnamefont{Donoghue}},
\emph{\bibinfo{title}{Leading quantum correction to the Newtonian potential}},
\bibinfo{journal}{Phys.\ Rev.\ Lett.\ }  \pmb{\bibinfo{volume}{72}},
\bibinfo{pages}{2996} (\bibinfo{year}{1994}).

\bibitem{s.weinberg} 
\bibinfo{author}{\bibnamefont{S.}~\bibfnamefont{Weinberg}}
\emph{\bibinfo{title}{Phenomenological Lagrangians}},
\bibinfo{journal}{Physica}  \pmb{\bibinfo{volume}{96A}},
\bibinfo{pages}{327} (\bibinfo{year}{1979}).

\bibitem{Gasser:1983yg} 
\bibinfo{author}{\bibnamefont{J.}~\bibfnamefont{Gasser}}  
\bibfnamefont{and}
\bibinfo{author}{\bibfnamefont{H.}~\bibnamefont{Leutwyler}}, 
\emph{\bibinfo{title}{Chiral Perturbation Theory to One Loop}},
\bibinfo{journal}{Annals \ Phys.}  \pmb{\bibinfo{volume}{158}},
\bibinfo{pages}{142} (\bibinfo{year}{1984}).

\bibitem{Gasser:1984gg}
\bibinfo{author}{\bibnamefont{J.}~\bibfnamefont{Gasser}}  
\bibfnamefont{and}
\bibinfo{author}{\bibfnamefont{H.}~\bibnamefont{Leutwyler}}, 
\emph{\bibinfo{title}{Chiral Perturbation Theory: Expansions in the Mass of the Strange Quark}},
\bibinfo{journal}{ Nucl.\ Phys.\ B}  \pmb{\bibinfo{volume}{250}},
\bibinfo{pages}{465} (\bibinfo{year}{1985}).

\bibitem{Ecker:1995zu} 
\bibinfo{author}{\bibnamefont{G.}~\bibfnamefont{Ecker}}
\emph{\bibinfo{title}{Low-energy QCD}},
\bibinfo{journal}{ Prog.\ Part.\ Nucl.\ Phys.\  }  \pmb{\bibinfo{volume}{36}},
\bibinfo{pages}{71} (\bibinfo{year}{1996}).

\bibitem{Halter:1993kj} 
\bibinfo{author}{\bibfnamefont{J.}~\bibnamefont{Halter}}, 
\emph{\bibinfo{title}{An Effective Lagrangian for photons}},
\bibinfo{journal}{Phys.\ Lett.\ B} \pmb{\bibinfo{volume}{316}},
\bibinfo{pages}{155} (\bibinfo{year}{1993}).

\bibitem{Kong:1998ic} 
\bibinfo{author}{\bibfnamefont{X.~W.}~\bibnamefont{Kong}}
\bibnamefont{and}
\bibinfo{author}{\bibfnamefont{F.}~\bibnamefont{Ravndal}}, 
\emph{\bibinfo{title}{Quantum corrections to the QED vacuum energy}},
\bibinfo{journal}{Nucl.\ Phys.\ B} \pmb{\bibinfo{volume}{526}},
\bibinfo{pages}{627} (\bibinfo{year}{1998}).

\bibitem{Dicus:1997ax} 
\bibinfo{author}{\bibfnamefont{D.~A.}~\bibnamefont{Dicus}},
\bibinfo{author}{\bibfnamefont{C.}~\bibnamefont{Kao}}
\bibnamefont{and}
\bibinfo{author}{\bibfnamefont{W.~W.}~\bibnamefont{Repko}}, 
\emph{\bibinfo{title}{Effective Lagrangians and low-energy photon-photon scattering}},
\bibinfo{journal}{Phys.\ Rev.\ D} \pmb{\bibinfo{volume}{57}},
\bibinfo{pages}{2443} (\bibinfo{year}{1998}).

\bibitem{alinapaper}
\bibinfo{author}{\bibfnamefont{S.}~\bibnamefont{Villalba-Ch\'avez}},
\bibinfo{author}{\bibfnamefont{A.}~\bibnamefont{Golub}},
\bibnamefont{and}
\bibinfo{author}{\bibfnamefont{C.}~\bibnamefont{M\"{u}ller}},
\emph{\bibinfo{title}{Axion-modified photon propagator, Coulomb potential and Lamb-shift}},
\bibinfo{journal}{submitted}; arXiv:1806.10940 [hep-ph].

\bibitem{Weinberg:1995mt}
\bibinfo{author}{\bibfnamefont{S.}~\bibnamefont{Weinberg}},
\emph{\bibinfo{book}{The Quantum theory of fields. Vol. 1: Foundations}},
\bibinfo{editor}{Cambridge, UK:  Univ.  Pr.},  (\bibinfo{year}{1995}).
 
\bibitem{Schwartz}
\bibinfo{author}{\bibfnamefont{M.~D.}~\bibnamefont{Schwartz}},
\emph{\bibinfo{book}{Quantum Field Theory and the Standard Model}},
\bibinfo{editor}{Cambridge:  Univ.  Pr.},  (\bibinfo{year}{2014}).

\bibitem{Arzt:1993gz} 
\bibinfo{author}{\bibnamefont{C.}~\bibfnamefont{Arzt}}, 
\emph{\bibinfo{title}{Reduced effective Lagrangians}},
\bibinfo{journal}{Phys.\ Lett.\ B }  \pmb{\bibinfo{volume}{342}},
\bibinfo{pages}{189} (\bibinfo{year}{1995}).

\bibitem{GrosseKnetter:1993td}
\bibinfo{author}{\bibnamefont{C.}~\bibfnamefont{Grosse-Knetter}}, 
\emph{\bibinfo{title}{Effective Lagrangians with higher derivatives and equations of motion}},
\bibinfo{journal}{ Phys.\ Rev.\ D}  \pmb{\bibinfo{volume}{49}},
\bibinfo{pages}{6709} (\bibinfo{year}{1994}).

\bibitem{Georgi:1991ch} 
\bibinfo{author}{\bibnamefont{H.}~\bibfnamefont{Georgi}}
\emph{\bibinfo{title}{On-shell effective field theory}},
\bibinfo{journal}{ Nucl.\ Phys.\ B }  \pmb{\bibinfo{volume}{361}},
\bibinfo{pages}{339} (\bibinfo{year}{1991}).

\bibitem{Weinberg:1996mt}
\bibinfo{author}{\bibfnamefont{S.}~\bibnamefont{Weinberg}},
\emph{\bibinfo{book}{The Quantum theory of fields. Vol. 2: Modern applications}},
\bibinfo{editor}{Cambridge, UK:  Univ.  Pr.},  (\bibinfo{year}{1996}).

\bibitem{batalin}
\bibinfo{author}{\bibfnamefont{I.~A.}~\bibnamefont{Batalin}} \bibnamefont{and}
\bibinfo{author}{\bibfnamefont{A.~E.} \bibnamefont{Shabad}}, \bibinfo{journal}{Zh. Eksp. Teo. Fiz} \textbf{\bibinfo{volume}{60}},
\bibinfo{pages}{894} (\bibinfo{year}{1971}). 
\emph{\bibinfo{title}{Green's function of a photon in a constant homogeneous electromagnetic field of general form.}}
[\bibinfo{journal}{Sov. Phys. JETP} \textbf{\bibinfo{volume}{33}},
\bibinfo{pages}{483} (\bibinfo{year}{1971})].

\bibitem{Shabad:1975ik}
\bibinfo{author}{\bibfnamefont{A.~E.}~\bibnamefont{Shabad}}, 
\emph{\bibinfo{title}{Photon Dispersion in a Strong Magnetic Field,}}
\bibinfo{journal}{ Annals Phys.\ }  \pmb{\bibinfo{volume}{90}},
\bibinfo{pages}{166} (\bibinfo{year}{1975}).

\bibitem{VillalbaChavez:2009ia}
\bibinfo{author}{\bibfnamefont{S.}~\bibnamefont{Villalba-Ch\'avez}}, 
\emph{\bibinfo{title}{Photon Magnetic Moment and Vacuum Magnetization in an Asymptotically Large Magnetic Field}},
\bibinfo{journal}{Phys.\ Rev.\ D }  \pmb{\bibinfo{volume}{81}},
\bibinfo{pages}{105019} (\bibinfo{year}{2010}).
  
\bibitem{VillalbaChavez:2012ea}
\bibinfo{author}{\bibfnamefont{S.}~\bibnamefont{Villalba-Ch\'avez}} \bibnamefont{and}
\bibinfo{author}{\bibfnamefont{A.~E.}~\bibnamefont{Shabad}}, 
\emph{\bibinfo{title}{QED with an external field: Hamiltonian treatment of Lorentz-non-invariant background as an anisotropic medium}},
\bibinfo{journal}{Phys.\ Rev.\ D }  \pmb{\bibinfo{volume}{86}},
\bibinfo{pages}{105040} (\bibinfo{year}{2012}).

\bibitem{PerezRojas:2008nx}
\bibinfo{author}{\bibfnamefont{H.}~\bibnamefont{Perez}~\bibnamefont{Rojas}} \bibnamefont{and}
\bibinfo{author}{\bibfnamefont{E.}~\bibnamefont{Rodriguez}~\bibnamefont{Querts}}, 
\emph{\bibinfo{title}{Is the photon paramagnetic?}},
\bibinfo{journal}{ Phys.\ Rev.\ D}  \pmb{\bibinfo{volume}{79}},
\bibinfo{pages}{093002} (\bibinfo{year}{2009}).

\bibitem{Rojas:2013zga}
\bibinfo{author}{\bibfnamefont{H.}~\bibnamefont{Perez}~\bibnamefont{Rojas}} \bibnamefont{and}
\bibinfo{author}{\bibfnamefont{E.}~\bibnamefont{Rodriguez}~\bibnamefont{Querts}}, 
\emph{\bibinfo{title}{The photon magnetic moment problem revisited}},
\bibinfo{journal}{Eur.\ Phys.\ J.\ C }  \pmb{\bibinfo{volume}{74}},
\bibinfo{pages}{2899} (\bibinfo{year}{2014}).

\bibitem{Borsanyi}
\bibinfo{author}{\bibfnamefont{S.}~\bibnamefont{Borsanyi}}, {\it et al.}, 
\emph{\bibinfo{Title}{Calculation of the axion mass based on high-temperature lattice quantum Chromodynamics}},
\bibinfo{journal}{Nature} \pmb{\bibinfo{volume}{539}},
\bibinfo{pages}{69} (\bibinfo{year}{2016}).

\bibitem{Chakrabarty}
\bibinfo{author}{\bibfnamefont{S.} \bibnamefont{Chakrabarty}},
\bibinfo{author}{\bibfnamefont{D.} \bibnamefont{Bandyopadhyay,}}  \bibnamefont{and}
\bibinfo{author}{\bibfnamefont{S.}~\bibnamefont{Pal}},
\emph{\bibinfo{Title}{Dense Nuclear Matter in a Strong Magnetic Field}},
\bibinfo{journal}{Phys. Rev.  Lett.}  \pmb{\bibinfo{volume}{78}},\bibinfo{pages}{2898} (\bibinfo{year}{1997}).

\bibitem{usovnature}
\bibinfo{author}{\bibfnamefont{V.~V.}~\bibnamefont{Usov}},
\emph{\bibinfo{Title}{Millisecond pulsars with extremely strong magnetic fields as a cosmological source of $\gamma$-ray bursts}},
\bibinfo{journal}{Nature} \pmb{\bibinfo{volume}{357}},
\bibinfo{pages}{472} (\bibinfo{year}{1992}). 

\bibitem{katz}
\bibinfo{author}{\bibfnamefont{J.~I.}~\bibnamefont{Katz}},
\emph{\bibinfo{Title}{Yet Another Model of Gamma-Ray Bursts}},
\bibinfo{journal}{Astrophys. J} \pmb{\bibinfo{volume}{490}},
\bibinfo{pages}{633} (\bibinfo{year}{1997}). 

\bibitem{ruderman}
\bibinfo{author}{\bibfnamefont{M.~A.}~\bibnamefont{Ruderman}},
\bibinfo{author}{\bibfnamefont{L.}~\bibnamefont{Tao}}, \bibnamefont{and}
\bibinfo{author}{\bibfnamefont{W.}~\bibnamefont{Kluzniak}},
\emph{\bibinfo{Title}{A Central Engine for Cosmic Gamma-Ray Burst Sources}},
\bibinfo{journal}{Astrophys. J} \pmb{\bibinfo{volume}{542}},
\bibinfo{pages}{243} (\bibinfo{year}{2000}).

\bibitem{Jaeckel:2015jla}
\bibinfo{author}{\bibfnamefont{J.}~\bibnamefont{Jaeckel}}  
\bibnamefont{and}
\bibinfo{author}{\bibfnamefont{M.}~\bibnamefont{Spannowsky}}, 
\emph{\bibinfo{Title}{Probing MeV to 90 GeV axion-like particles with LEP and LHC}},
\bibinfo{journal}{Phys. Lett. B} \pmb{\bibinfo{volume}{753}},
\bibinfo{pages}{482} (\bibinfo{year}{2016}).

\bibitem{Dobrich:2015jyk}
\bibinfo{author}{\bibfnamefont{B.}~\bibnamefont{D\"{o}brich}}, 
\bibinfo{author}{\bibfnamefont{J.}~\bibnamefont{Jaeckel}},
\bibinfo{author}{\bibfnamefont{F.}~\bibnamefont{Kahlhoefer}},
\bibinfo{author}{\bibfnamefont{A.}~\bibnamefont{Ringwald}},
\bibnamefont{and} 
\bibinfo{author}{\bibfnamefont{K.}~\bibnamefont{Schmidt-Hoberg}}, 
\emph{\bibinfo{Title}{ALPtraum: ALP production in proton beam dump experiment}},
\bibinfo{journal}{JHEP} \pmb{\bibinfo{volume}{1602}},
\bibinfo{pages}{018} (\bibinfo{year}{2016}).

\bibitem{Leinson1}
\bibinfo{author}{\bibfnamefont{L.~B.}~\bibnamefont{Leinson}}, 
\bibnamefont{and} 
\bibinfo{author}{\bibfnamefont{V.~N.}~\bibnamefont{Oraevskii}},
\emph{\bibinfo{Title}{Positronium-photon and photon-positronium quantum transitions in strong magnetic fields}},
\bibinfo{journal}{Sov. J. Nucl. Phys.} \pmb{\bibinfo{volume}{42}},
\bibinfo{pages}{245} (\bibinfo{year}{1985}).

\bibitem{Leinson2}
\bibinfo{author}{\bibfnamefont{L.~B.}~\bibnamefont{Leinson}}, 
\bibnamefont{and} 
\bibinfo{author}{\bibfnamefont{V.~N.}~\bibnamefont{Oraevskii}},
\emph{\bibinfo{Title}{Gamma-positronium in strong magnetic fields}},
\bibinfo{journal}{Phys. Lett. B} \pmb{\bibinfo{volume}{165}},
\bibinfo{pages}{422} (\bibinfo{year}{1985}).


\bibitem{Potekhin}
\bibinfo{author}{\bibfnamefont{A.~Y.}~\bibnamefont{Potekhin}},
\emph{\bibinfo{title}{Atmospheres and radiating surfaces of neutron stars}},
\bibinfo{journal}{Phys. Usp.},
\textbf{\bibinfo{volume}{57:8}},
\bibinfo{pages}{735}, (\bibinfo{year}{2014}).

\bibitem{Marx2011}
\bibinfo{author}{\bibfnamefont{B.}~\bibnamefont{Marx}} \emph{et al.},
\emph{\bibinfo{Title}{Determination of high-purity polarization state of x-rays}},
\bibinfo{journal}{Opt. Commun.} \pmb{\bibinfo{volume}{284}},
\bibinfo{pages}{915} (\bibinfo{year}{2011}).

\bibitem{Marx2013}
\bibinfo{author}{\bibfnamefont{B.}~\bibnamefont{Marx}} \emph{et al.},
\emph{\bibinfo{Title}{High-Precision X-Ray Polarimetry}},
\bibinfo{journal}{Phys. Rev. Lett.} \pmb{\bibinfo{volume}{110}},
\bibinfo{pages}{254801} (\bibinfo{year}{2013}).

\bibitem{HIBEF} See: http://www.hzdr.de/db/Cms?pNid=427$\&$pOid=35325

\bibitem{Schlenvoigt} 
\bibinfo{author}{\bibfnamefont{H.~P.~}\bibnamefont{Schlenvoigt}},
\bibinfo{author}{\bibfnamefont{T.~}\bibnamefont{Heinzl}},
\bibinfo{author}{\bibfnamefont{U.~}\bibnamefont{Schramm}},
\bibinfo{author}{\bibfnamefont{T.~E}~\bibnamefont{Cowan}},
\bibnamefont{and}
\bibinfo{author}{\bibfnamefont{R.~} \bibnamefont{Sauerbrey}},
\emph{\bibinfo{title}{Detecting vacuum birefringence with x-ray free electron lasers and optical laser: feasibility study}},
\bibinfo{journal}{Phys.\ Scr.} \pmb{\bibinfo{volume}{91}},
\bibinfo{pages}{023010} (\bibinfo{year}{2016}).


\bibitem{Soffitta:2013hla}
\bibinfo{author}{\bibfnamefont{P.~}\bibnamefont{Soffita}}, {\it et al.},  
\emph{\bibinfo{title}{XIPE: the X-ray Imaging Polarimetry Explorer}},
\bibinfo{journal}{Exper.\ Astron.} \pmb{\bibinfo{volume}{36}},
\bibinfo{pages}{523} (\bibinfo{year}{2013}).

\bibitem{Bernard:2013jea} 
\bibinfo{author}{\bibfnamefont{D.~}\bibnamefont{Bernard}}, {\it HARPO Collaboration}
\emph{\bibinfo{title}{Polarimetry of cosmic gamma-ray sources above $e^+e^-$ pair creation threshold}},
\bibinfo{journal}{ Nucl.\ Instrum.\ Meth.\ A} \pmb{\bibinfo{volume}{729}},
\bibinfo{pages}{765} (\bibinfo{year}{2013}).

\bibitem{Iwakiri:2016ioo} 
\bibinfo{author}{\bibfnamefont{W.~B.}~\bibnamefont{Iwakiri}}, {\it et al.}
\emph{\bibinfo{title}{Performance of the PRAXyS X-ray Polarimeter}},
\bibinfo{journal}{ Nucl.\ Instrum.\ Meth.\ A} \pmb{\bibinfo{volume}{838}},
\bibinfo{pages}{89} (\bibinfo{year}{2016}).
 
\bibitem{Caiazzo:2018evl}
\bibinfo{author}{\bibfnamefont{I.~}\bibnamefont{Caiazzo}},  
 \bibnamefont{and}
\bibinfo{author}{\bibfnamefont{J.}~\bibnamefont{Heyl}},
\emph{\bibinfo{title}{Vacuum birefringence and the X-ray polarization from black-hole accretion disks}},
\bibinfo{journal}{Phys.\ Rev.\ D} \pmb{\bibinfo{volume}{97}},
\bibinfo{pages}{083001} (\bibinfo{year}{2018}).

\bibitem{Wang:2009sg} 
\bibinfo{author}{\bibfnamefont{C.~}\bibnamefont{Wang}},   
\bibnamefont{and}
\bibinfo{author}{\bibfnamefont{D.}~\bibnamefont{Lai}}, 
\emph{\bibinfo{title}{Polarization Evolution in A Strongly Magnetized Vacuum: QED Effect and Polarized X-ray Emission from Magnetized Neutron Stars}},
\bibinfo{journal}{ Mon.\ Not.\ Roy.\ Astron.\ Soc. } \pmb{\bibinfo{volume}{398}}, 
\bibinfo{pages}{515} (\bibinfo{year}{2009}).

\bibitem{Mignani:2016fwz} 
\bibinfo{author}{\bibfnamefont{R.~P.}~\bibnamefont{Mignani}},  
\bibinfo{author}{\bibfnamefont{V.~}\bibnamefont{Testa}},  
\bibinfo{author}{\bibfnamefont{D.~G.~}\bibnamefont{Caniulef}},  
\bibinfo{author}{\bibfnamefont{R.~}\bibnamefont{Taverna}},  
\bibinfo{author}{\bibfnamefont{R.~}\bibnamefont{Turolla}},  
\bibinfo{author}{\bibfnamefont{S.~}\bibnamefont{Zane}}, \bibnamefont{and}
\bibinfo{author}{\bibfnamefont{K.}~\bibnamefont{Wu}},
\emph{\bibinfo{title}{Evidence for vacuum birefringence from the first optical-polarimetry measurement of the isolated neutron star RX J1856.5−3754}},
\bibinfo{journal}{ Mon.\ Not.\ Roy.\ Astron.\ Soc. } \pmb{\bibinfo{volume}{465}},
\bibinfo{pages}{492} (\bibinfo{year}{2017}).

\bibitem{Gabrielli:2006im}
\bibinfo{author}{\bibfnamefont{E.} \bibnamefont{Gabrielli}},
\bibinfo{author}{\bibfnamefont{K.} \bibnamefont{Huitu}}  \bibnamefont{and}
\bibinfo{author}{\bibfnamefont{S.}~\bibnamefont{Roy}},
\emph{\bibinfo{Title}{Photon propagation in magnetic and electric fields with scalar/pseudoscalar couplings: A New look?}},
\bibinfo{journal}{Phys. Rev.  D.}  \pmb{\bibinfo{volume}{74}},\bibinfo{pages}{073002} (\bibinfo{year}{2006}).

\bibitem{NIST}
\bibinfo{author}{\bibfnamefont{F.~W.~J.}~\bibnamefont{Olver}}, 
\bibinfo{author}{\bibfnamefont{D.~W.}~\bibnamefont{Lozier}}, 
\bibinfo{author}{\bibfnamefont{R.~F.}~\bibnamefont{Boisvert}}, 
\bibnamefont{and}
\bibinfo{author}{\bibfnamefont{C.~W.}~\bibnamefont{Clark}},
\emph{\bibinfo{title}{NIST Handbook of Mathematical Functions}},
\bibinfo{editor}{Cambridge University Press, England},  (\bibinfo{year}{2010}).

\bibitem{Gradshteyn}
\bibinfo{author}{\bibfnamefont{I.}~\bibfnamefont{S.}~\bibnamefont{Gradshteyn}}
\bibnamefont{and}
\bibinfo{author}{\bibfnamefont{I.}~\bibfnamefont{M.}~\bibnamefont{Ryzhik}},
\emph{\bibinfo{book}{Table of Integrals, Series and Products}},
\textrm{Seventh Edition}, \bibinfo{editor}{Elsevier}, San Diego, (\bibinfo{year}{2007}).

\bibitem{Jentschura:2014yla}
\bibinfo{author}{\bibfnamefont{U.~D.} \bibnamefont{Jentschura}},
\emph{\bibinfo{title}{Muonic bound systems, virtual particles and proton radius}},
\bibinfo{journal}{Phys.\ Rev.\ A} \textbf{\bibinfo{volume}{92}},
\bibinfo{pages}{012123} (\bibinfo{year}{2015}).

\bibitem{skobelev}
\bibinfo{author}{\bibfnamefont{V.~V.}~\bibnamefont{Skobelev}},
\bibinfo{journal}{Izv. Vyssh. Uchebn. Zaved. Fiz. No.} \pmb{\bibinfo{volume}{10}},
\bibinfo{pages}{142} (\bibinfo{year}{1975}).

\bibitem{shabad-lebedev}
\bibinfo{author}{\bibfnamefont{A.~E.}~\bibnamefont{Shabad}},
\bibinfo{journal}{Kratkie Soobshcheniapo  Fizike (Sov. Phys. - Lebedev Inst. Reps.)} \pmb{\bibinfo{volume}{3}},
\bibinfo{pages}{13} (\bibinfo{year}{1976}).

\bibitem{melrose} 
\bibinfo{author}{\bibfnamefont{D.~B.}~\bibnamefont{Melrose}},
\bibnamefont{and}
\bibinfo{author}{\bibfnamefont{R.~J.}~\bibnamefont{Stoneham}}, 
\emph{\bibinfo{title}{Vacuum polarization and photon propagation in a magnetic field}},
\bibinfo{journal}{Nuovo Cimento A} \pmb{\bibinfo{volume}{32}},
\bibinfo{pages}{435} (\bibinfo{year}{1976}).

\bibitem{heyl} 
\bibinfo{author}{\bibfnamefont{J.~S.}~\bibnamefont{Heyl}},
\bibnamefont{and}
\bibinfo{author}{\bibfnamefont{L.}~\bibnamefont{Hernquist}},
\emph{\bibinfo{Title}{Birefringence and dichroism of the QED vacuum}},
\bibinfo{journal}{J. Phys. A} \pmb{\bibinfo{volume}{30}},
\bibinfo{pages}{6485} (\bibinfo{year}{1997}).

\end{thebibliography}
\end{document}